\documentclass{aa}
\usepackage{textcomp}
\usepackage{graphicx} %
\usepackage{amsmath} %
\usepackage{amssymb} %
\usepackage{bm} %
\usepackage{upgreek} %
\usepackage{IEEEtrantools} %
\usepackage{multirow}
\usepackage[dvipsnames]{xcolor}
\usepackage[colorlinks=true,allcolors=blue,urlcolor=blue]{hyperref}
\usepackage{txfonts}
\usepackage[normalem]{ulem} 

\newcommand{\sect}[1]{\text{Sect.~\ref{#1}}}
\newcommand{\fig}[1]{\text{Fig.~\ref{#1}}}
\newcommand{\tab}[1]{\text{Table~\ref{#1}}}

\newcommand{\mtd}{\textlangle3D\textrangle}
\newcommand{\multitd}{\texttt{Multi3D}}
\newcommand{\balder}{\texttt{Balder}}

\newcommand{\marcs}{\texttt{MARCS}}
\newcommand{\stagger}{\texttt{STAGGER}}

\newcommand{\atmo}{\texttt{ATMO}}

\newcommand{\gbs}{\texttt{GBS}}
\newcommand{\sme}{\texttt{SME}}
\newcommand{\turbo}{\texttt{Turbospectrum}}
\newcommand{\synspec}{\texttt{Synspec}}

\newcommand{\kelvin}{\mathrm{K}}
\newcommand{\kms}{\mathrm{km\,s^{-1}}}
\newcommand{\teff}{T_{\mathrm{eff}}}
\newcommand{\lgg}{\log{g}}
\newcommand{\lggu}{\log{g / \mathrm{cm\,s^{-2}}}}

\newcommand{\lggf}{\log{gf}}
\newcommand{\feh}{\mathrm{\left[Fe/H\right]}}

\newcommand{\xfe}[1]{\mathrm{\left[#1/Fe\right]}}
\newcommand{\xm}[1]{\mathrm{\left[#1/M\right]}}
\newcommand{\xh}[1]{\mathrm{\left[#1/H\right]}}
\newcommand{\lgeps}[1]{A\left({\mathrm{#1}}\right)}

\newcommand{\dex}{\mathrm{dex}}

\newcommand{\nm}{\mathrm{nm}}
\newcommand{\K}{\mathrm{K}}

\newcommand{\eV}{\mathrm{eV}}
\newcommand{\vmic}{\xi_{\text{mic;1D}}}

\newcommand{\reqw}{\log W/\lambda}
\newcommand{\elo}{E_{\text{low}}}
\newcommand{\eup}{E_{\text{up}}}
\newcommand{\abdl}{\Delta A_{\text{1L-3N}}}
\newcommand{\abdn}{\Delta A_{\text{1N-3N}}}

\begin{document} 

\title{3D non-LTE iron abundances in FG-type dwarfs}
\author{A.~M.~Amarsi\inst{\ref{uu1}}
\and
S.~Liljegren\inst{\ref{su}}
\and
P.~E.~Nissen\inst{\ref{sac}}}
\institute{\label{uu1}Theoretical Astrophysics, 
Department of Physics and Astronomy,
Uppsala University, Box 516, SE-751 20 Uppsala, Sweden\\
\email{anish.amarsi@physics.uu.se} 
\and
\label{su}The Oskar Klein Centre, Department of Astronomy, 
Stockholm University, Albanova 10691, Stockholm, Sweden
\and 
\label{sac}Stellar Astrophysics Centre,
Department of Physics and Astronomy, Aarhus
University, Ny Munkegade 120, DK-8000 Aarhus C, Denmark}

\abstract{Iron is one of the
most important elements in stellar astrophysics.
However, spectroscopic measurements of its abundance are prone to
systematic modelling errors.  We present 
three dimensional non-local thermodynamic equilibrium
(3D non-LTE) calculations across $32$ 
\stagger{}-grid models with effective temperatures from $5000\,\K$ to
$6500\,\K$, surface gravities of $4.0\,\dex$ 
and $4.5\,\dex$, and metallicities from
$-3\,\dex$ to $0\,\dex$, and we study the effects on $171$ \ion{Fe}{I} and $12$
\ion{Fe}{II} optical lines.  
In warm metal-poor stars,
the 3D non-LTE abundances are up to $0.5\,\dex$
larger than 1D LTE abundances inferred from
\ion{Fe}{I} lines of an intermediate
excitation potential.
In contrast, the 3D non-LTE abundances can be $0.2\,\dex$ smaller
in cool metal-poor stars when using \ion{Fe}{I} lines of
a low excitation potential.
The corresponding abundance differences between 3D non-LTE and 
1D non-LTE are generally less severe but can still reach $\pm0.2\,\dex$.
For \ion{Fe}{II} lines, the 
3D abundances range from up to $0.15\,\dex$ larger
to $0.10\,\dex$ smaller than 1D abundances,
with negligible departures from 3D LTE except for the warmest
stars at the lowest metallicities.  The results were used to correct 1D LTE
abundances of the Sun and Procyon (HD~61421), and of the metal-poor stars
HD~84937 and HD~140283, using an interpolation routine based on neural
networks.  
The 3D non-LTE models achieve an improved ionisation balance
in all four stars.  In the two metal-poor stars, they removed excitation
imbalances amounting to $250\,\K$ to $300\,\K$ errors in effective
temperature.  For Procyon, the 3D non-LTE models suggest $\feh=0.11\pm0.03$,
which is significantly larger than literature values based on simpler models.
We make the 3D non-LTE interpolation routine for FG-type dwarfs publicly
available, in addition to 1D non-LTE departure coefficients for standard
\marcs{} models of FGKM-type dwarfs and giants.  These tools, together with an
extended 3D LTE grid for \ion{Fe}{II} from 2019, can help improve the accuracy
of stellar parameter and iron abundance determinations for late-type stars.}

\keywords{atomic processes --- radiative transfer --- line: formation --- 
stars: atmospheres --- stars: fundamental parameters --- stars: abundances}

\date{Received 19 July 2022 / Accepted 26 September 2022}
\maketitle
\section{Introduction}
\label{introduction}

Iron is one of the most discussed elements in astrophysics.
It is the seventh most abundant element in the Sun
\citep{2021A&A...653A.141A}
and an important source of opacity 
throughout the solar interior \citep{2015Natur.517...56B,2015ApJS..220....2M}.
Its high cosmic abundance and open $3d$ shell 
conspire to make iron dominate
the spectra of FG-type dwarfs 
such as the Sun, and so it is
the usual proxy of
the stellar metal mass fraction ($Z$).
It is also a precise and illuminating tracer 
of the evolution of galaxies, 
being formed copiously by both core-collapse supernovae on short timescales,
and Type Ia supernovae at later cosmic epochs
\citep{2017ApJ...848...25M}.
In spectroscopic surveys, including extremely large ones 
such as APOGEE \citep{2020ApJS..249....3A}
and GALAH \citep{2021MNRAS.506..150B},
iron is often used to infer other fundamental stellar 
parameters such as the effective temperature 
($\teff$) and surface gravity ($\lgg$),
via excitation and ionisation equilibria
of \ion{Fe}{I} and \ion{Fe}{II} lines
\citep{2013ApJ...769...57F,2013MNRAS.429..126R,
2013A&A...555A.150T,2017ApJ...847..142E,2022arXiv220709415L}.

However, measurements of iron abundances
($\lgeps{Fe}$\footnote{$\lgeps{Fe}=\log_{10}
\left(N_{\mathrm{Fe}}/N_{\mathrm{H}}\right)+12$} 
or $\feh$\footnote{$\feh=\lgeps{Fe}_{\text{star}}-\lgeps{Fe}_{\text{Sun}}$})
can suffer from systematic modelling errors. 
Most spectroscopic analyses carried out today do the following: a) employ 
one-dimensional (1D) hydrostatic model stellar atmospheres that 
are necessarily in radiative equilibrium at the upper boundary;
and b) invoke the assumption of local thermodynamic equilibrium (LTE).  
Both assumptions can alter \ion{Fe}{I} and \ion{Fe}{II} line strengths
\citep{2001ApJ...550..970S,2012A&A...547A..46H,2013A&A...558A..20H}.
The quantitative 1D LTE systematic errors typically depend on the
excitation potential ($\elo$), and are different for 
\ion{Fe}{I} and \ion{Fe}{II}. Thus they
may introduce errors in spectroscopic determinations of
$\teff$ and $\lgg$ as well.

The 1D LTE systematic errors 
are particularly prominent for warm, metal-poor stars,
where $\lgeps{Fe}$ determinations are usually underestimated
\citep{2016MNRAS.463.1518A,2017A&A...597A...6N}.
Here, the reduced metal opacity results in
a larger UV flux escaping the deeper photosphere,
which drives a severe over-ionisation of \ion{Fe}{I}
and a slight over-excitation of \ion{Fe}{II}
\citep{2005ApJ...618..939S}.
This non-LTE effect is enhanced by the steeper temperature
gradients within metal-poor 3D model stellar atmospheres
as efficient adiabatic cooling
makes them much cooler in the upper layers
than what is expected from 1D models that are close to
radiative equilibrium
\citep{2006ApJ...644L.121C,2011A&A...528A..32C}.

In the last decade, 3D model stellar atmospheres
combined with 3D non-LTE post-processing radiative transfer calculations
have increasingly been used 
for spectroscopic abundance determinations of different elements.
For iron this approach has not been utilised much,
other than in studies of the Sun
\citep{2017MNRAS.468.4311L,2021A&A...653A.141A}
and of a small number of metal-poor stars
\citep{2016MNRAS.463.1518A,2017A&A...597A...6N}.

Here we describe recent 3D non-LTE calculations for iron
for FG-type dwarfs (\sect{method}).
To our knowledge, this is the first grid of its kind.
We briefly explore the systematic effects
on 1D LTE and 1D non-LTE measurements of $\lgeps{Fe}$ across parameter space
(\sect{results}),
focussing on the $171$ \ion{Fe}{I} 
and $12$ \ion{Fe}{II} optical lines 
used in the analysis of the \textit{Gaia} 
benchmark stars
(\gbs{}; \citealt{2014A&A...564A.133J}). We present a
routine based on neural networks
that interpolates the predicted abundance differences
(\sect{interpolator}),
and we describe the impact on four standard stars:
the Sun, Procyon (HD~61421), HD~84937, and HD~140283 (\sect{gbs}).
We end with a summary of our overall conclusions,
and provide grids of 1D non-LTE departure
coefficients for FGKM-type dwarfs and 
giants\footnote{\url{https://zenodo.org/record/7088951}}
as well as the interpolation routine for 1D LTE versus 3D non-LTE abundance
differences\footnote{\url{https://github.com/sliljegren/1L-3NErrors}} 
which is publicly available (\sect{conclusion}).

\section{Method}
\label{method}

\begin{table*}
\begin{center}
    \caption{Inelastic collision cross sections included in the model atom.}
\label{tab:collisions}
    \begin{tabular}{l l l r}
\hline
\hline
\noalign{\smallskip}
        \multicolumn{1}{l}{Process} & 
        \multicolumn{1}{l}{Equation} & 
        \multicolumn{1}{l}{Method} & 
        \multicolumn{1}{l}{Ref.} \\
\noalign{\smallskip}
\hline
\hline
\noalign{\smallskip}
        \multirow{2}{*}{Electron impact excitation} &
        $\mathrm{\ion{Fe}{I}}(i)+\mathrm{e^{-}}
        \leftrightarrow\mathrm{\ion{Fe}{I}}(j)+\mathrm{e^{-}}$ & 
        $B$-spline $R$-matrix &
        a \\
        &
        $\mathrm{\ion{Fe}{II}}(i)+\mathrm{e^{-}}
        \leftrightarrow\mathrm{\ion{Fe}{II}}(j)+\mathrm{e^{-}}$ & 
        $R$-matrix & b, c \\
\noalign{\smallskip}
        \multirow{2}{*}{Electron impact ionisation} &
        $\mathrm{\ion{Fe}{I}}(i)+\mathrm{e^{-}}
        \leftrightarrow\mathrm{\ion{Fe}{II}}(j)+2\mathrm{e^{-}}$ & 
        \multirow{2}{*}{Semi-empirical} & \multirow{2}{*}{d} \\
        &
        $\mathrm{\ion{Fe}{II}}(i)+\mathrm{e^{-}}
        \leftrightarrow\mathrm{\ion{Fe}{III}}(j)+2\mathrm{e^{-}}$ & 
        & \\
\noalign{\smallskip}
        \multirow{4}{*}{Hydrogen impact excitation} &
        \multirow{2}{*}{$\mathrm{\ion{Fe}{I}}(i)+\mathrm{H(n=1)}
        \leftrightarrow\mathrm{\ion{Fe}{I}}(j)+\mathrm{H(n=1)}$} & 
        Asymptotic model (LCAO) & e \\
        &
        &
        Free electron model & f \\
        &
        \multirow{2}{*}{$\mathrm{\ion{Fe}{II}}(i)+\mathrm{H(n=1)}
        \leftrightarrow\mathrm{\ion{Fe}{II}}(j)+\mathrm{H(n=1)}$} & 
        Asymptotic model (Simplified) & g \\
        &
        &
        Drawin & h \\
\noalign{\smallskip}
        \multirow{2}{*}{Charge transfer} & 
        $\mathrm{\ion{Fe}{I}}(i)+\mathrm{H(n=1)}
        \leftrightarrow\mathrm{\ion{Fe}{II}}(j)+\mathrm{H^{-}}$ & 
        Asymptotic model (LCAO) & e \\
        &
        $\mathrm{\ion{Fe}{II}}(i)+\mathrm{H(n=1)}
        \rightarrow\mathrm{\ion{Fe}{III}}(j)+\mathrm{H^{-}}$ & 
        Asymptotic model (Simplified) & g \\
\noalign{\smallskip}
\hline
\hline
\end{tabular}
\end{center}
    \tablebib{(a)~\citet{2018ApJ...867...63W};
    (b)~\citet{2015ApJ...808..174B};
    (c)~\citet{1995A&A...293..953Z};
    (d)~\citet{1973asqu.book.....A};
    (e)~\citet{2018A&A...612A..90B};
    (f)~\citet{1991JPhB...24L.127K};
    (g)~\citet{2019MNRAS.483.5105Y};
    (h)~\citet{1993PhST...47..186L}}
\end{table*}

\begin{figure}
    \begin{center}
        \includegraphics[scale=0.325]{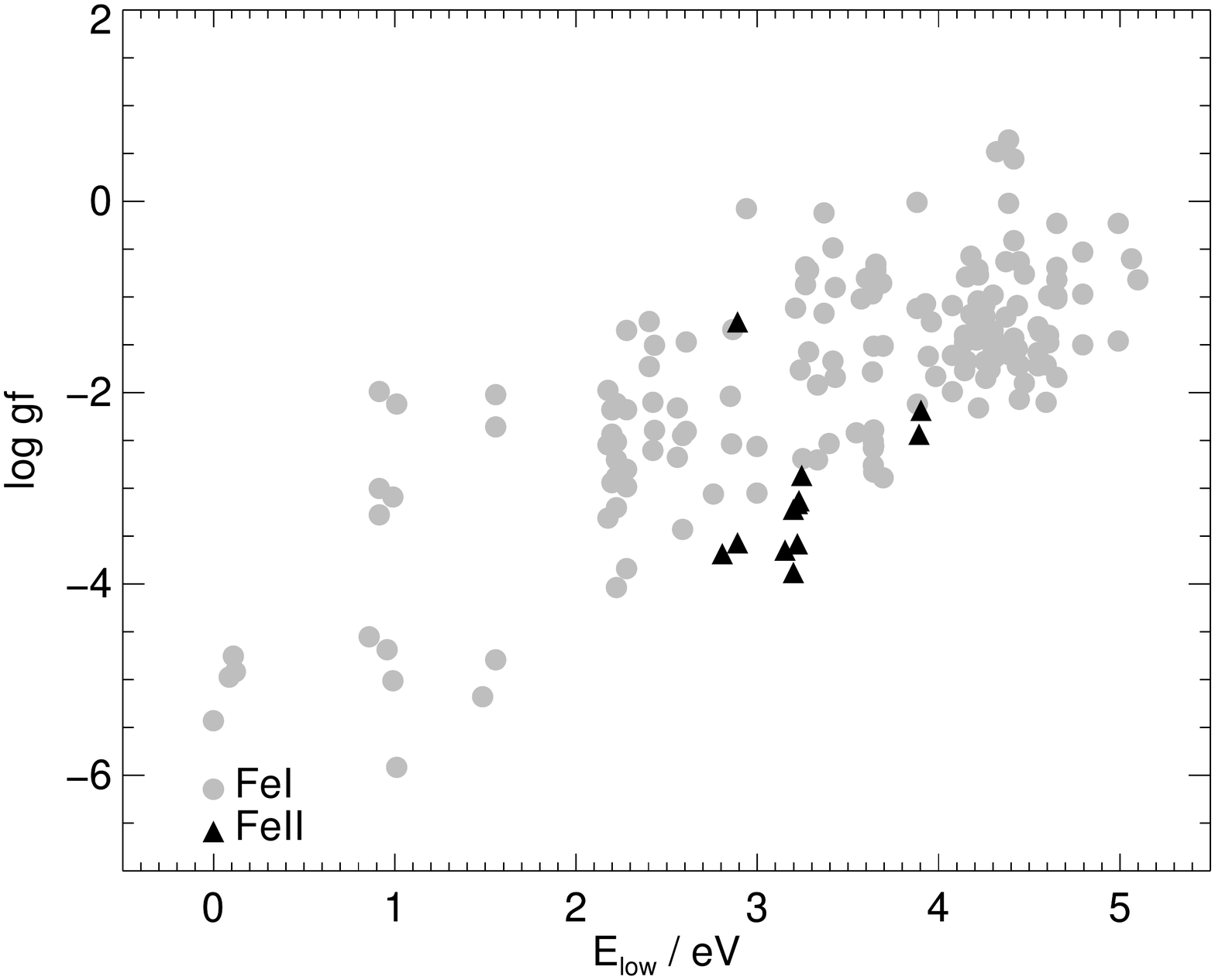}
        \caption{$\elo$ and $\lggf$
        for the \citet{2014A&A...564A.133J}
        set of \ion{Fe}{I} and \ion{Fe}{II} lines
        considered in this study.}
        \label{fig:golden}
    \end{center}
\end{figure}

\begin{figure*}
    \begin{center}
        \includegraphics[scale=0.325]{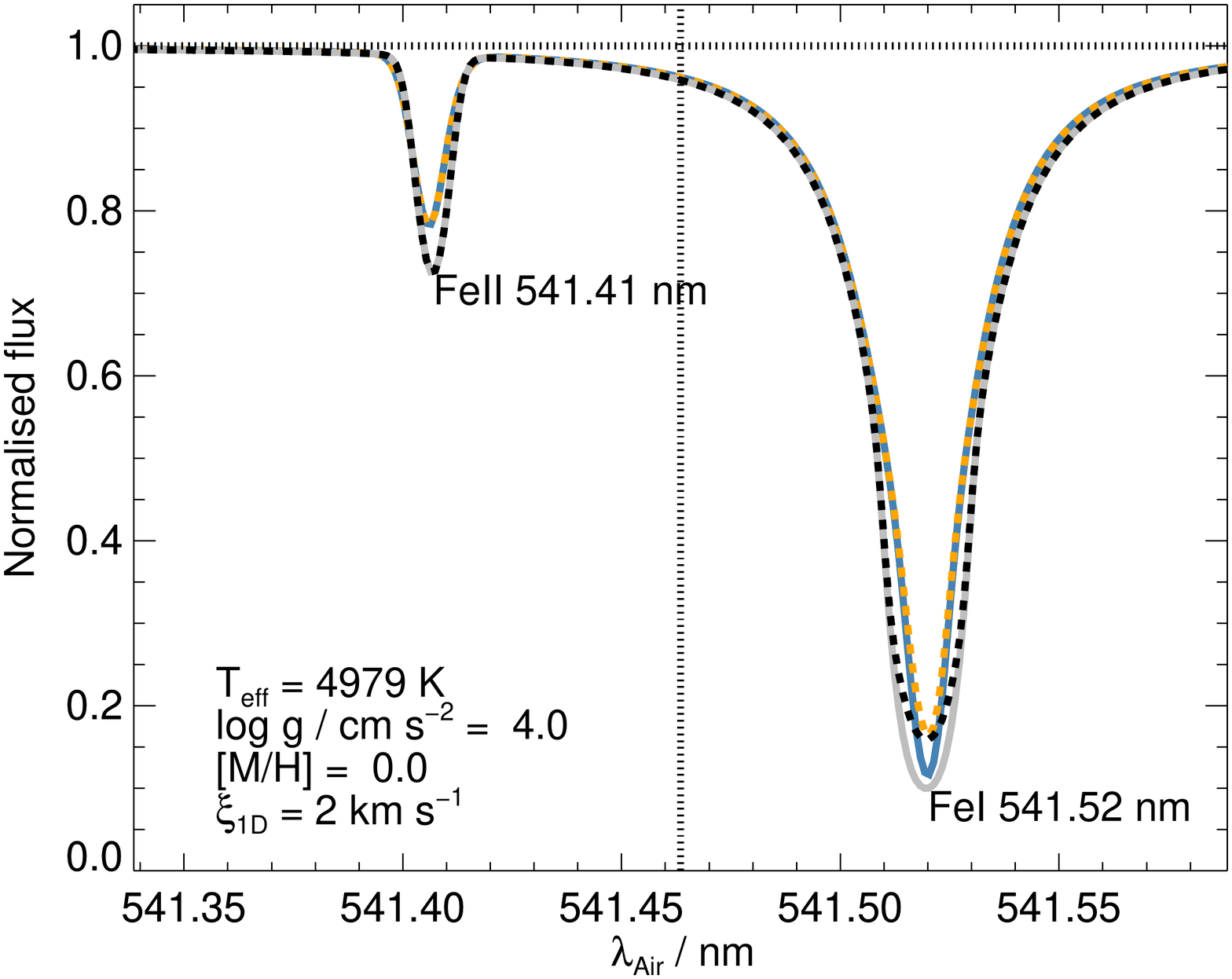}\includegraphics[scale=0.325]{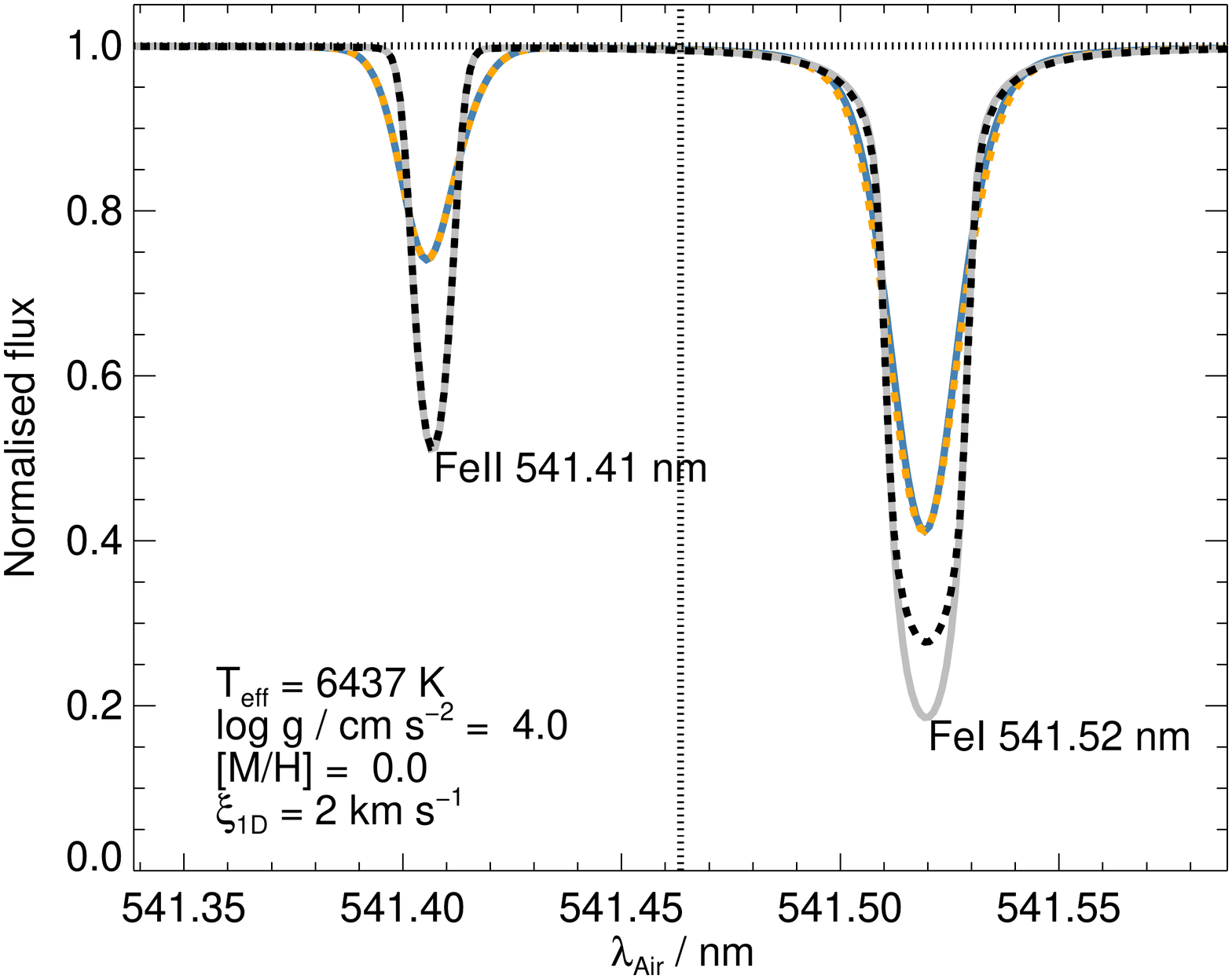}
        \includegraphics[scale=0.325]{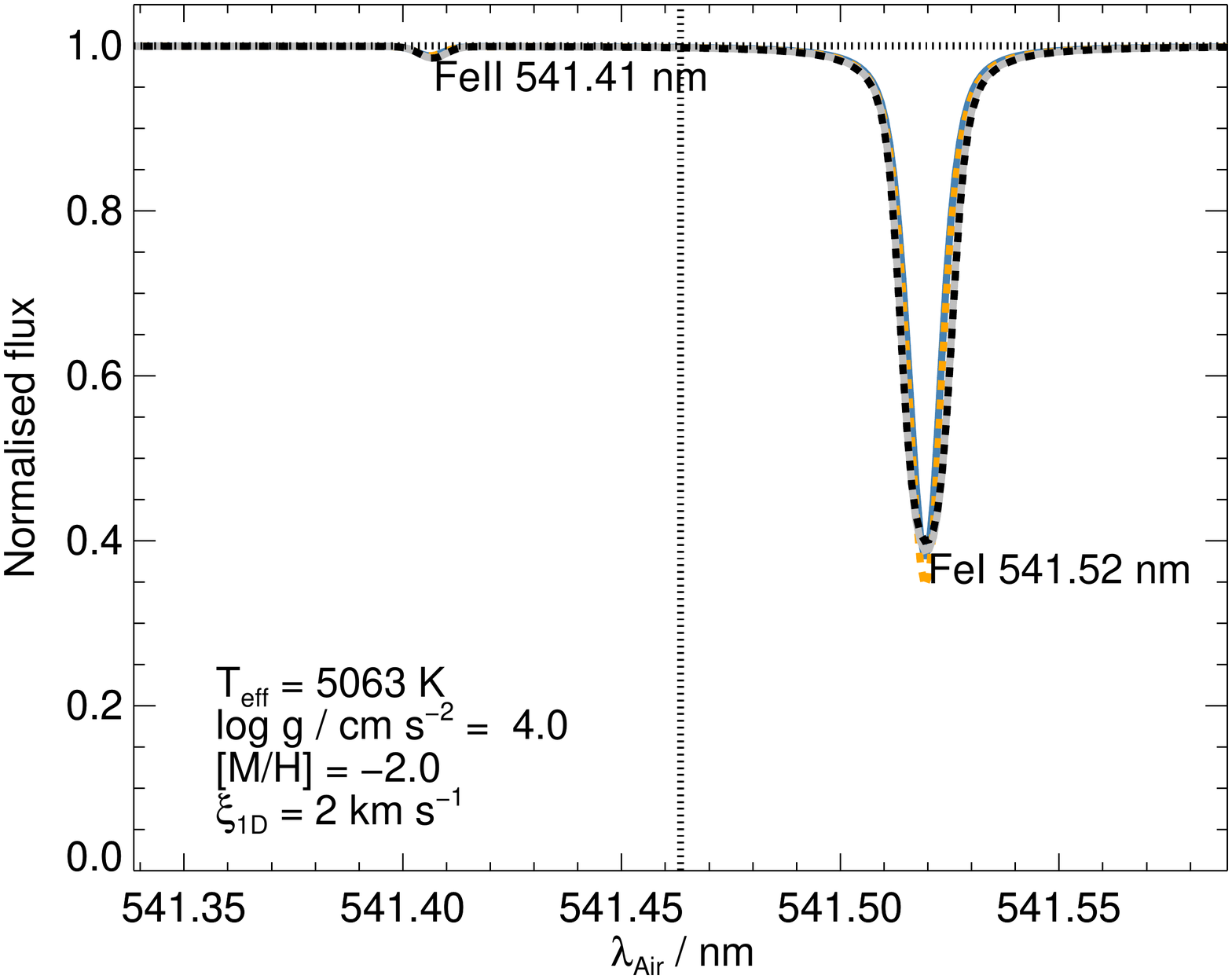}\includegraphics[scale=0.325]{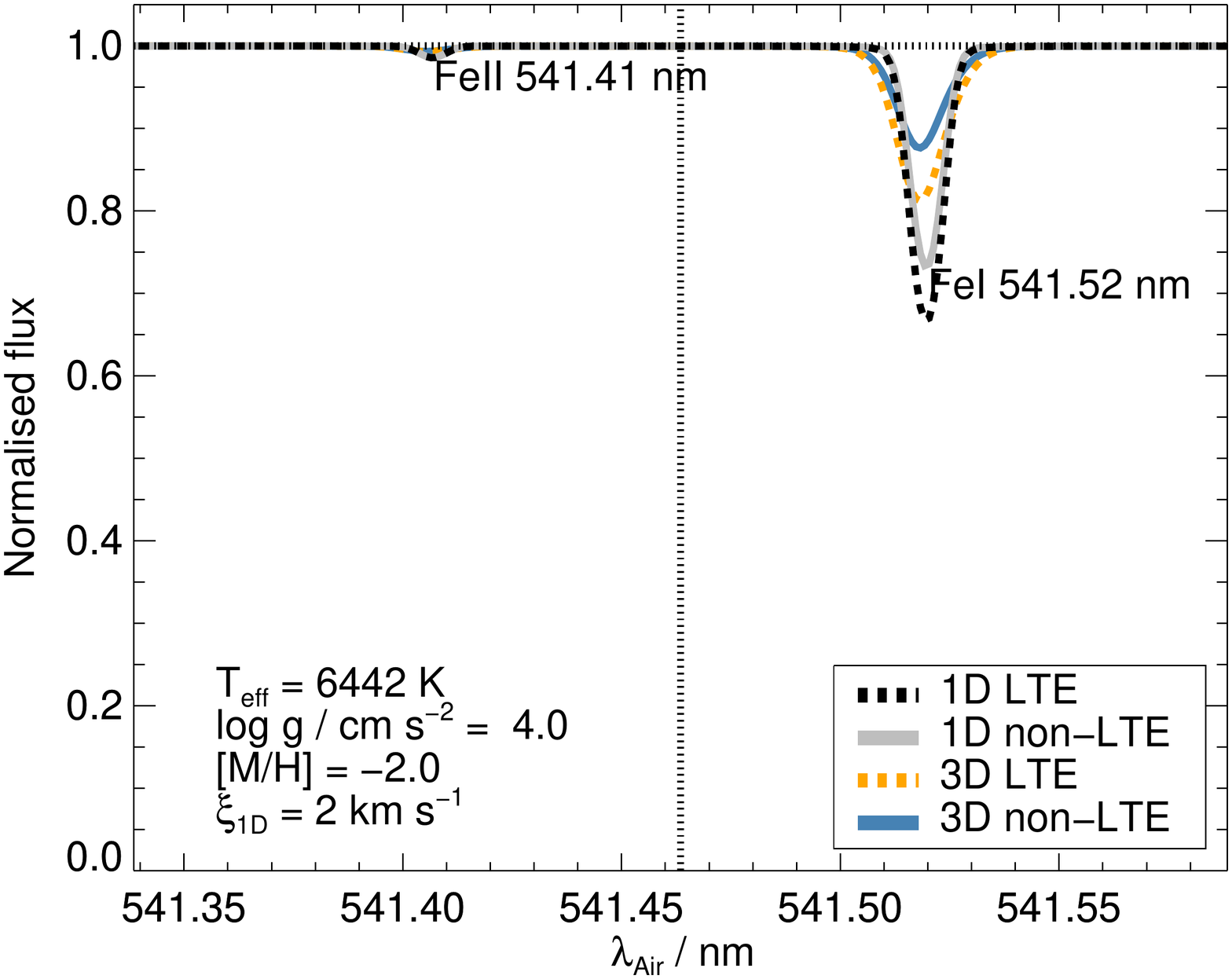}
        \caption{Synthetic spectra in LTE (dashed) and 
        non-LTE (solid) from 1D \atmo{} and 3D \stagger{} models with
        different $\teff$ (columns)
        and $\xh{M}$ (rows).
        The 1D spectra shown here
        were calculated with $\vmic=2\,\kms$ and without
        any macroturbulent broadening.
        An integration limit is illustrated as a vertical dashed line.}
        \label{fig:spectra}
    \end{center}
\end{figure*}

\subsection{Model stellar atmospheres}
\label{method_atmosphere}

Post-processing radiative transfer 
calculations were performed across a subset of the \stagger{} grid
of 3D hydrodynamic model stellar atmospheres
\citep{2013A&A...557A..26M}.  The selected models
have four values of $\teff$ between $5000\,\kelvin$ and $6500\,\kelvin$,
two values of $\lgg$ of
$4.0\,\dex$ and $4.5\,\dex$ (in units of logarithmic $\mathrm{cm\,s^{-2}}$),
and four values of 
$\feh$ between $-3.0\,\dex$ and $0.0\,\dex$.
The models were computed with the solar chemical composition of 
\citet{2009ARA&A..47..481A}, scaled by $\feh$,
and with an enhancement to the canonical $\upalpha$ elements
(O, Ne, Mg, Si, S, Ar, Ca, and Ti)
of $+0.4\,\dex$ for $\feh\leq-1.0$.
To avoid confusion in the subsequent discussion, 
$\xh{M}$ is used in place of $\feh$ 
to represent the overall metallicity of the model stellar atmospheres.

Prior to carrying out the calculations in this work,
the models were trimmed and re-gridded to both reduce the horizontal
resolution (to speed up the calculations),
and to increase the resolution in the steep
continuum-forming regions (to improve accuracy),
in the manner described in Sect.~2.1.1 of \citet{2018A&A...615A.139A}.
The calculations were carried out on five snapshots of
each model.

Post-processing radiative transfer
calculations were also performed on two families of
1D hydrostatic model stellar atmospheres:
\atmo{} models (\citealt{2013A&A...557A..26M}, Appendix A)
and an extended grid of \marcs{} models \citep{2008A&A...486..951G}.
The \atmo{} models adopt the same equation of state, raw opacity data,
opacity bins, and formal solver 
as their companion 3D hydrodynamic \stagger{} models.
These calculations permit a differential 
determination of 3D versus 1D effects.
The calculations on \marcs{} models were only
used for the analysis of Procyon (HD~61421; \sect{gbs_analysis}),
and also to generate a grid of 1D non-LTE departure coefficients
to be made publicly available (\sect{conclusion}).
The grid extends over dwarfs and giants of spectral types
FGKM; further details \citet{2020A&A...642A..62A}.

\subsection{Model atom}
\label{method_atom}

The model atom is based on the model described by
\citet{2017MNRAS.468.4311L}, 
with a number of updates and modifications outlined
in \citet{2021A&A...653A.141A}.
These updates are summarised here for completeness.

The energies of $LSJ$ levels and the 
bound-bound radiative transition probabilities 
($\lggf$) were taken from
the compilation of \citet{2017MNRAS.468.4311L}.
Natural line broadening parameters were calculated here consistently
with these radiative data.
Pressure broadening parameters
due to elastic hydrogen collisions
were computed by interpolating the tables
of \citet{1995MNRAS.276..859A},
\citet{1997MNRAS.290..102B}, and
\citet{1998MNRAS.296.1057B}.
Bound-free radiative transition probabilities were taken from
\citet{2017A&A...606A.127B}, based on the R-matrix method. 
We note that the B-spline R-matrix
calculations of \citet{2017A&A...606A.127B}
are likely to be superior and should be
adopted in future work.  
Inelastic collision cross sections remain a significant source
of uncertainty \citep{2019A&A...631A..43M}.
These were taken from a number of different studies,
as summarised in \tab{tab:collisions}.

The complete data set is too large to permit 3D non-LTE calculations
on a reasonable timescale: it consists of
$2980$ $LSJ$ levels of \ion{Fe}{I} (up to $0.003\,\eV$ below the
ionisation limit),
$2880$ $LSJ$ levels of \ion{Fe}{II} (up to $0.007\,\eV$ below the
ionisation limit),
plus the ground state of \ion{Fe}{III}.
To make the calculations tractable,
fine structure $LSJ$ levels were collapsed into $LS$ terms. Moreover,
since highly excited states are rather efficiently coupled via 
inelastic hydrogen collisions 
in our model (according to the description of
\citealt{1985JPhB...18L.167K,kaulakys1986free,1991JPhB...24L.127K}),
levels with energies
greater than $6\,\mathrm{eV}$ above the ground state in \ion{Fe}{I} and 
$8.7\,\mathrm{eV}$ above the ground state in \ion{Fe}{II},
as well as within the same spin state were collapsed into super-levels
with widths of up to $\pm0.25\,\eV$ for \ion{Fe}{I}
and $\pm2.5\,\eV$ for \ion{Fe}{II}
following the same approach tested for silicon \citep{2017MNRAS.464..264A},
carbon and oxygen \citep{2019A&A...630A.104A},
nitrogen \citep{2020A&A...636A.120A},
and magnesium and calcium \citep{2021A&A...653A.141A}.
This resulted in $15$ super-levels for \ion{Fe}{I}
and
seven super-levels for \ion{Fe}{II}.
The highest \ion{Fe}{I} super-level is $0.1\,\eV$ below the ionisation limit,
which previous studies suggest is sufficient for providing a realistic
collisional coupling to \ion{Fe}{II} \citep{2011A&A...528A..87M}.
Lines involving collapsed levels were also
collapsed into super-lines \citep{2017MNRAS.464..264A}.
The reduced atom consists of $177$ $LS$ levels and super-levels,
of which $100$ are
made up of \ion{Fe}{I} and $76$ are 
made up of \ion{Fe}{II}.

\subsection{Non-LTE radiative transfer in 3D and 1D}
\label{method_code}

The post-processing radiative transfer calculations were performed
with \balder{} \citep{2018A&A...615A.139A}.
This code has its roots in \multitd{} 
\citep{2009ASPC..415...87L_short}
with updates in particular to the opacity package
and formal solver \citep{2016MNRAS.463.1518A,2019A&A...624A.111A}.
Continuum Rayleigh scattering from hydrogen as well as Thomson
scattering from electrons were included
as described in \citet{2020A&A...642A..62A},
but, in contrast to that work, background lines
were treated in pure absorption. 
The calculations employed the $8$-point Lobatto 
quadrature over $\mu=\cos{\theta}$ 
and the $4$-point trapezoidal integration over $\phi$
across the unit sphere,
as described in Sect.~2.2 of \citet{2018A&A...615A.139A}
(see also \citealt{2017MNRAS.464..264A}, Sect.~2.1).
The emergent synthetic spectra (\sect{method_abundance})
were based on a formal solution 
using the $7$-point Lobatto quadrature over $\mu=\cos{\theta}$ 
and $8$-point trapezoidal integration over $\phi$ across the unit 
hemisphere.
The non-LTE solution was deemed to have converged when the 
emergent intensities changed by less than a factor of $10^{-3}$
between successive iterations.

The calculations on the 1D model stellar atmospheres
(\atmo{} and \marcs{} models) were performed in almost
the same way as the calculations on the 3D models.
In particular they employed the same code, \balder{}, and the same model atom. 
The key difference is that the 1D approach 
included a depth-independent, isotropic microturbulence
($\vmic$). This parameter 
broadens spectral lines; the physical interpretation
is that it reflects
a temperature and velocity gradients on scales much smaller
than one optical depth (\citealt{2008oasp.book.....G}, Chapter 17).
These effects are naturally accounted for in the 3D models,
and thus no analogous broadening was adopted for those calculations.
Therefore, the 1D versus 3D abundance differences
(\sect{method_abundance})
are functions of the $\vmic$ parameter adopted in the 1D models.

The depth-independent $\vmic$ has been 
treated as an additional dimension in the theoretical
grids presented in \sect{results} and in \sect{interpolator};
in addition, it is fixed to a representative literature
value in \sect{gbs}.
To reduce the dimensionality of the parameter space, 
one could attempt to calibrate $\vmic$ on the 3D models
\citep{2013MSAIS..24...37S,2018A&A...611A..19V}.
However, a depth-independent, isotropic $\vmic$ is an imperfect descriptor 
of the dynamical properties of stellar atmospheres. This is in part 
because temperature and velocity gradients decrease with increasing
geometrical height in the line-forming regions,
which implies that a smaller $\vmic$ is appropriate for lines
that form higher up in the atmosphere
(\citealt{2013MSAIS..24...37S}, Fig.~2).
At the same time, light from the stellar limb 
originates from stellar granules that are seen edge-on, 
characterised by large temperature and velocity gradients,
which implies
a larger $\vmic$ when sampling more light from the stellar limb
\citep{2022SoPh..297....4T}. Thus, the calibrated $\vmic$
is a function of the formation depths
and centre-to-limb variations of the particular set of 
lines used for the analysis.
Consequently, a proper calibration would require increasing the
dimensionality of the parameter space, rather than decreasing it.  As such, for simplicity, a 3D-calibrated
$\vmic$ has not been adopted here.  In any case, in the present study, the 1D results
mainly serve as an intermediate step to obtain 3D non-LTE abundances through the
application of abundance differences.

In the post-processing calculations for the 3D models, $\lgeps{Fe}$ was
kept consistent with the value used to construct the model
stellar atmospheres: in other words, $\xm{Fe}=0.0$.
This was also the case for the calculations on  1D \marcs{} models.
For the 1D \atmo{} models, iron was treated as a trace element
such that, for a given $\xh{M}$, calculations were performed for nine values of $\lgeps{Fe}$, instead of just one value as for the
3D models.  These values correspond to $\xm{Fe}$ between
$-1.0\,\dex$ and $+1.0\,\dex$ in uniform steps
of $0.25\,\dex$.

\subsection{Synthetic spectra and abundance differences}
\label{method_abundance}

The emergent synthetic spectra
from the model stellar atmospheres
were calculated after the non-LTE populations had converged.
The non-LTE populations of the reduced $LS$ model atom 
were first redistributed onto the full $LSJ$ data set (\sect{method_atom}).
This was done by multiplying the $LSJ$ LTE populations 
by the $LS$ departure coefficients
\citep[e.g.][]{2016MNRAS.463.1518A}.
This is valid in the limit of large collisions
within fine-structure levels as well as within super-levels.
Following this, a formal solution in the outgoing direction
was performed to obtain the emergent intensity in different
directions (\sect{method_code}), 
from which the emergent disc-averaged intensity 
(the emergent stellar flux) was determined.

Emergent spectra were only calculated for a small number of 
\ion{Fe}{I} and \ion{Fe}{II} lines. This study
employs the `golden' line list presented in
\citet{2014A&A...564A.133J}, adopting the 
energies, $\lggf$, and pressure broadening parameters 
given in their Tables 4 and 5.
The $\lggf$ and $\elo$ for this
set of lines are illustrated in \fig{fig:golden}.
Since this line list was constructed to be useful
for stellar spectroscopy,
lines of larger $\elo$, thus corresponding to
lower absorption due to the Boltzmann factor,
tend to have larger $\lggf$, giving
rise to a linear trend in \fig{fig:golden}.
All of the lines shown are in the optical regime:
$\lambda_{\text{Air}}=478.783\,\nm$ to $681.026\,\nm$.

Equivalent widths were determined via direct integration.
A complication arose
because the line profiles were calculated 
simultaneously, rather than in a line-by-line fashion.
This means that different lines may overlap
with each other. Moreover, the extent of the overlap 
is a function of stellar parameters, as illustrated in \fig{fig:spectra}.
When the overlapping lines are of different 
$\elo$ or of different species, this can
add some noise to the analysis and complicate the interpretation.
Thus, the discussion of the abundance differences
across parameter space (\sect{results})
is restricted to lines that do not overlap with each other,
such that the flux depression
at the two integration limits 
 -- which are generally
set at the wavelengths that are in between the neighbouring line cores --
is less than $1\%$ from the continuum.

Given 1D LTE, 1D non-LTE, and 3D non-LTE equivalent widths as a function of
$\lgeps{Fe}$, 1D LTE versus 3D non-LTE abundance differences ($\abdl$) and 1D
non-LTE versus 3D non-LTE abundance differences ($\abdn$) were then evaluated.
These quantities are formally abundance errors: a negative value indicates a
positive abundance correction, or in other words that the 3D non-LTE model
predicts a larger $\lgeps{Fe}$ than the 1D LTE or 1D non-LTE model.

\section{Theoretical abundance differences across parameter space}
\label{results}


\begin{figure*}
    \begin{center}
        \includegraphics[scale=0.325]{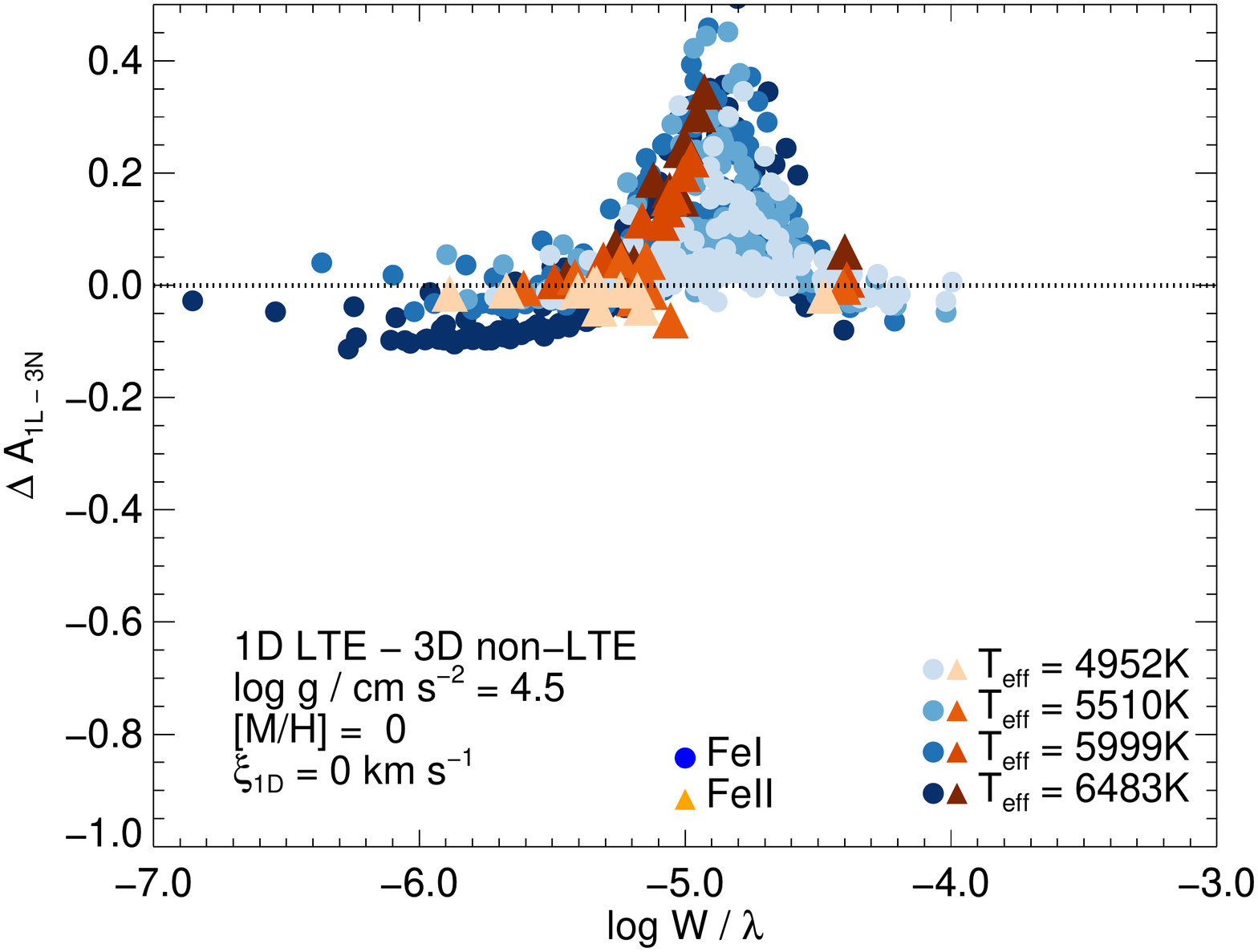}\includegraphics[scale=0.325]{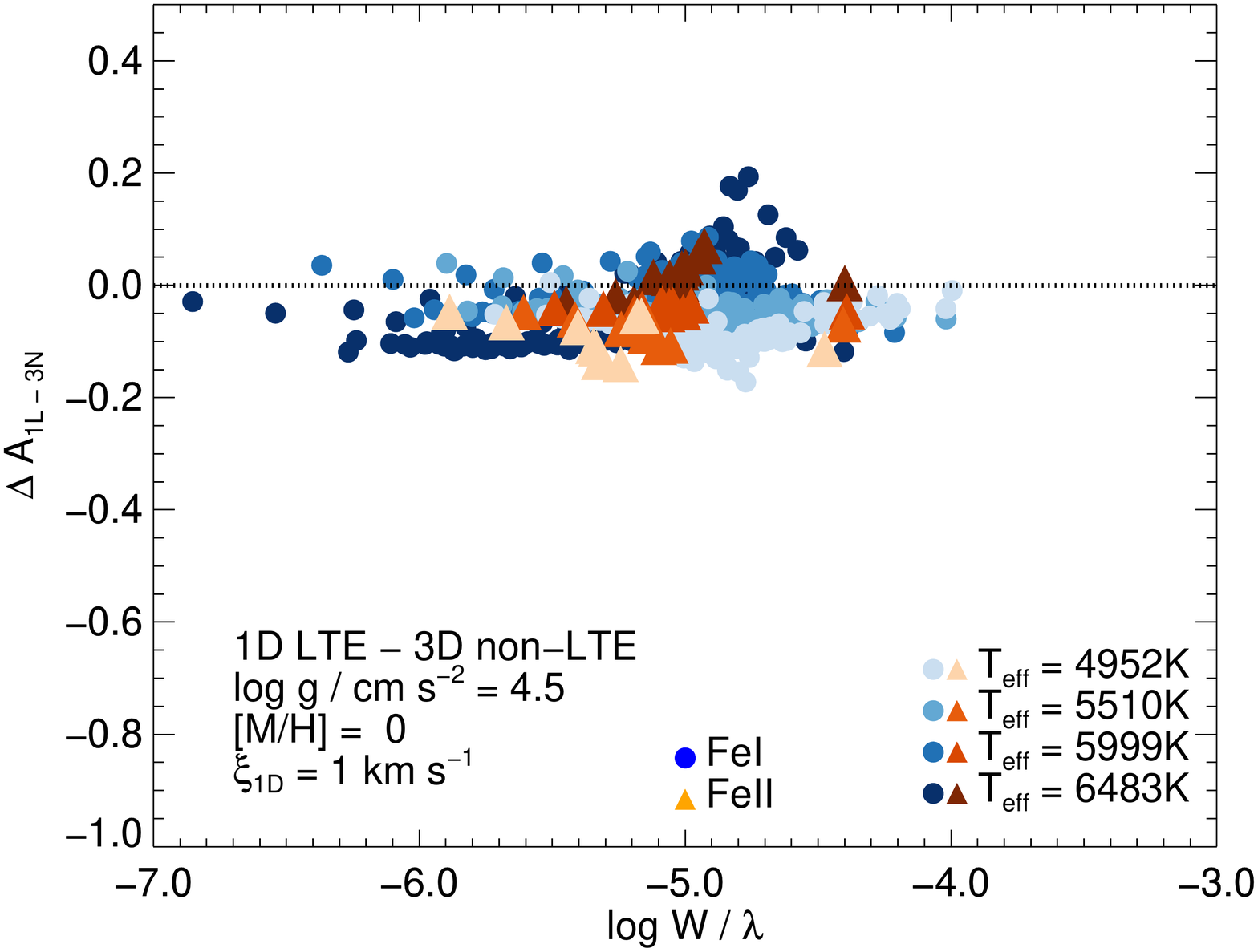}
        \includegraphics[scale=0.325]{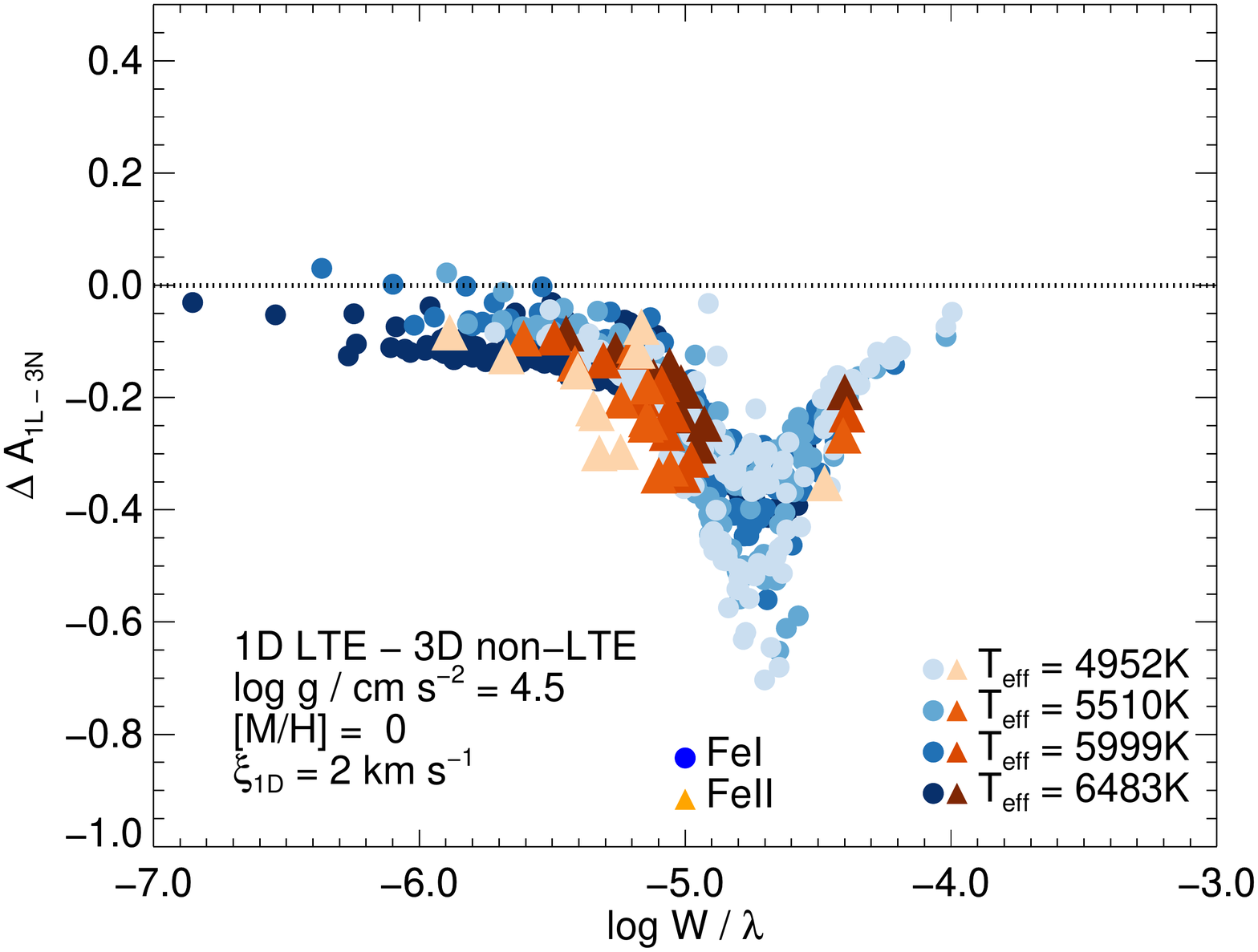}\includegraphics[scale=0.325]{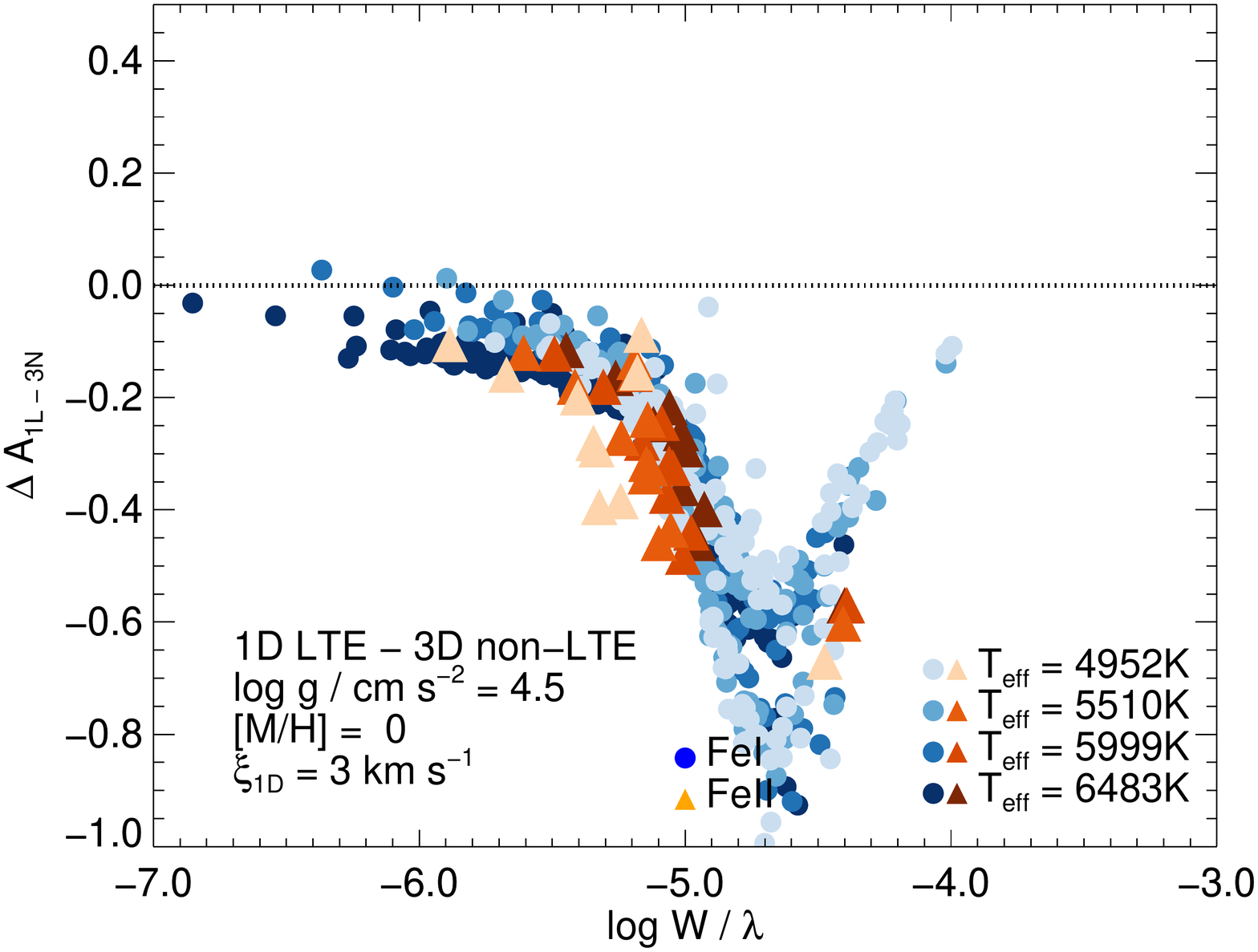}
        \caption{$\abdl$ against $\reqw{}$,
        for different lines of $\ion{Fe}{I}$ and 
        $\ion{Fe}{II}$ (symbols),
        and different values of $\vmic$ (panels).
        Lighter colours represent smaller $\teff$.
        The results are shown for $\lgg=4.5$ and $\xh{M}=0$.}
        \label{fig:aberr_reqw1}
    \end{center}
\end{figure*}

\begin{figure*}
    \begin{center}
        \includegraphics[scale=0.325]{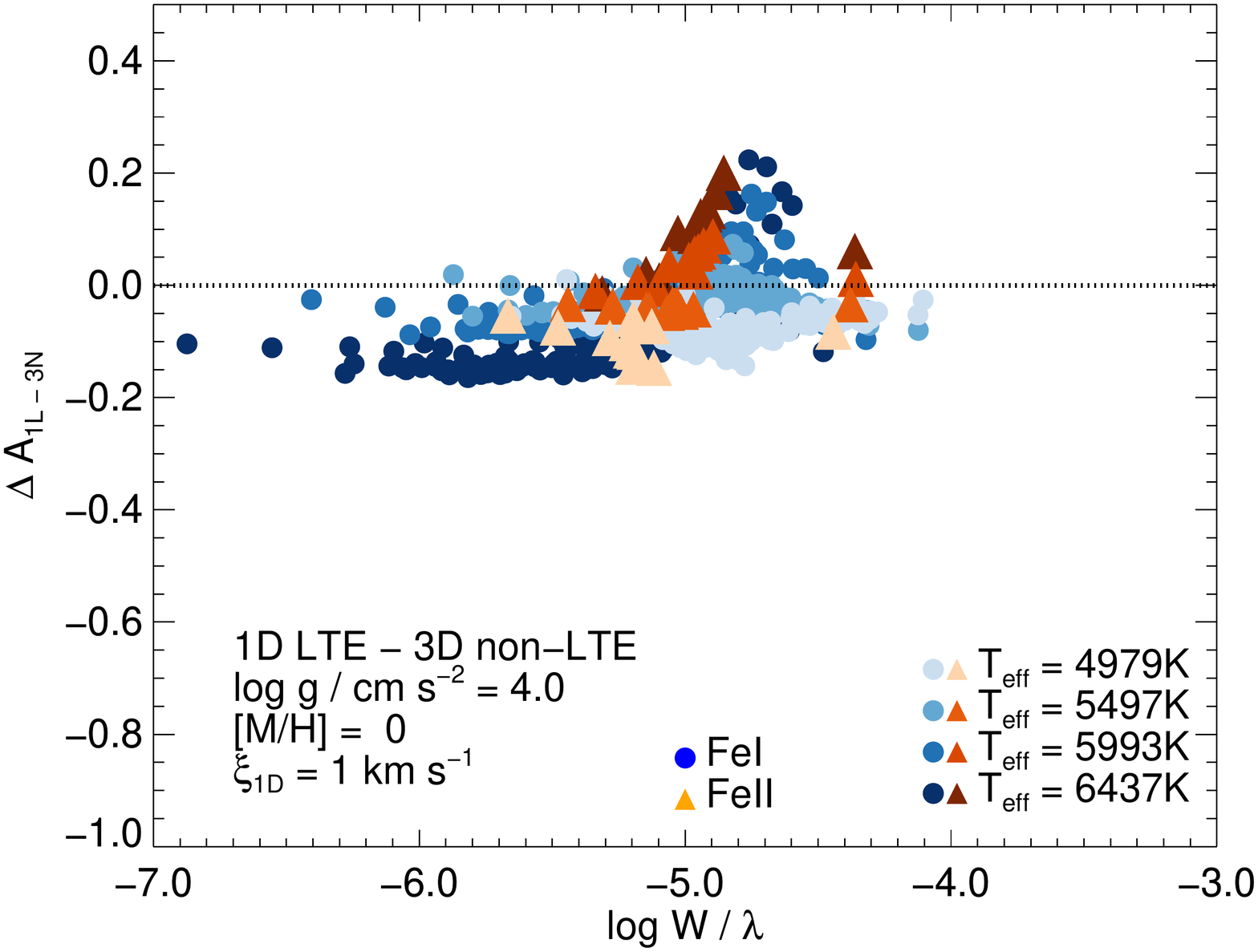}\includegraphics[scale=0.325]{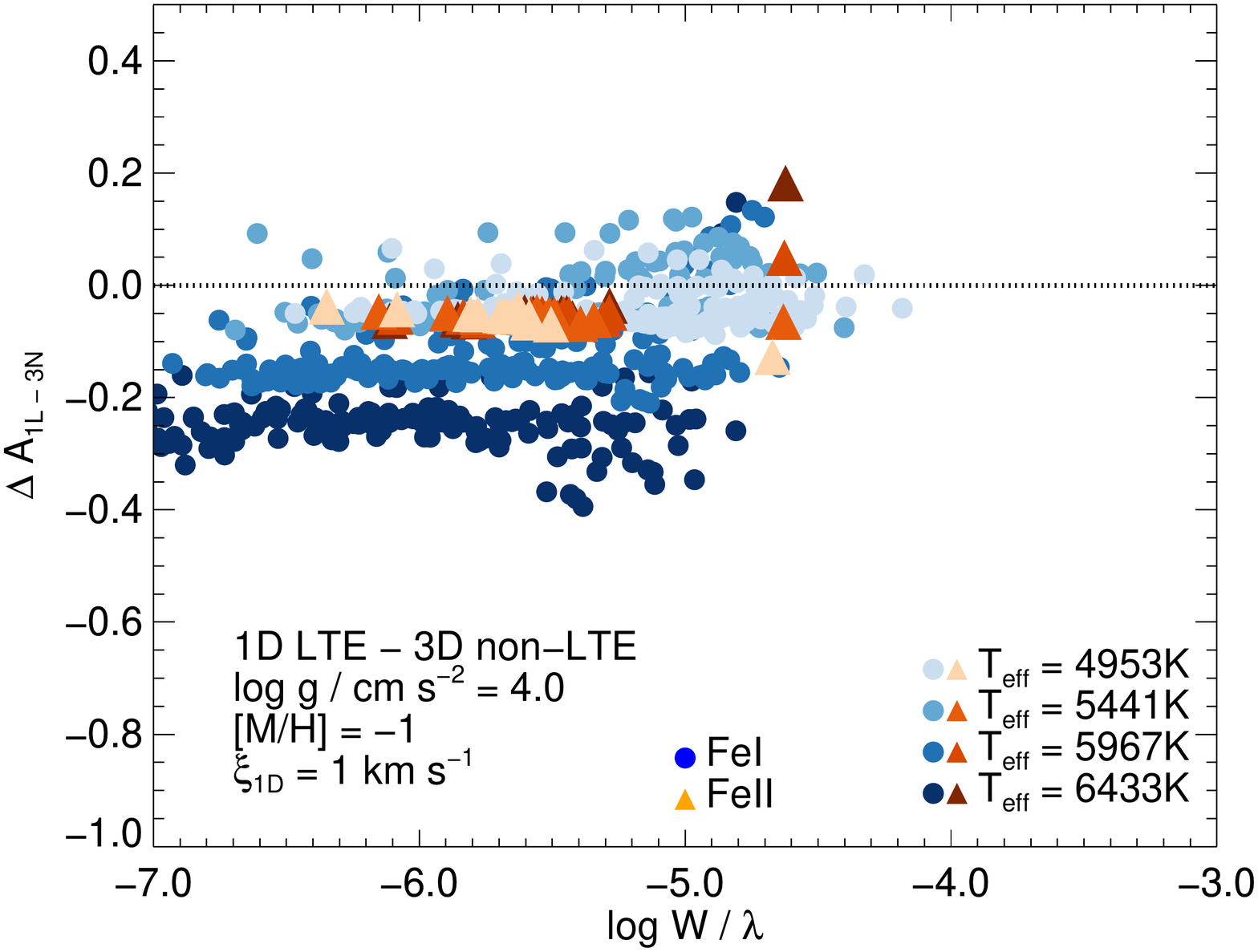}
        \includegraphics[scale=0.325]{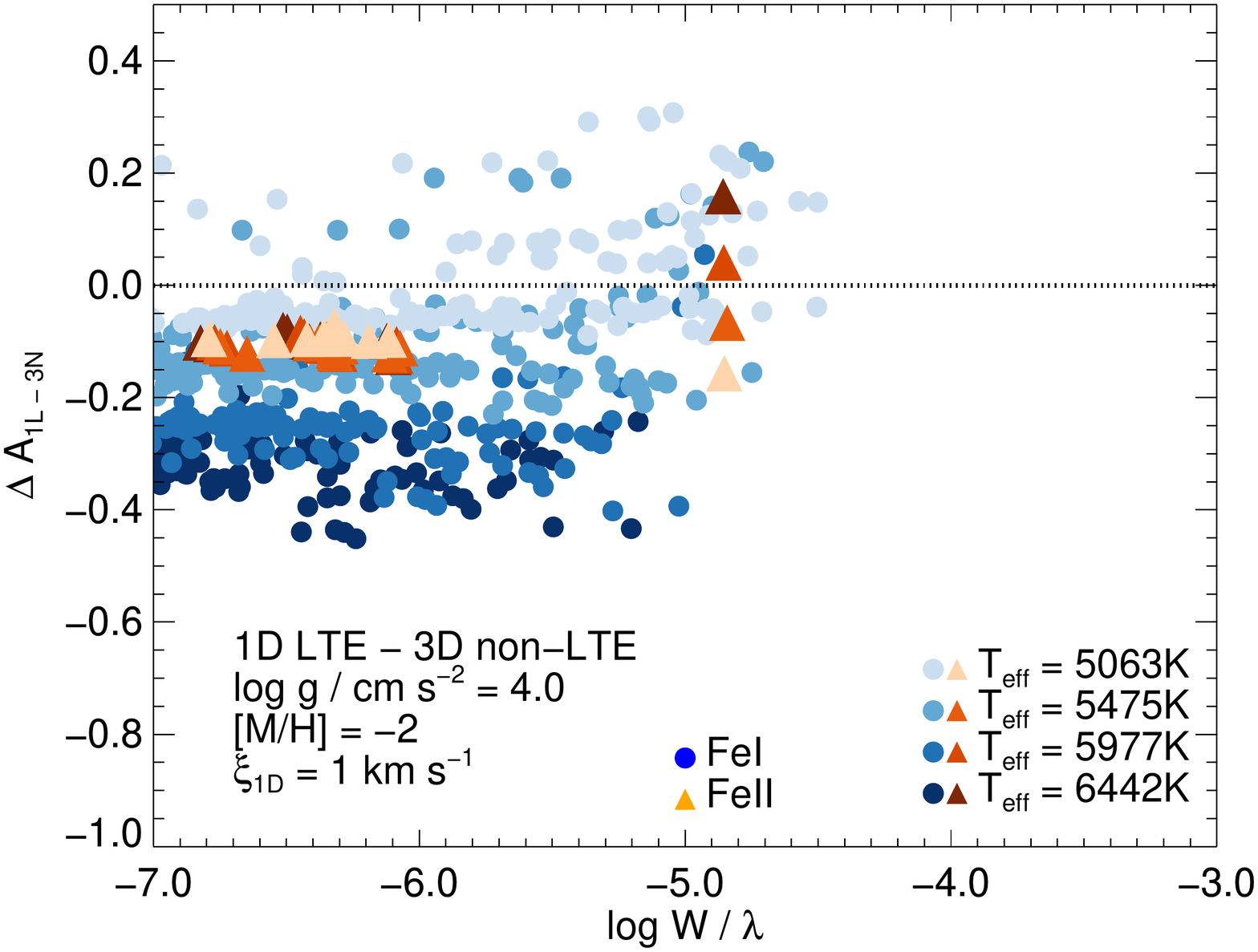}\includegraphics[scale=0.325]{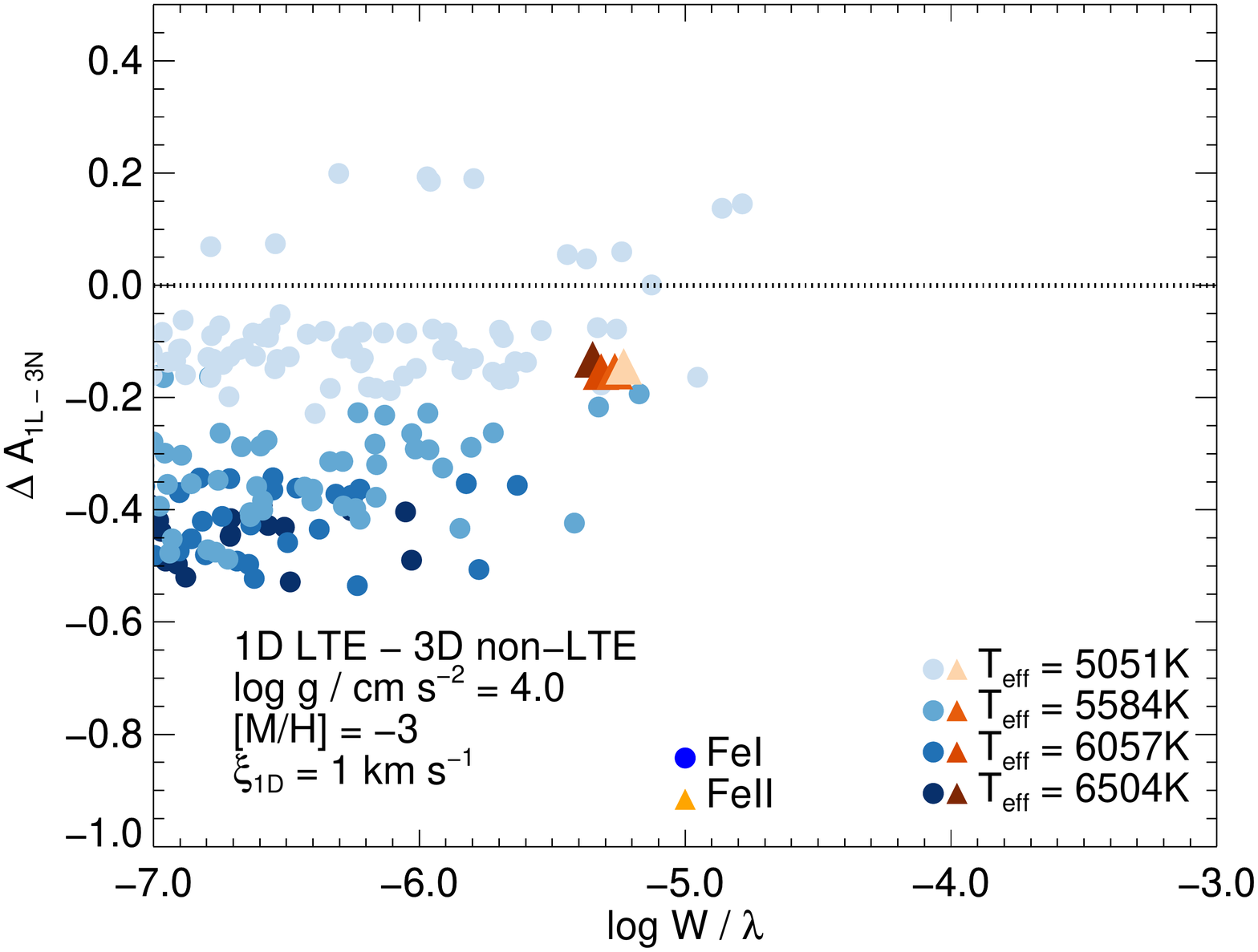}
        \caption{$\abdl$ against $\reqw{}$,
        for different lines of $\ion{Fe}{I}$ and 
        $\ion{Fe}{II}$ (symbols),
        and different values of $\xh{M}$ (panels).
        Lighter colours represent smaller $\teff$.
        The results are shown for $\lgg=4.0$ and $\vmic=1$.}
        \label{fig:aberr_reqw2}
    \end{center}
\end{figure*}

\begin{figure*}
    \begin{center}
        \includegraphics[scale=0.325]{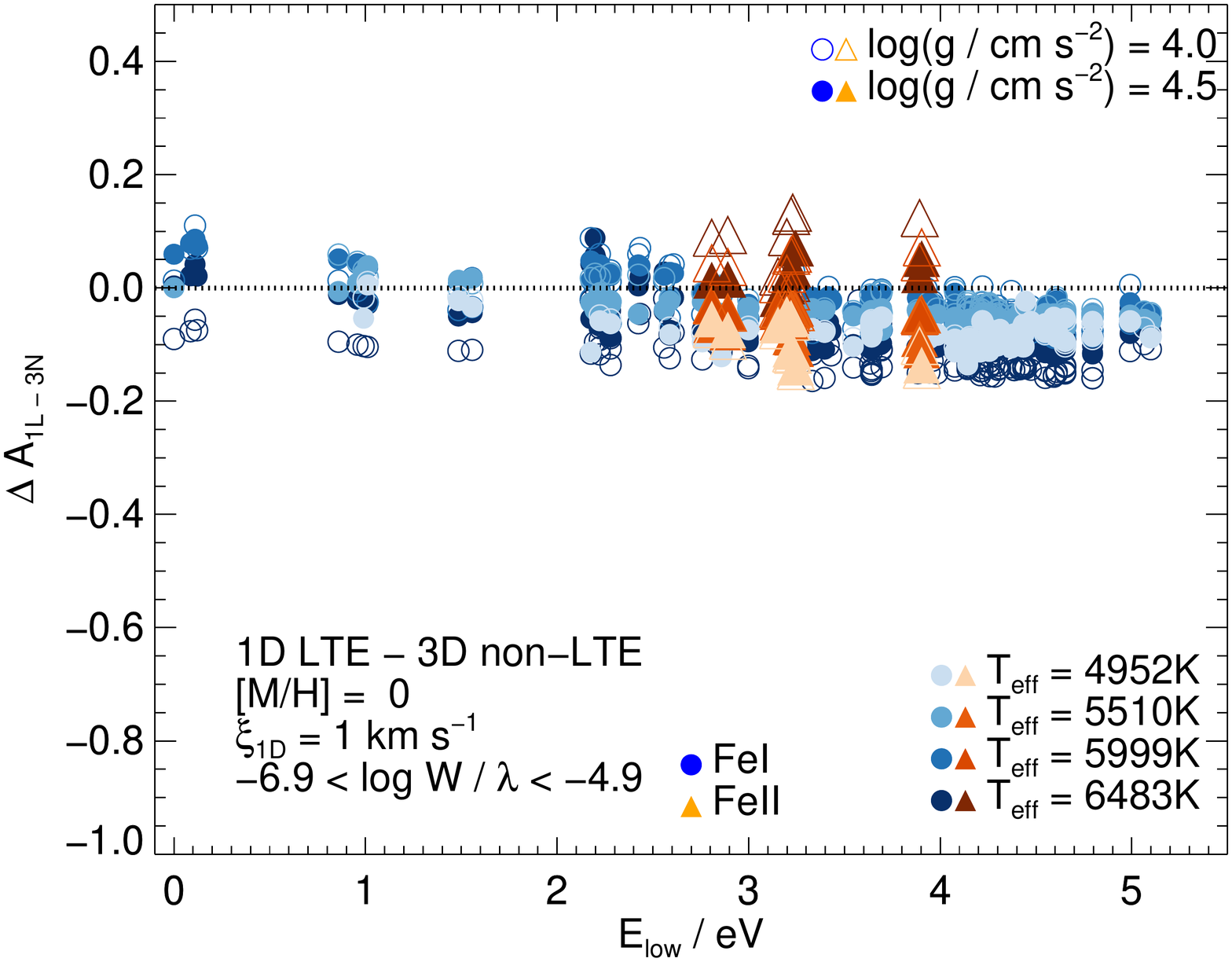}\includegraphics[scale=0.325]{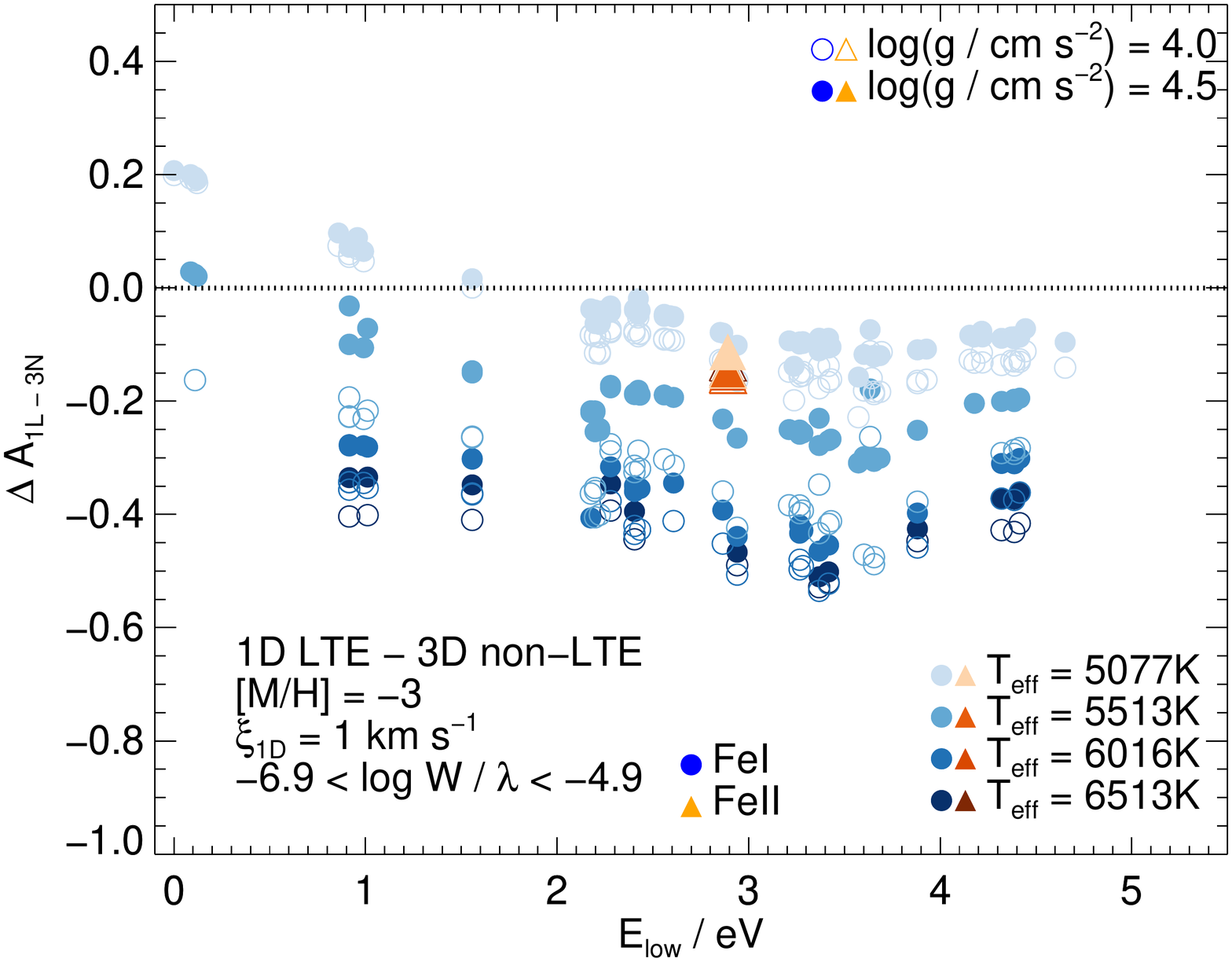}
        \includegraphics[scale=0.325]{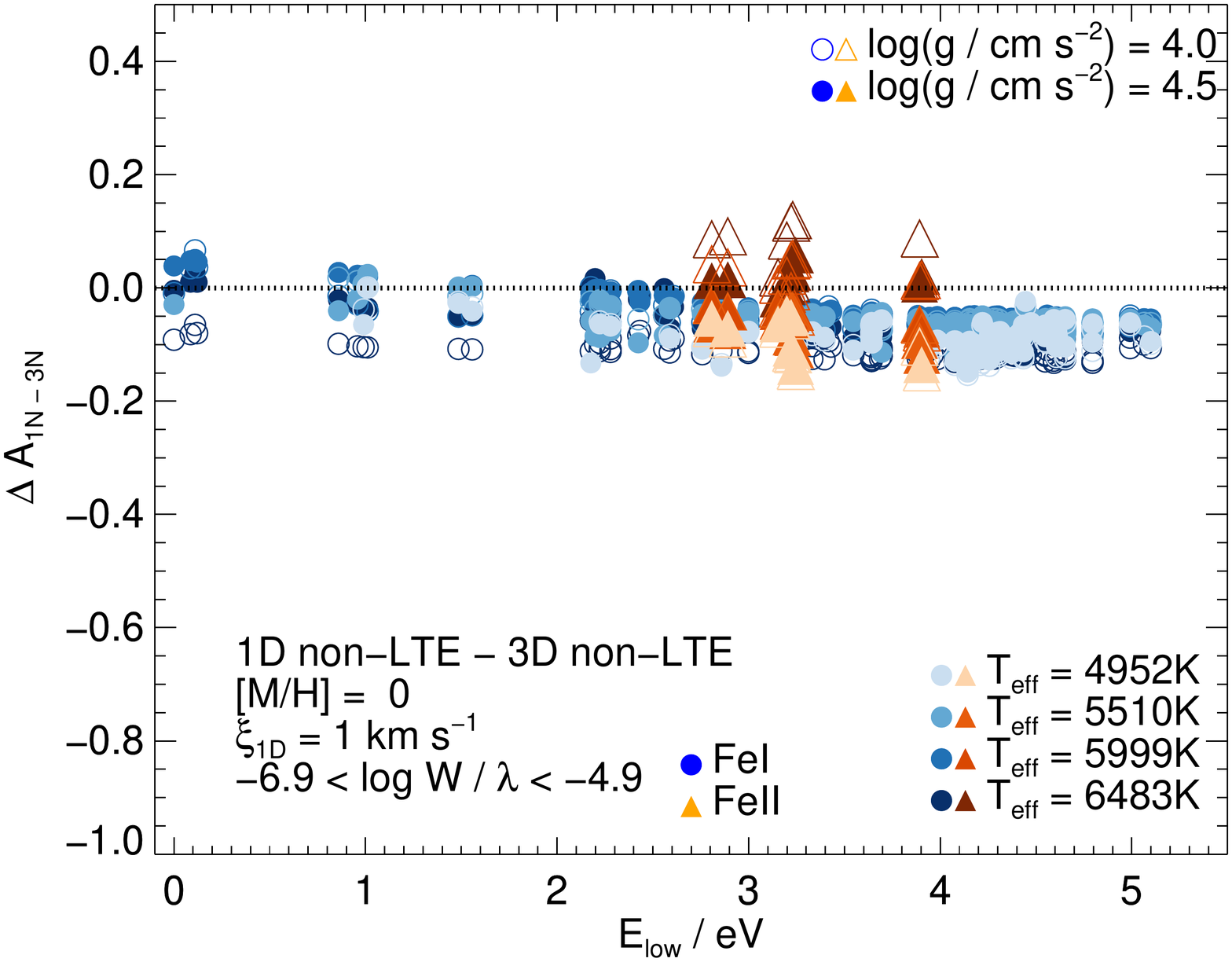}\includegraphics[scale=0.325]{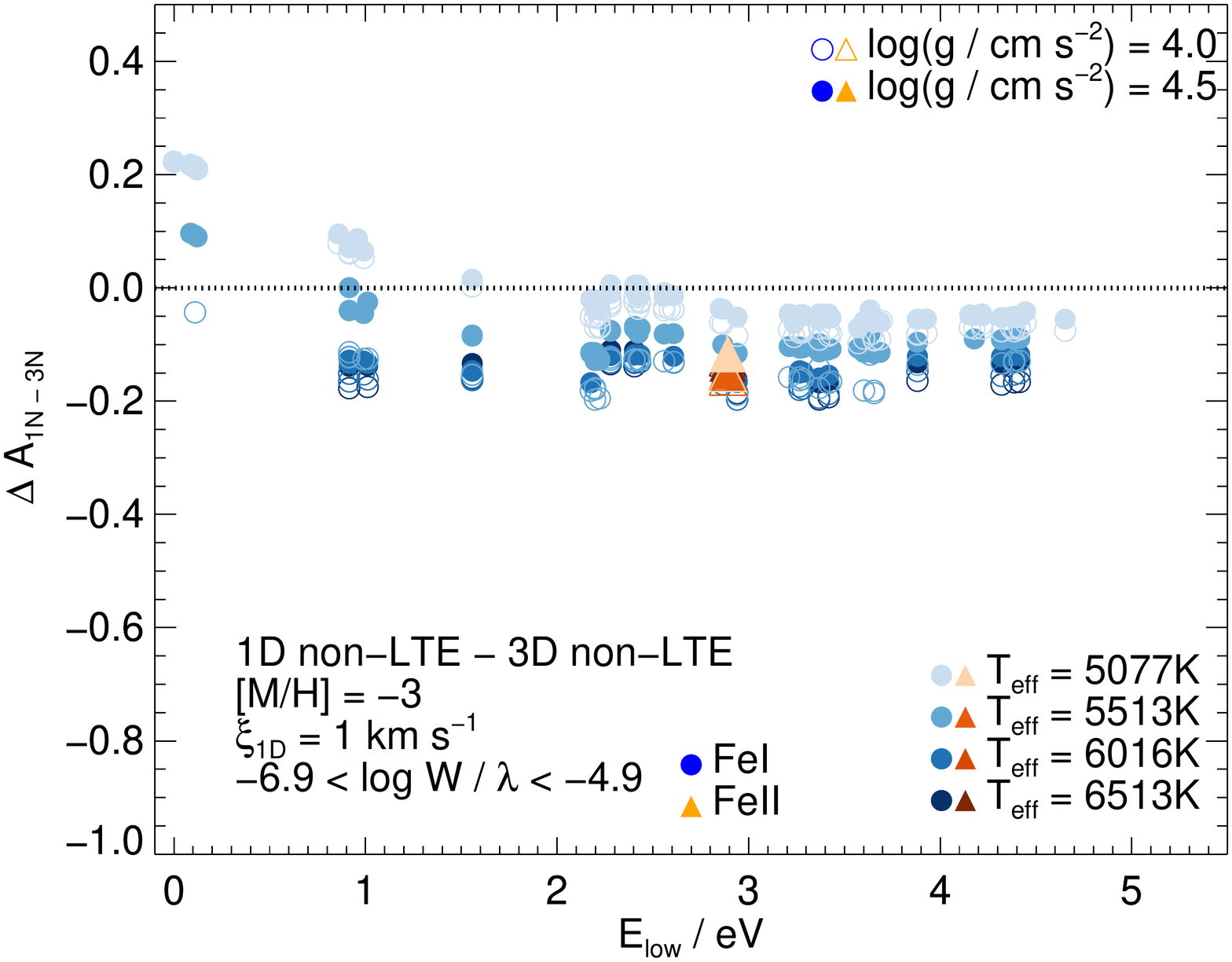}
        \caption{$\abdl$ (upper panels) and $\abdn$
        (lower panels) against $\elo{}$, for
        different lines of $\ion{Fe}{I}$ and 
        $\ion{Fe}{II}$ (symbols),
        and different values of $\xh{M}$ (columns).
        Lighter colours represent smaller $\teff$.
The results are shown for $\vmic=1\,\kms$ and $-6.9<\reqw<-4.9$.}
        \label{fig:aberr_elo}
    \end{center}
\end{figure*}

\begin{figure*}
    \begin{center}
        \includegraphics[scale=0.325]{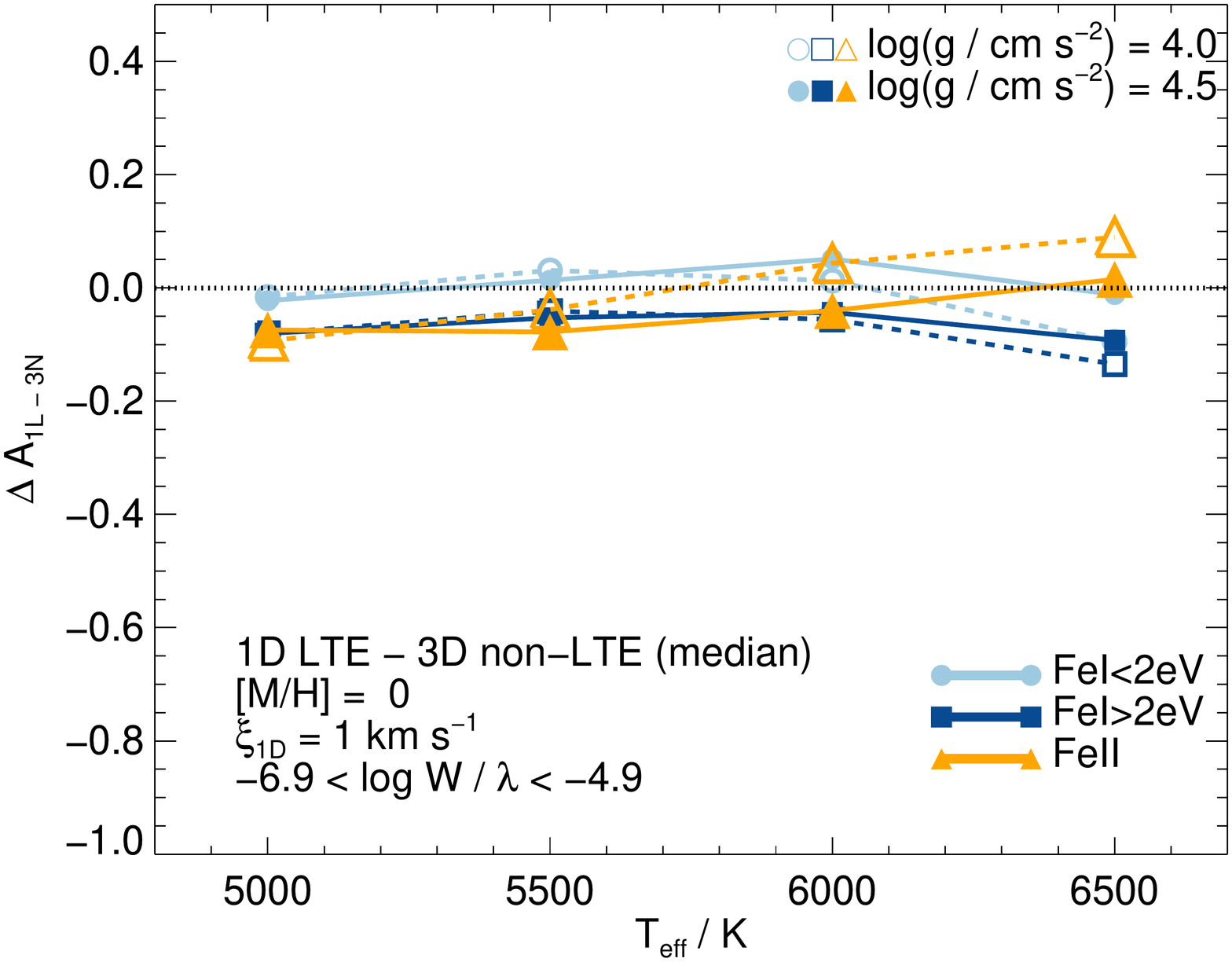}\includegraphics[scale=0.325]{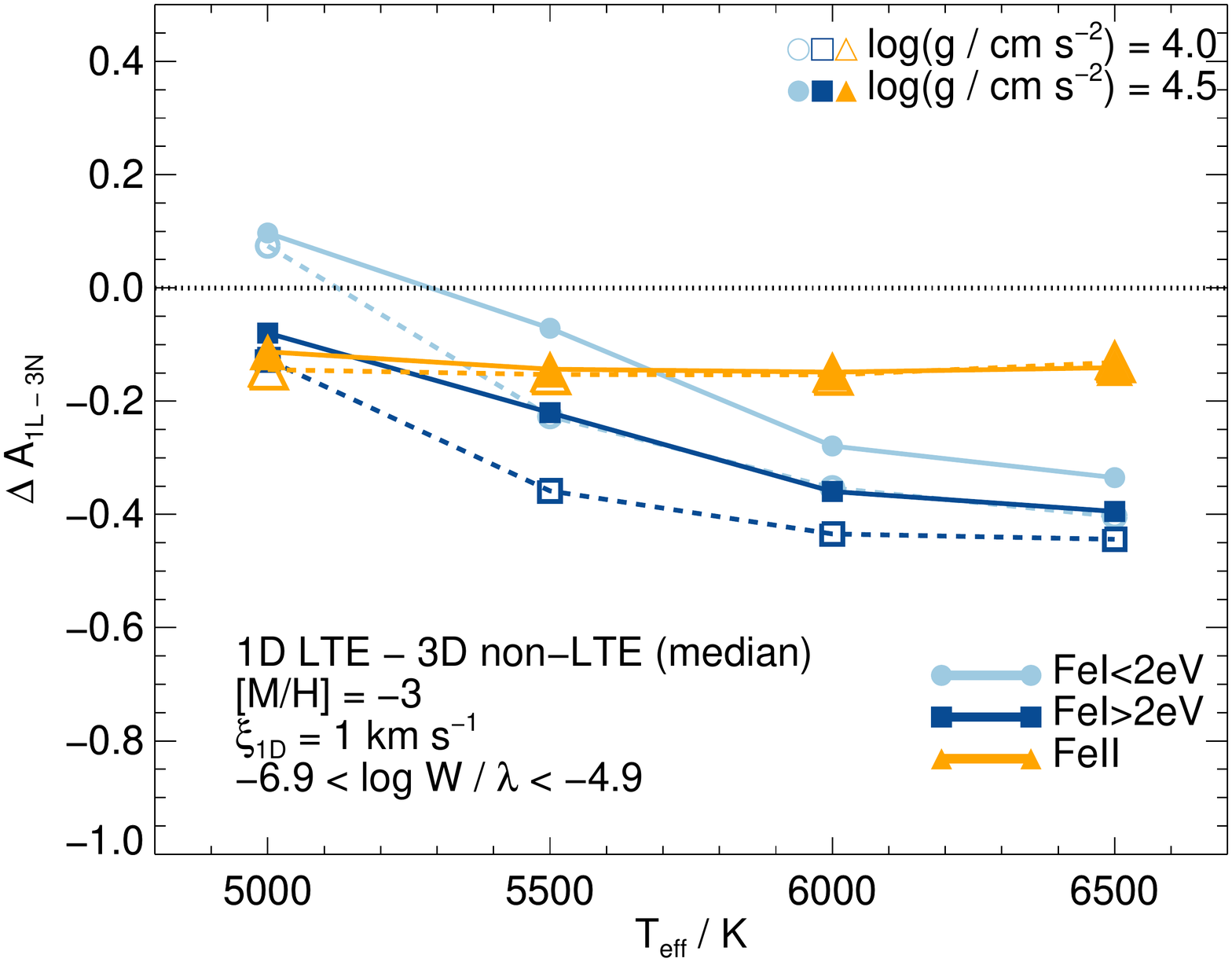}
        \includegraphics[scale=0.325]{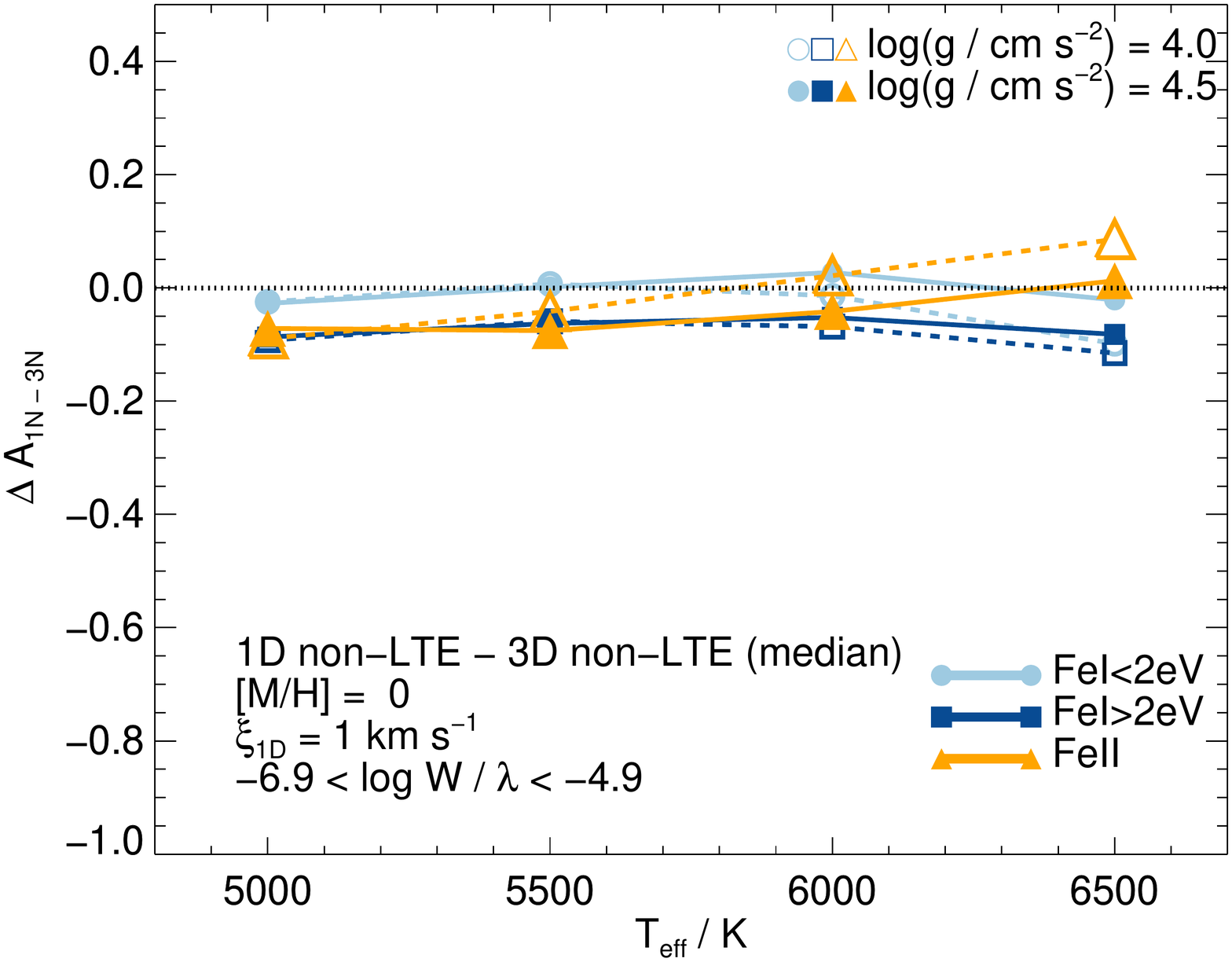}\includegraphics[scale=0.325]{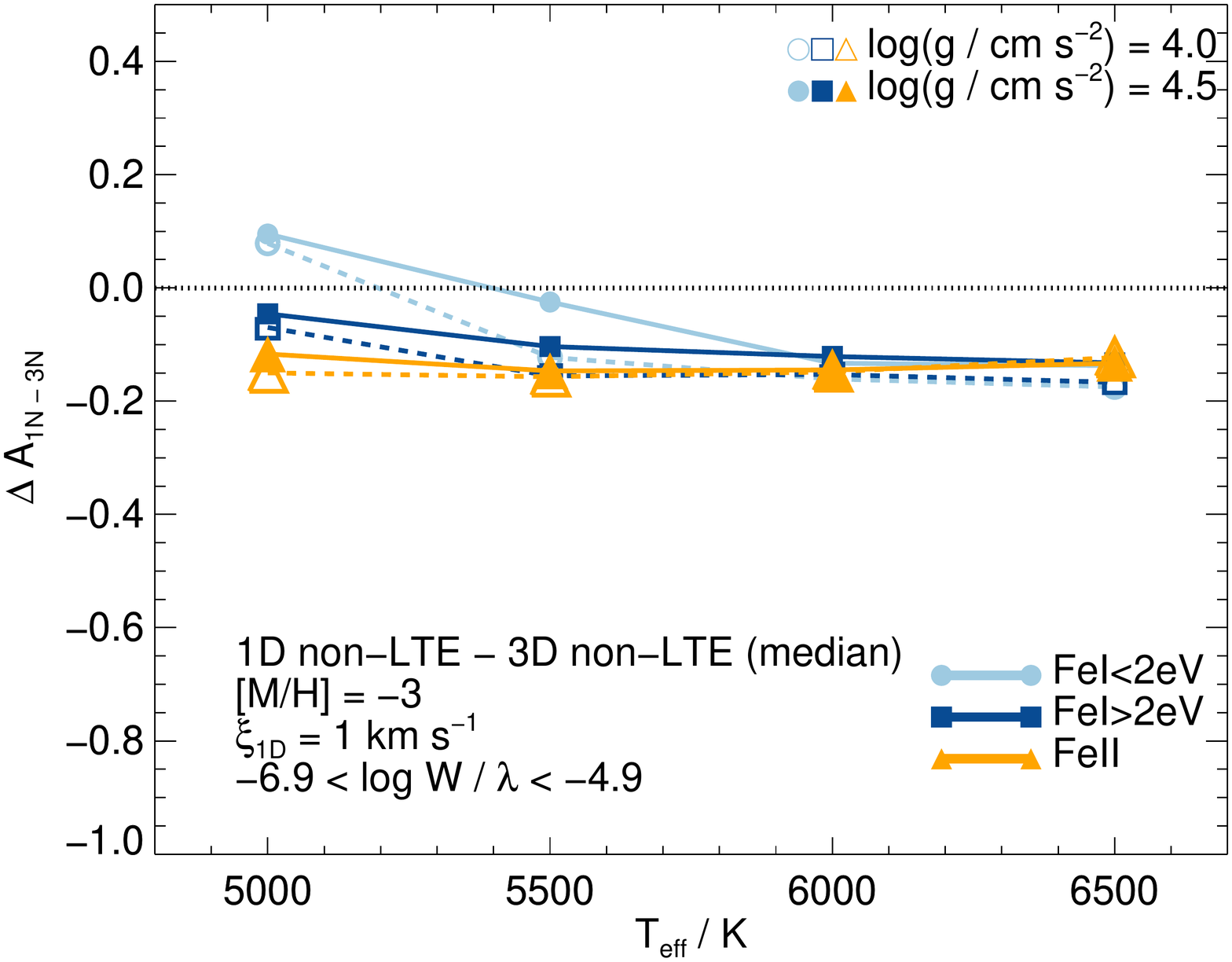}
        \caption{$\abdl$ (upper panels) and 
        $\abdn$ (lower panels) against $\teff{}$,
        averaged over lines of $\ion{Fe}{I}$ with 
        $\elo<2\,\eV$ and $\elo>2\,\eV$ as well as 
        over lines of $\ion{Fe}{II}$,
        for different values of $\xh{M}$ (columns).
        The results are shown for $\vmic=1\,\kms$ and $-6.9<\reqw<-4.9$.}
        \label{fig:aberr_teff}
    \end{center}
\end{figure*}

\begin{figure*}
    \begin{center}
        \includegraphics[scale=0.325]{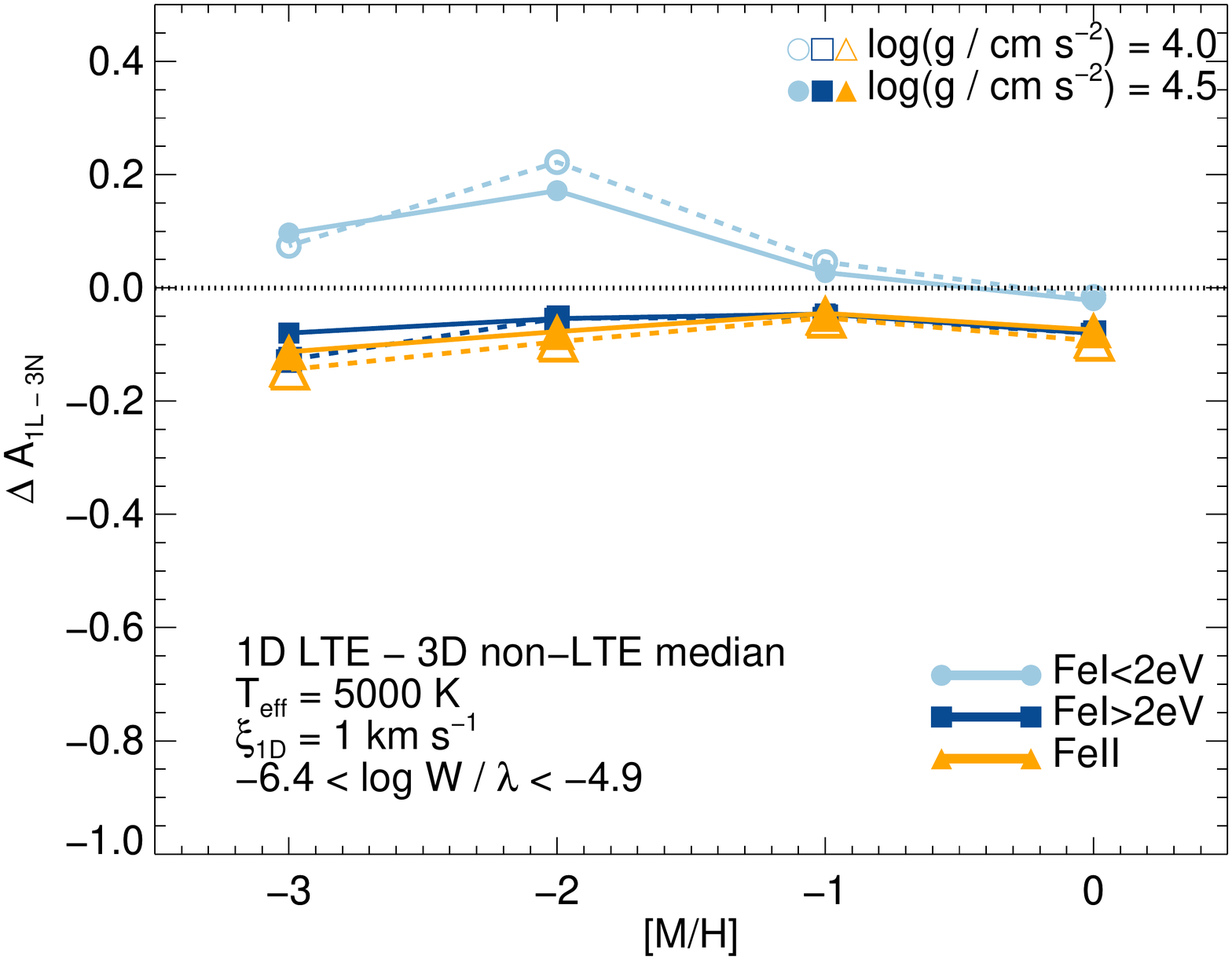}\includegraphics[scale=0.325]{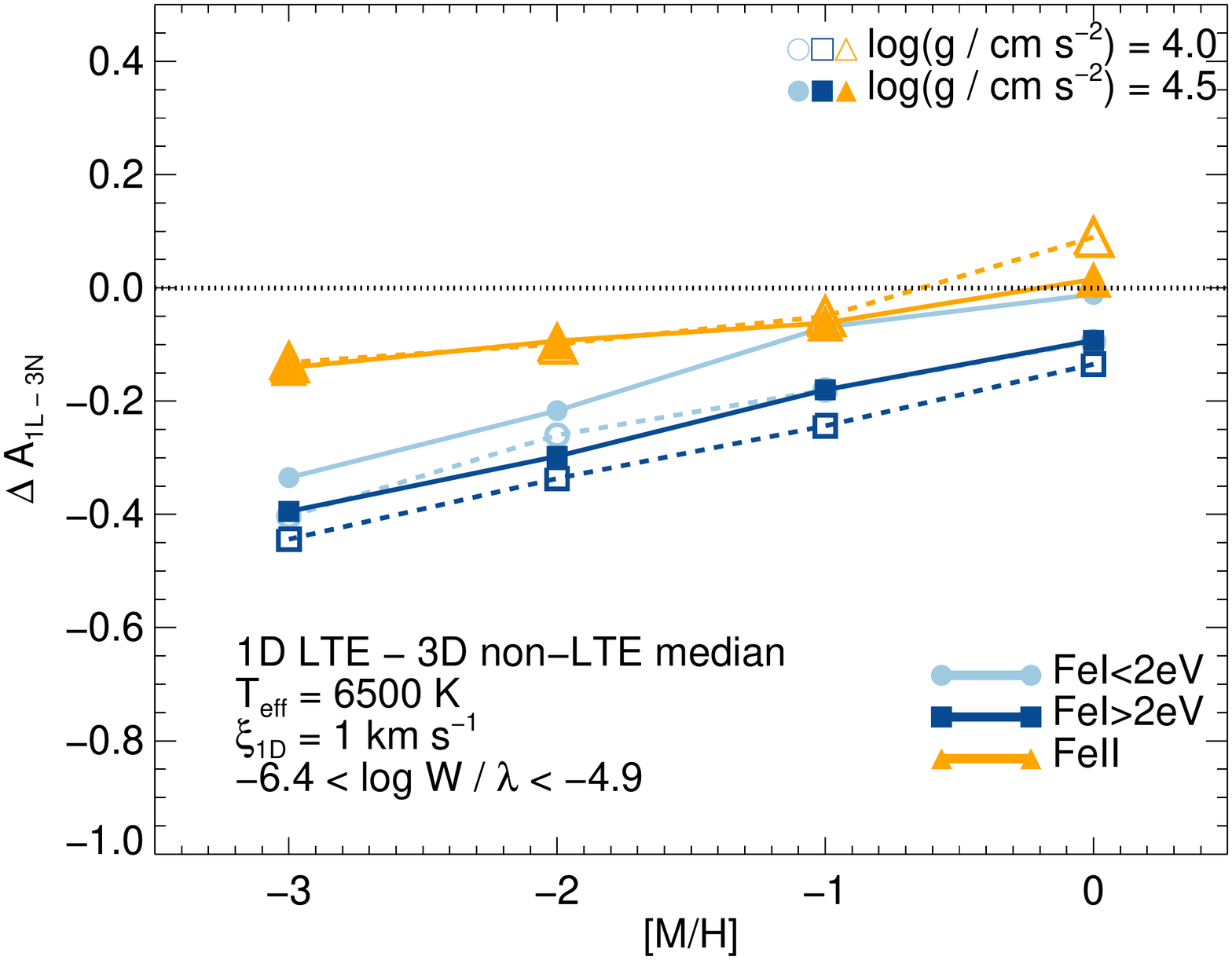}
        \includegraphics[scale=0.325]{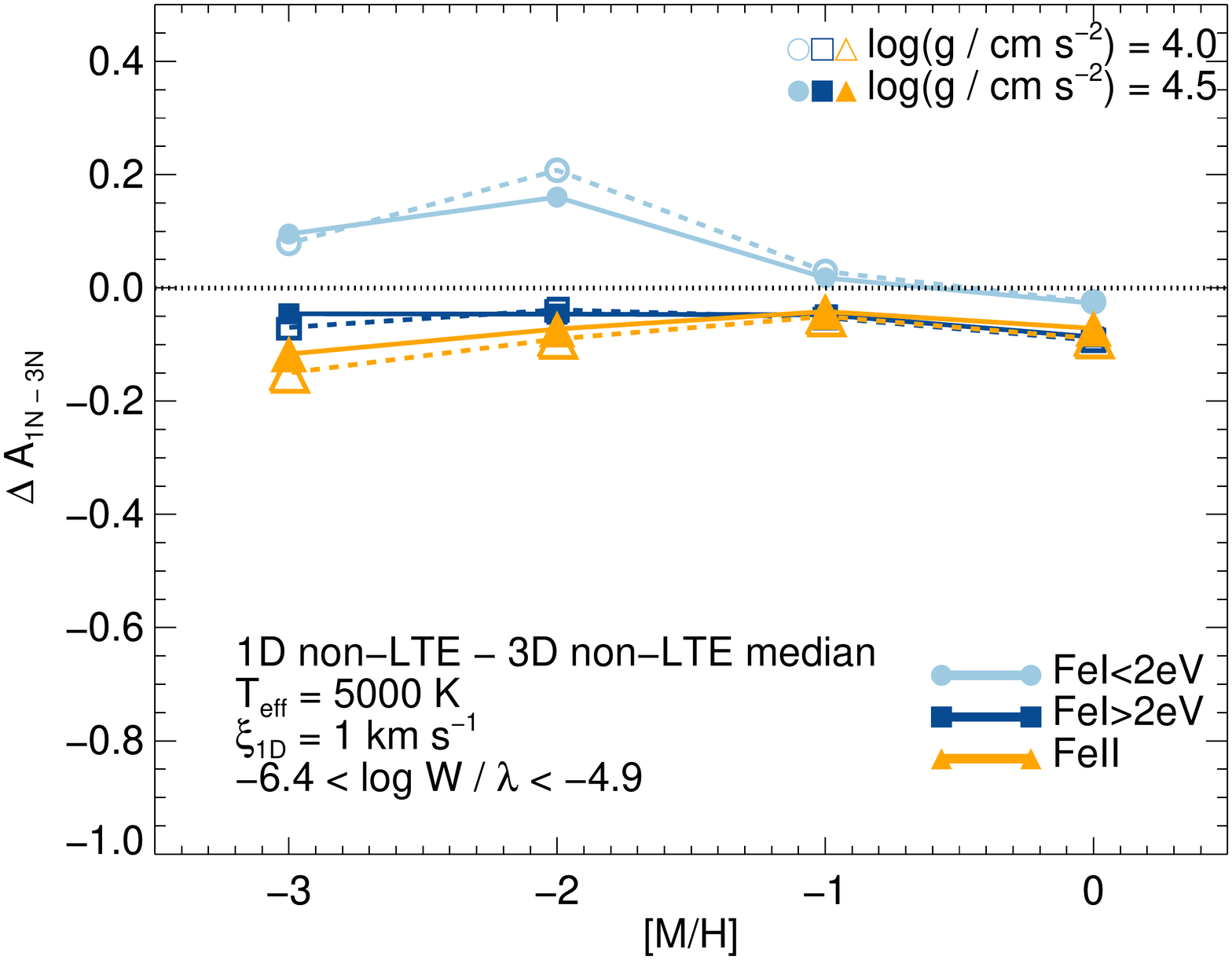}\includegraphics[scale=0.325]{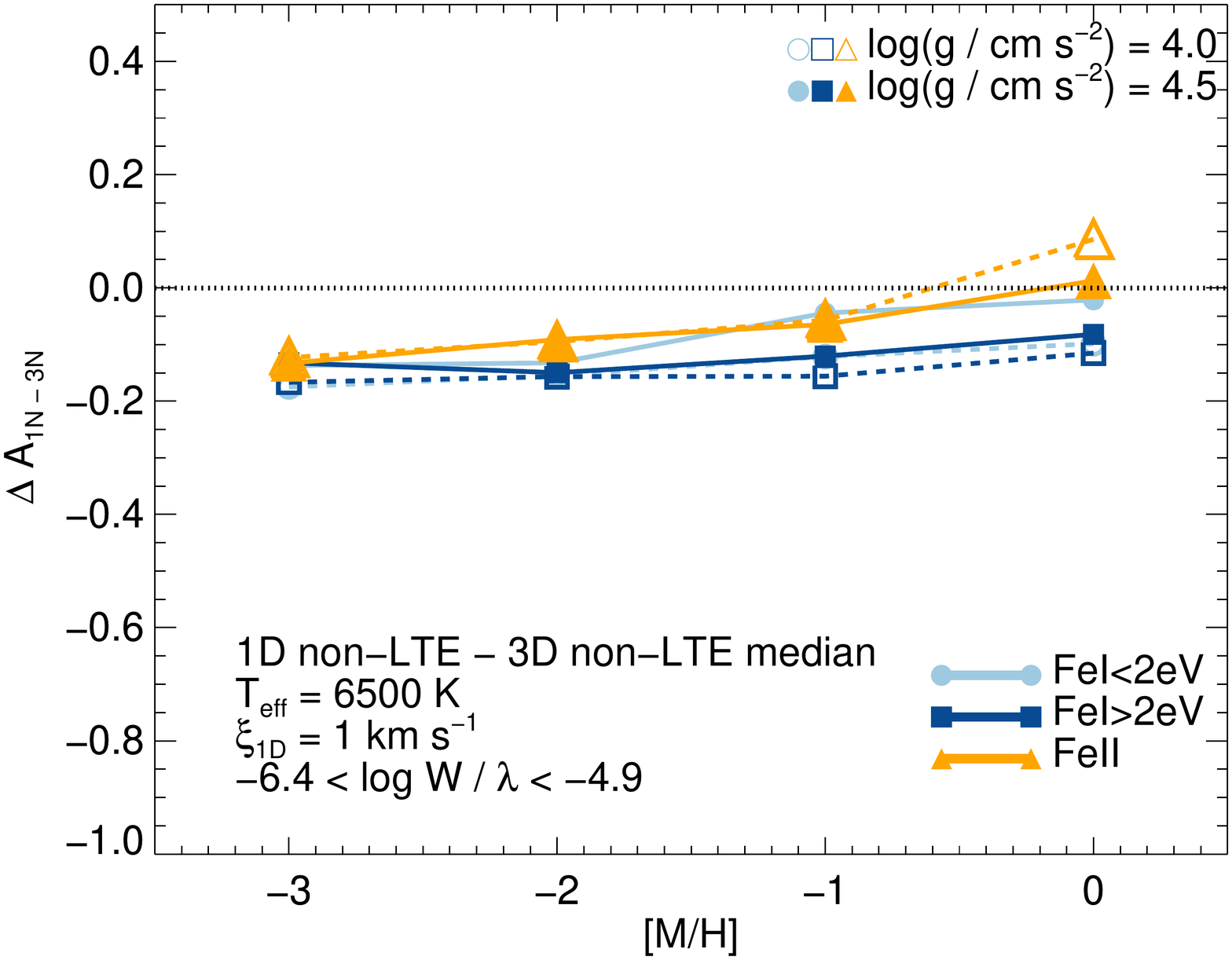}
        \caption{$\abdl$ (upper panels) and 
        $\abdn$ (lower panels) against $\xh{M}$,
        averaged over lines of $\ion{Fe}{I}$ with 
        $\elo<2\,\eV$ and $\elo>2\,\eV$ as well as 
        over lines of $\ion{Fe}{II}$,
        for different values of $\teff{}$ (columns).
        The results are shown for $\vmic=1\,\kms$ and $-6.9<\reqw<-4.9$.}
        \label{fig:aberr_feh}
    \end{center}
\end{figure*}

\subsection{Partially saturated lines}
\label{results_reqw_vturb}

\fig{fig:aberr_reqw1} illustrates the behaviour of $\abdl$ with $\reqw$
for $\lgg=4.5$ and $\xh{M}=0$, and 
$\vmic$ between $0\,\kms$ and $3\,\kms$. 
Depending on $\vmic$, the plots either 
have a peak or a trough at $\reqw\approx-4.8$:
for small $\vmic$, $\abdl$ reaches up to $+0.4\,\dex$,
while for large $\vmic$,
$\abdl$ reaches down to $-1.0\,\dex$.
\fig{fig:aberr_reqw2} illustrates this same
behaviour for $\lgg=4.0$, $\vmic=1\,\kms$,
and different values of $\xh{M}$.
The peak position is roughly the same in these plots;
it becomes less apparent towards smaller $\xh{M}$ as
the studied lines become weaker.

The peaked appearance of $\abdl$ is typical of abundance differences.
For example, it can be seen for \ion{Na}{I} in Fig.~5 of
\citet{2011A&A...528A.103L}, for \ion{K}{I} in Fig.~13 of
\citet{2019A&A...627A.177R}, and for \ion{C}{I} and \ion{O}{I} in Fig.~7 of
\citet{2019A&A...630A.104A}.  It is related to lines leaving the linear part of
the curve of growth and becoming partially saturated as $\reqw$ increases.
When this happens first in 1D LTE, a large increase in $\lgeps{Fe}_{\text{1L}}$
is needed to match $\lgeps{Fe}_{\text{3N}}$.  This results in large, positive
values of $\abdl$.  Vice versa, when partial saturation first occurs in the 3D
non-LTE model, one finds large, negative values of $\abdl$.  At even larger
$\reqw\gtrsim-4.8$, both 1D LTE and 3D non-LTE lines are partially saturated,
and the behaviour of the abundance differences again reflect those in the
weak-line regime.  Further discussion of this effect may be found in Sect.~3.1
of \citet{2011A&A...528A.103L} and Sect.~3.2.3 of \citet{2019A&A...630A.104A}.

The main effect of $\vmic$ in \fig{fig:aberr_reqw1}
can be understood in terms of the behaviour of
$\abdl$ with $\reqw$.  Adopting a smaller $\vmic$ would mean 
that the 1D LTE lines become partially saturated and
leave the linear part of the curve of growth 
before the 3D non-LTE lines (first panel of \fig{fig:aberr_reqw1}).
Conversely, adopting a larger $\vmic$ desaturates the 1D LTE lines;
they leave the
linear part of the curve of growth and become partially saturated after the
3D non-LTE lines (lower two panels
of \fig{fig:aberr_reqw1}).  The effect of $\vmic$
is smaller for lines with $\reqw\ll-4.8$ and $\reqw\gg-4.8$.
In the former case (the weak-line regime),
the line core is not yet saturated and so does not suffer so strongly
from $\vmic$; while in the latter case (strong lines), the equivalent widths are dominated
by the area spanned by extended pressure broadened wings.

The effects described above are 3D effects.
As such, $\abdn$ displays a qualitatively similar behaviour
to $\abdl$.  The behaviour is also qualitatively similar for 
other values of $\lgg$ and $\xh{M}$ not shown in the figures.

For clarity and brevity, the remaining parts of \sect{results} are based on
lines with $-6.9<\reqw<-4.9$, 
and the choice of $\vmic=1\,\kms$.  For this choice of
$\vmic$, \fig{fig:aberr_reqw1} indicates that the 1D LTE and 3D non-LTE model
lines leave the linear part of the curve of growth at roughly the same point.
By adopting this $\vmic$, and only considering weak lines, the discussion and
interpretation of the abundance differences is greatly simplified.

\subsection{Weak \ion{Fe}{I} lines} 
\label{results_elo_fe1}

At $\xh{M}=0$, for $\vmic=1\,\kms$, the upper left panels of 
\fig{fig:aberr_elo} show that for weak \ion{Fe}{I} lines, $\abdl$
ranges from $-0.15\,\dex$ to $+0.10\,\dex$.
At $\xh{M}=-3$, the right panels
show that $\abdl$ reaches $-0.5$
at high $\teff$ and low $\lgg$, and $+0.2\,\dex$ at low $\teff$,
at least for the selection of \ion{Fe}{I} lines considered here.
The overall range of results is similar for $\abdn$ at $\xh{M}=0$;
although, interestingly, the scatter in the plots is smaller 
than for $\abdl$.
At low $\xh{M}$, however, the lines with the most negative values of $\abdl$
show less severe values of
$\abdn$, which only reaches down to $-0.2\,\dex$ at $\xh{M}=-3$.

These abundance differences
are due to the combination of 3D and non-LTE effects
that are rather complicated to disentangle
(Sect.~3.2 and 3.3 of \citealt{2016MNRAS.463.1518A}).
The 3D effects can be due to
differences in the mean model stellar atmosphere stratifications,
the presence of inhomogeneities, and horizontal radiative transfer.
These various phenomena are complicated,
with the 3D models being either shallower or steeper on average
than the 1D models depending on the stellar parameters
\citep{2012MNRAS.427...27B},
and with the sign of $\abdn$ and $\abdl$
varying across the granulation pattern and line formation height
\citep{2013A&A...558A..20H}.
They also depend on the particular choice of $\vmic$,
with larger values shifting $\abdl$ and $\abdn$ downwards.
The non-LTE effects on \ion{Fe}{I}, on the other hand, can be due to
the over-ionisation of the minority species
that would act to increase the inferred $\lgeps{Fe}$,
or photon losses in the line cores that act to
reduce it, broadly speaking \citep{1996A&A...305..325K}.
This competition can be seen in the comparison of the 3D LTE and
3D non-LTE profiles in the upper right panel of \fig{fig:spectra},
where the 3D non-LTE profile is narrower due to over-ionisation,
but has a similar core flux depression due to photon losses.

The 3D and non-LTE effects may couple to each
other in complex ways.
Nevertheless, the overall trend is that $\abdl$ and $\abdn$ are usually 
the most negative for the
high $\teff$, low $\lgg$, and, at least for higher
$\teff$, low $\xh{M}$ models
(as expected from previous studies; \sect{introduction}).
These trends are more clearly visible in
\fig{fig:aberr_teff} and \fig{fig:aberr_feh},
which show the median results for \ion{Fe}{I}
lines of low and high excitation potential.
The left panels of 
\fig{fig:aberr_feh} show that there are peaks in $\abdl$ and 
$\abdn$ at $\xh{M}=-2$ and $\teff=5000\,\K$, 
for \ion{Fe}{I} lines of low excitation potential.
This reflects the competition between 
(reverse) granulation effects (driving weaker lines in 3D
corresponding to more positive $\abdl$ and $\abdn$)
and the effect of the lower temperatures in the upper layers of the 3D models
(driving stronger lines in 3D 
corresponding to more negative $\abdl$ and $\abdn$), 
with the latter effect becoming
much more pronounced at $\xh{M}=-3$.

Over-ionisation is certainly larger for such stars, because
the escaping UV flux increases for such stars,
and also because more compact stars have more efficient collisional
rates that help counteract these effects
\citep{2012MNRAS.427...50L}. At smaller $\xh{M}$,
photon losses are also typically weaker due to the smaller
strengths of the \ion{Fe}{I} lines.
The 3D effects might in general also be 
expected to be more severe for such stars,
with larger granulation contrasts and thus 
larger fluctuations in the atmosphere \citep{2013A&A...557A..26M}.
In general, as noted above, this does not immediately
imply more negative values of $\abdn$ and $\abdl$;
although, at low $\xh{M}$ the steeper mean stratifications of the 
3D models act to enhance the non-LTE over-ionisation 
\citep{2017A&A...597A...6N}.

\fig{fig:aberr_elo} indicates that the abundance differences 
for weak \ion{Fe}{I} lines depend on $\elo$.
For the cooler models,
$\abdl$ and $\abdn$ even change sign with increasing $\elo$.
At $\xh{M}=0$ there is a roughly linear trend
such that $\abdl$ and $\abdn$ decrease by
around $-0.05\,\dex$ to $-0.1\,\dex$
as $\elo$ increases from $0\,\eV$ to $5\,\eV$,
which reflects a combination of differences
in the mean stratifications and inhomogeneities
at different depths.
At $\xh{M}=-3$ the relationship becomes more complicated,
with an upturn at $\elo\approx3\,\eV$. 
This value corresponds to the broad peak in
ionisation and recombination events in
Fig.~3 of \citet{2012MNRAS.427...50L};
these levels are the ones most likely to absorb hot UV photons 
and undergo over-ionisation directly.

These trends with $\elo$ imply that 3D non-LTE effects
are implicit in 1D LTE spectroscopic determinations of $\teff$
that are based on the excitation balance of \ion{Fe}{I} lines.
It is beyond the scope of the present study
to attempt to derive grids of $\teff$ corrections here. 
Such an effort would be complicated by
the non-linear behaviour of $\abdl$ with $\elo$
and the strong dependence on $\reqw$ 
(\sect{results_reqw_vturb}),
such that the extent of the $\teff$ errors
would depend on the particular set of \ion{Fe}{I} lines
under consideration.

\subsection{Weak \ion{Fe}{II} lines} 
\label{results_elo_fe2}

At $\xh{M}=0$, for $\vmic=1\,\kms$, the left panels of 
\fig{fig:aberr_elo} show that for weak \ion{Fe}{II} lines, $\abdl$
ranges from $-0.15\,\dex$ to $+0.10\,\dex$,
similar to \ion{Fe}{I} (\sect{results_elo_fe1}).
There is a trend of more positive values
towards larger $\teff$ and smaller $\lgg$,
as can also be seen in the upper left panel
of \fig{fig:aberr_teff}.
At $\xh{M}=-3$, 
unfortunately there is just one \ion{Fe}{II} line 
in the current study with $-6.9<\reqw<-4.9$.
For this line the value of $\abdl$ is 
between $-0.11\,\dex$ and $-0.15\,\dex$,
changing mildly with $\teff$ or $\lgg$;
it gives rise to flat trends in the right 
panels of \fig{fig:aberr_teff}.
The non-LTE effects in \ion{Fe}{II} lines
are not significant at $\xh{M}=0$, and so $\abdn\approx\abdl$.
These abundance differences are thus primarily driven by 3D LTE effects,
which are discussed in \citet{2019A&A...630A.104A}.

There are, however, non-LTE effects for the warmest stars at
the lowest $\xh{M}$.
This can be seen via the subtle differences between the
upper and lower right panels of \fig{fig:aberr_elo}, which
indicate small 1D non-LTE effects.
We note that 
\ion{Fe}{II} lines can suffer from over-excitation 
and become weaker relative to 1D LTE (translating to
larger inferred $\lgeps{Fe}$ in non-LTE).
The non-LTE effects on \ion{Fe}{II} lines are even
larger in the metal-poor 3D models due to their
steeper gradients \citep{2005ApJ...618..939S}.
The 3D LTE versus 3D non-LTE abundance differences 
can be estimated by using
the logarithmic ratios of equivalent widths.
They are the most negative for the
high $\teff$, low $\lgg$, and low $\xh{M}$ models
(for similar reasons given in \sect{results_elo_fe1}).
At $\teff\approx6500\,\K$, $\lgg=4.0$, and
$\xh{M}=-3$, these differences are $-0.09\pm0.05\,\dex$, averaging
over all the \ion{Fe}{II} lines.
The differences are already much less severe for 
the model with similar $\teff$ and $\lgg$ but with $\xh{M}=-2\,\dex$,
where they amount to $-0.02\pm0.01$.

\fig{fig:aberr_elo} illustrates that 
for the cooler models, $\abdl$ typically has the same sign
for \ion{Fe}{II} lines as for \ion{Fe}{I} lines of 
$2\lesssim\elo/\eV\lesssim5$;
usually they are both negative.
The consequence of this is that 1D LTE analyses of 
lines of either of these species may underestimate $\lgeps{Fe}$.
On the other hand, the fact that the abundance differences
for \ion{Fe}{I} and \ion{Fe}{II}
usually have the same sign helps to counteract the systematic
errors on 1D LTE spectroscopic determinations of $\lgg$ to some extent,
at least for $\teff\lesssim6000\,\K$.

\section{Interpolation in stellar and line parameters}
\label{interpolator}

\begin{figure}
    \begin{center}
        \includegraphics[scale=0.71]{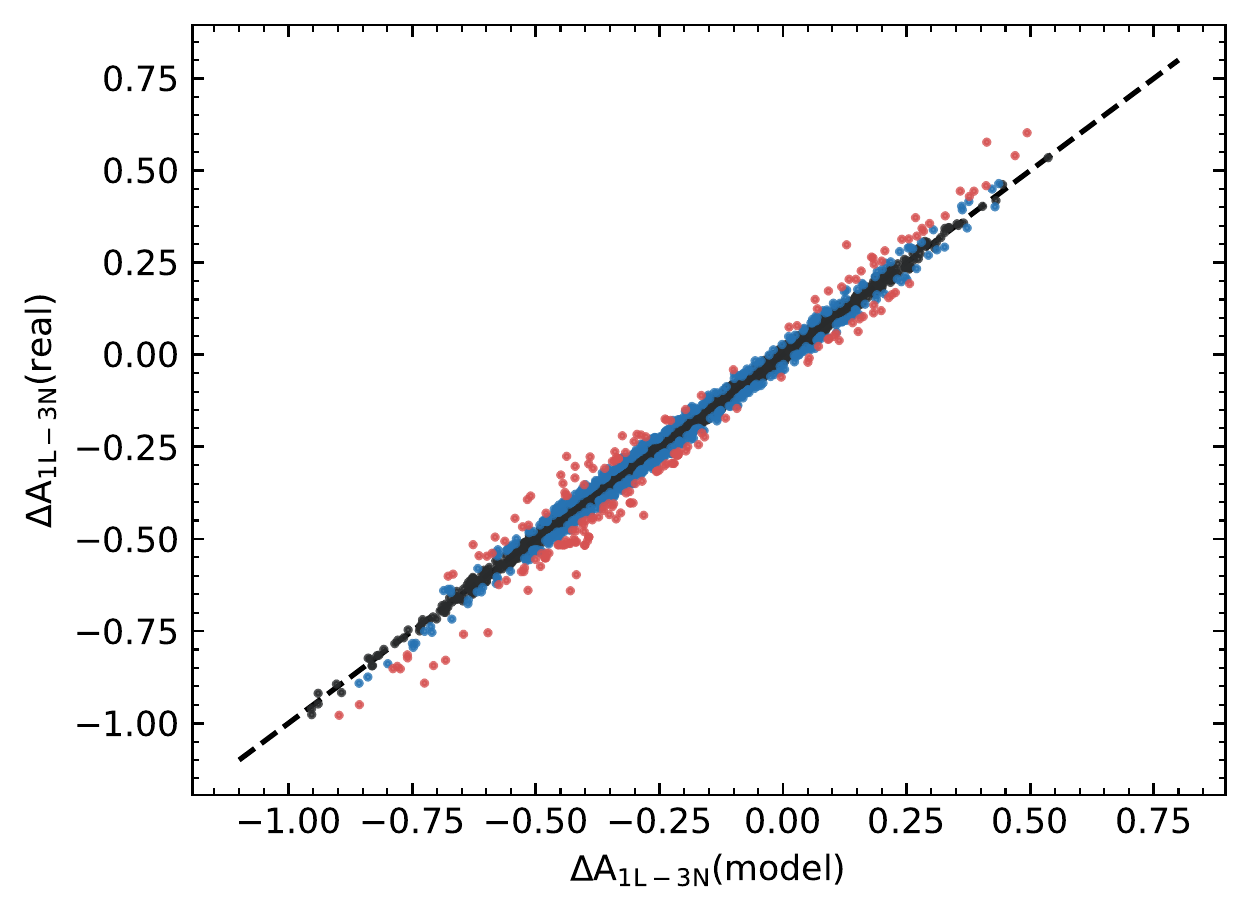}
        \includegraphics[scale=0.71]{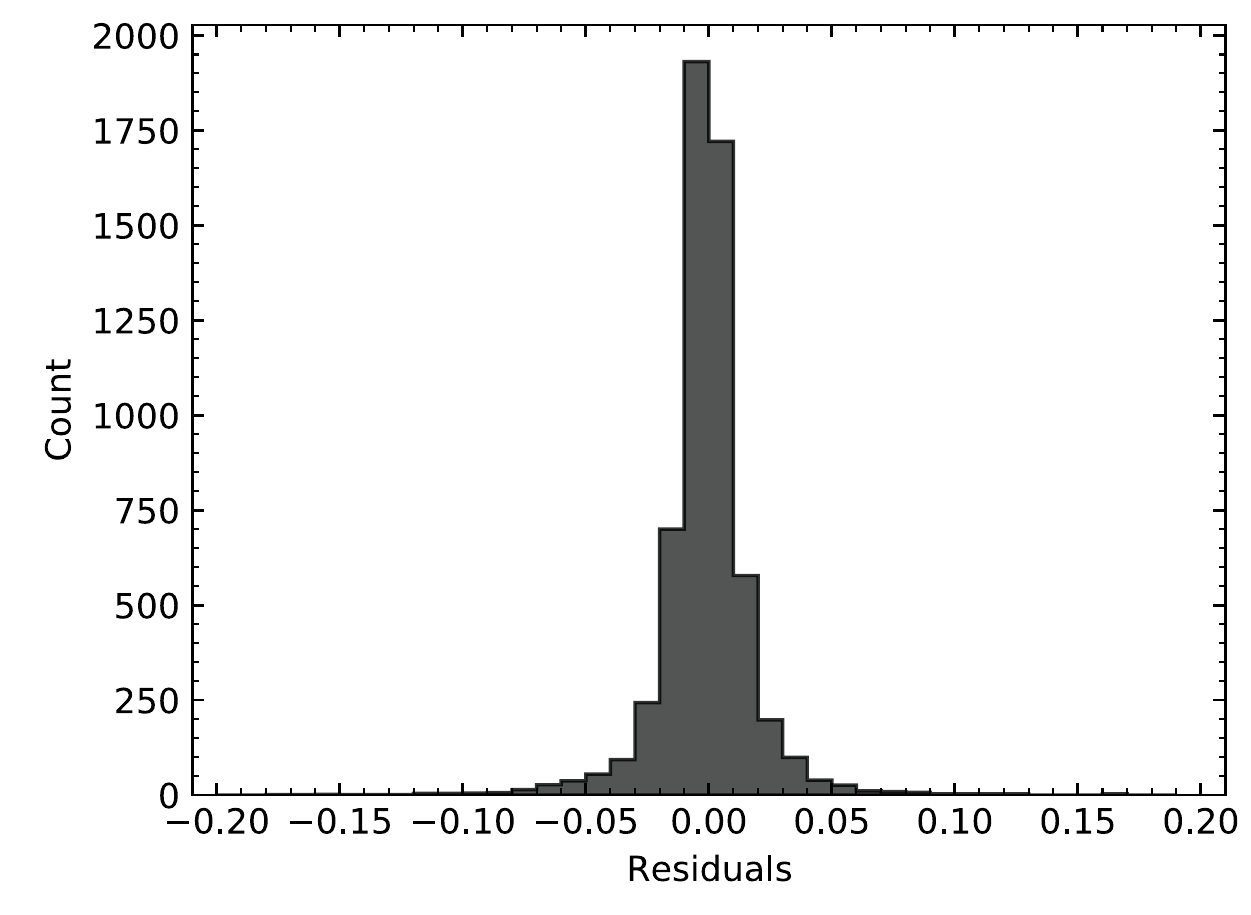}
        \caption{Predicted versus actual $\abdl$ colour-coded
        by the size of the interpolation error (upper panel),
        and distribution of interpolation errors in $\dex$ (lower panel)
        for the routine presented in \sect{interpolator}.
        Interpolation was carried out over stellar and line parameters 
        simultaneously.}
    \label{fig:interperror}
    \end{center}
\end{figure}

The derived values of $\abdl$ illustrated above (\sect{results})
were used to analyse 
standard stars, the results of which are presented in the following
section (\sect{gbs}).
To do so, it was necessary to interpolate these abundance
differences as a function of stellar parameters
($\teff$, $\lgg$, $\lgeps{Fe}_{\text{3N}}$, and $\vmic$)
and line parameters ($\elo$, $\eup$, and $\lggf$).

Several different approaches based on simple linear and spline
interpolation and more sophisticated machine-learning algorithms
were explored.
Ultimately, following \citet{2021MNRAS.500.2159W}, 
the interpolation model was based on
multi-layer perceptrons (MLP), which is a feed-forward fully connected
neural network that connects the input and output layers through $n_l$ number
of hidden layers. The hidden layers contain $n$ neurons, and each neuron is
connected to all the neurons in the previous layers with weights fitted through
back-propagation training. The models are constructed using the 
\texttt{MLPRegressor} class from the \textit{scikit-learn} python library
\citep{2012arXiv1201.0490P}.

Three different MLP models were built, separating 
the data into three groups: \ion{Fe}{I} lines with $\elo<2\,\eV$;
\ion{Fe}{I} lines with $\elo>2\,\eV$;
and \ion{Fe}{II} lines.  This was found to give more reliable
results than building a single MLP model for these three groups.
The tolerance was set to $10^{-5}$,
and the activation function used was the Rectified Linear Unit function
(ReLU). Via a $5$-fold cross-validation scheme,
the network size was set to $n_l = 3$ and $n =200$,
and the L2 regularisation
term (that prevents overfitting) was set to $\alpha = 5 \times 10^{-4}$,
$7 \times 10^{-4}$, and $10^{-2}$ for the three different
MLP models, respectively.
Via a $k$-fold cross-validation scheme using the MLP models,
a 
select number of severe outliers,
reflecting the difficulties in determining equivalent widths
(\sect{method_abundance}), were identified;
the MLP models were then rebuilt, with these
data removed.

To evaluate the performance of the MLP models, the datasets were randomly
divided into a training set ($80\%$ of the data) and a test set ($20\%$ of the
data).  The three MLP models were trained on their respective training sets,
and then applied to their respective test sets.  The overall performance of the
interpolation routine is illustrated in \fig{fig:interperror}, where
the histogram shows that the residuals of the three MLP models
have standard deviations of 
$0.024$, $0.017$, and $0.016\,\dex$, respectively.
They reflect the expected $1\,\sigma$ interpolation error when looking
at one \ion{Fe}{I} or \ion{Fe}{II} line in one star.
These interpolation errors translate to scatter
in plots of line-by-line $\lgeps{Fe}$ and average out when using
sufficiently large numbers of \ion{Fe}{I} and \ion{Fe}{II} lines.
Moreover the scatter plot in \fig{fig:interperror}
shows that the largest interpolation errors tend 
to occur where $\abdl$ is larger.

The input parameters of the interpolation routine are the stellar
parameters and line parameters listed above:
$\teff$, $\lgg$, $\lgeps{Fe}_{\text{3N}}$, and $\vmic$
as well as $\elo$, $\eup$, and $\lggf$.
The interpolation routine also returns results for stars outside
of the grid of stellar parameters 
($\teff$ between $5000\,\kelvin$ and $6500\,\kelvin$,
$\lgg$ between $4.0\,\dex$ and $4.5\,\dex$, and
$\lgeps{Fe}_{\text{3N}}$ between $4.5\,\dex$ and $7.5\,\dex$)
and line parameters 
(see \fig{fig:golden} and \sect{method_abundance}).
In such cases the returned results are intrinsically more uncertain,
and should be treated with caution. 
The possible impact of interpolation and extrapolation in line parameters 
is discussed more in \sect{gbs_extrapolation}
in the context of the application to standard stars.

The input parameter
$\lgeps{Fe}_{\text{3N}}$ is the 3D non-LTE iron abundance.
Since this is usually unknown, in principle
this routine needs to be applied iteratively.
In practice, when the abundance differences are small
or do not strongly vary with $\xh{M}$,
it is safe to make the approximation
$\lgeps{Fe}_{\text{3N}}\approx\lgeps{Fe}_{\text{1L}}$,
thus avoiding any iterations and allowing for a quick
estimate of the abundance correction 
(which is given by the negative of $\abdl$).
When the abundance differences are larger, there can be a notable
difference
in the final result.
For example, in the analysis of the star HD~84937 (\sect{gbs}),
the results for individual \ion{Fe}{I} lines changed by up to $0.05\,\dex$
when iterations were neglected,
and the mean difference changed by $0.016\,\dex$.
These differences are, however, within the overall uncertainty
in $\lgeps{Fe}$ for this star.
Nevertheless, in the present work, the routine was always applied iteratively
to avoid these small systematic errors.

To compare between 1D non-LTE and 3D non-LTE $\lgeps{Fe}$,
an analogous interpolation routine was constructed
to correct for 1D non-LTE effects. 
To facilitate a fair comparison, this was based on
the calculations from the $32$ \atmo{} models, rather than the 
finer grid of \marcs{} models (\sect{method_atmosphere}).
This 1D non-LTE interpolation routine uses
the same input parameters
as the 3D non-LTE one discussed above.
When applying this interpolation routine, it was assumed that
the 1D LTE and 1D non-LTE values of $\vmic{}$ were identical.
In practice, calibrations of $\vmic{}$ based on flattening
trends with a (reduced) equivalent width can be sensitive to non-LTE 
effects up to the order of $10\%$ 
\citep{2016MNRAS.463.1518A}.

\section{Application to standard stars}
\label{gbs}

\begin{table*}
\begin{center}
\caption{Adopted values and uncertainties for $\teff$, $\lgg$, and (for the 1D models) $\vmic$, as well as 1D non-LTE $\feh$ from the literature. Final columns show the values of [Fe/H] derived in this study in 1D LTE, 1D non-LTE, and 3D non-LTE (mean of $\feh_{\ion{Fe}{I}}$ and $\feh_{\ion{Fe}{II}}$ values weighted by statistical and stellar parameter uncertainties).}
\label{tab:stars}
\begin{tabular}{l |r r r r r | r r r}
\hline
\hline
\noalign{\smallskip}
Name & 
\multicolumn{1}{c}{$\teff / \kelvin$} & 
\multicolumn{1}{c}{$\lggu$} & 
\multicolumn{1}{c}{$\vmic / \kms$} & 
\multicolumn{1}{c}{$\feh_{\text{Lit.}}$} & 
\multicolumn{1}{c|}{Ref.} & 
\multicolumn{1}{c}{$\feh_{\text{1L}}$} & 
\multicolumn{1}{c}{$\feh_{\text{1N}}$} & 
\multicolumn{1}{c}{$\feh_{\text{3N}}$} \\ 
\noalign{\smallskip}
\hline
\hline
\noalign{\smallskip}
Sun & $5772$ &$ 4.44$ &$ 1.00$ &$ 0.00$ & a & $ 0.00$ & $ 0.00$ & $ 0.00$ \\ 
Procyon & $6554\pm  84$ &$ 4.00\pm 0.02$ &$ 1.66\pm 0.11$ &$-0.02\pm  0.08$ & b & $ 0.04\pm  0.03$ & $ 0.05\pm  0.03$ & $ 0.11\pm  0.03$ \\ 
HD 84937 & $6356\pm  97$ &$ 4.13\pm 0.03$ &$ 1.39\pm 0.24$ &$-2.06\pm  0.08$ & b,c & $-2.04\pm  0.03$ & $-2.03\pm  0.03$ & $-1.96\pm  0.03$ \\ 
HD 140283 & $5792\pm  55$ &$ 3.65\pm 0.02$ &$ 1.56\pm 0.20$ &$-2.29\pm  0.11$ & b,d & $-2.42\pm  0.03$ & $-2.38\pm  0.03$ & $-2.28\pm  0.02$ \\ 
\noalign{\smallskip}
\hline
\hline
\end{tabular}
\end{center}
\tablefoot{\tablefoottext{a}{Solar $\teff$ and $\lgg$ from \citet{2016AJ....152...41P}.} \tablefoottext{b}{Standard \gbs{} parameters from \citet{2015A&A...582A..49H} and \citet{2015A&A...582A..81J}; $\feh_{\text{Lit.}}$ shifted by $-0.03\,\dex$ such that the solar value is zero.} \tablefoottext{c}{Updated $\lgg$ from \citet{2021A&A...650A.194G}.} \tablefoottext{d}{Updated $\teff$, $\lgg$, and $\feh_{\text{Lit.}}$ from \citet{2020A&A...640A..25K}.}} 
\end{table*}
 
\begin{table*}
\begin{center}
\caption{Breakdown of uncertainties propagated onto $\lgeps{Fe}$. The uncertainties in solar parameters are neglected.}
\label{tab:errors}
\begin{tabular}{l | r r r r | r r r r | r r r r}
\hline
\hline
\noalign{\smallskip}
\multirow{2}{*}{Name} & 
\multicolumn{4}{c|}{1D LTE} & 
\multicolumn{4}{c|}{1D non-LTE} & 
\multicolumn{4}{c}{3D non-LTE} \\ 
 & 
\multicolumn{1}{c}{$\sigma_{\text{stat.}}$} & 
\multicolumn{1}{c}{$\sigma_{\teff}$} & 
\multicolumn{1}{c}{$\sigma_{\lgg}$} & 
\multicolumn{1}{c|}{$\sigma_{\vmic}$} & 
\multicolumn{1}{c}{$\sigma_{\text{stat.}}$} & 
\multicolumn{1}{c}{$\sigma_{\teff}$} & 
\multicolumn{1}{c}{$\sigma_{\lgg}$} & 
\multicolumn{1}{c|}{$\sigma_{\vmic}$} & 
\multicolumn{1}{c}{$\sigma_{\text{stat.}}$} & 
\multicolumn{1}{c}{$\sigma_{\teff}$} & 
\multicolumn{1}{c}{$\sigma_{\lgg}$} & 
\multicolumn{1}{c}{$\sigma_{\vmic}$} \\ 
\noalign{\smallskip}
\hline
\hline
\noalign{\smallskip}
\multicolumn{1}{c}{ } & \multicolumn{4}{c}{\ion{Fe}{I}} & 
 \multicolumn{4}{c}{\ion{Fe}{I}} \\ 
\noalign{\smallskip}
Sun & $ 0.011$ & $ 0.000$ & $ 0.000$ & $ 0.000$ & $ 0.010$ & $ 0.000$ & $ 0.000$ & $ 0.000$ & $ 0.011$ & $ 0.000$ & $ 0.000$ & $ 0.000$ \\ 
Procyon & $ 0.005$ & $ 0.050$ & $ 0.000$ & $ 0.013$ & $ 0.006$ & $ 0.055$ & $ 0.001$ & $ 0.011$ & $ 0.006$ & $ 0.064$ & $ 0.003$ & $ 0.001$ \\ 
HD 84937 & $ 0.008$ & $ 0.068$ & $ 0.000$ & $ 0.015$ & $ 0.005$ & $ 0.065$ & $ 0.003$ & $ 0.011$ & $ 0.007$ & $ 0.074$ & $ 0.003$ & $ 0.002$ \\ 
HD 140283 & $ 0.010$ & $ 0.045$ & $ 0.000$ & $ 0.018$ & $ 0.007$ & $ 0.049$ & $ 0.002$ & $ 0.014$ & $ 0.008$ & $ 0.053$ & $ 0.003$ & $ 0.002$ \\ 
\noalign{\smallskip}
\multicolumn{1}{c}{ } & \multicolumn{4}{c}{\ion{Fe}{II}} & 
 \multicolumn{4}{c}{\ion{Fe}{II}} \\ 
\noalign{\smallskip}
Sun & $ 0.016$ & $ 0.000$ & $ 0.000$ & $ 0.000$ & $ 0.017$ & $ 0.000$ & $ 0.000$ & $ 0.000$ & $ 0.018$ & $ 0.000$ & $ 0.000$ & $ 0.000$ \\ 
Procyon & $ 0.024$ & $ 0.005$ & $ 0.008$ & $ 0.024$ & $ 0.024$ & $ 0.005$ & $ 0.008$ & $ 0.024$ & $ 0.023$ & $ 0.001$ & $ 0.009$ & $ 0.003$ \\ 
HD 84937 & $ 0.014$ & $ 0.014$ & $ 0.011$ & $ 0.011$ & $ 0.013$ & $ 0.012$ & $ 0.011$ & $ 0.012$ & $ 0.014$ & $ 0.010$ & $ 0.010$ & $ 0.000$ \\ 
HD 140283 & $ 0.020$ & $ 0.008$ & $ 0.008$ & $ 0.017$ & $ 0.021$ & $ 0.008$ & $ 0.010$ & $ 0.016$ & $ 0.017$ & $ 0.007$ & $ 0.007$ & $ 0.007$ \\ 
\noalign{\smallskip}
\hline
\hline
\end{tabular}
\end{center}
\end{table*}

\begin{figure*}
    \begin{center}
        \includegraphics[scale=0.325]{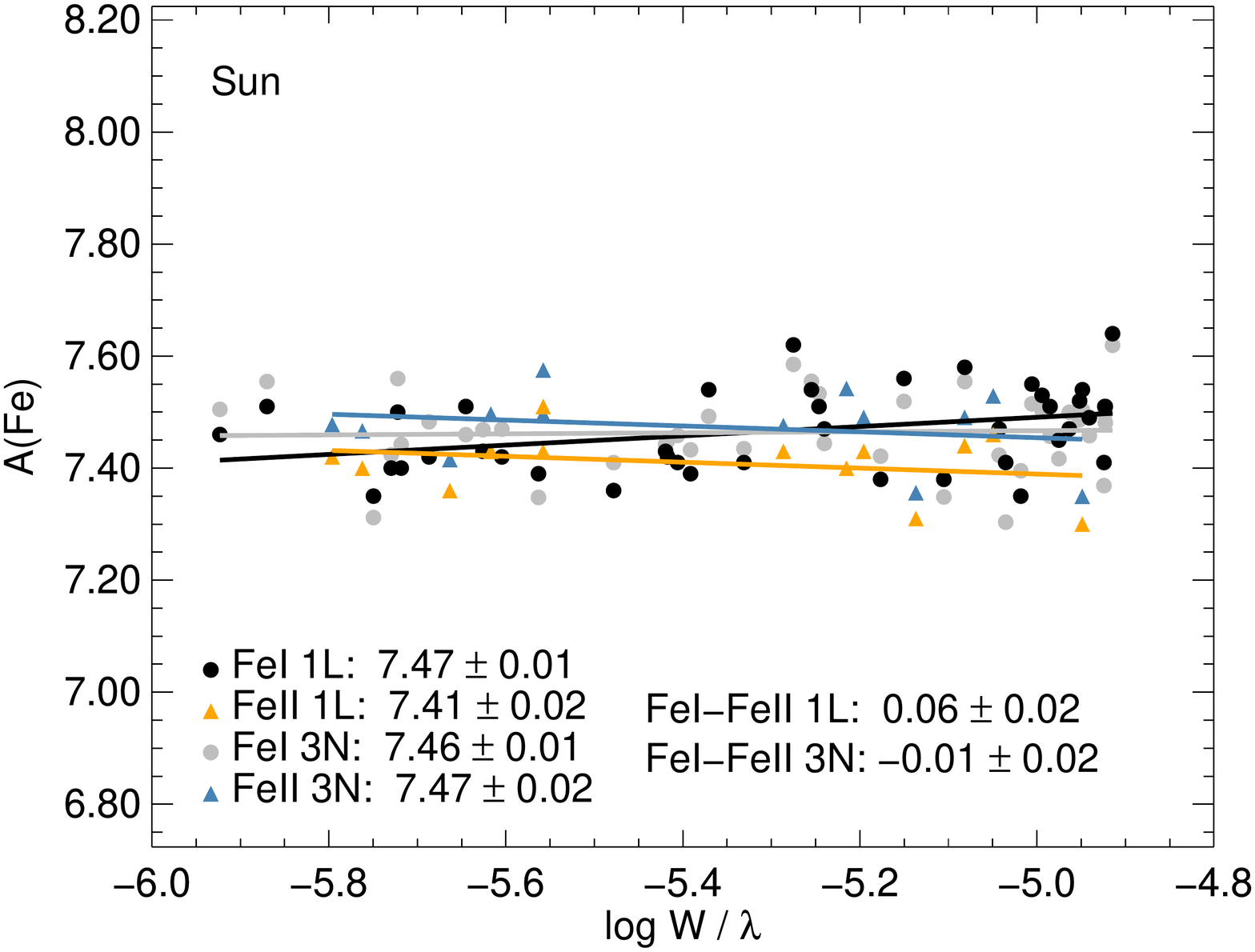}\includegraphics[scale=0.325]{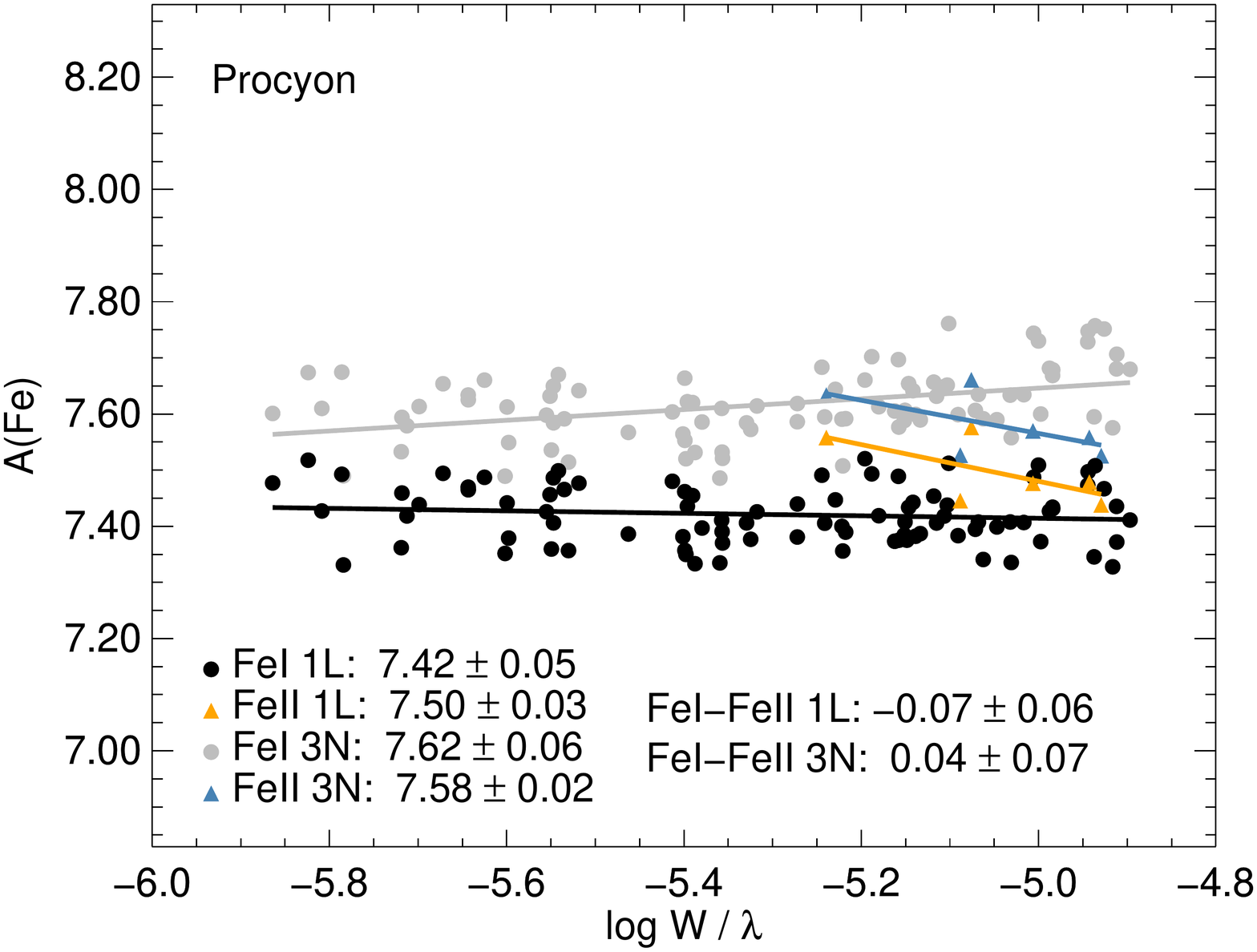}
        \includegraphics[scale=0.325]{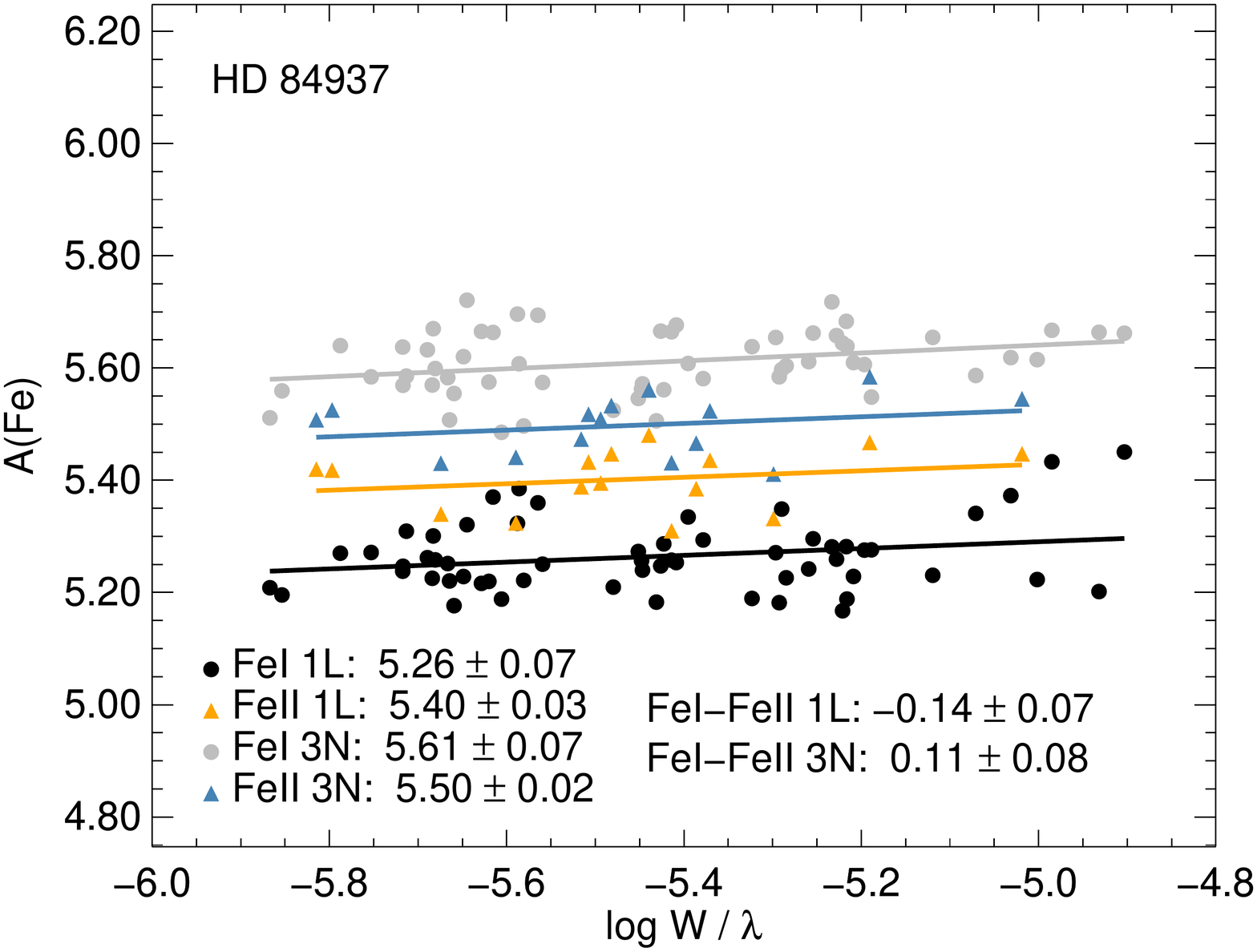}\includegraphics[scale=0.325]{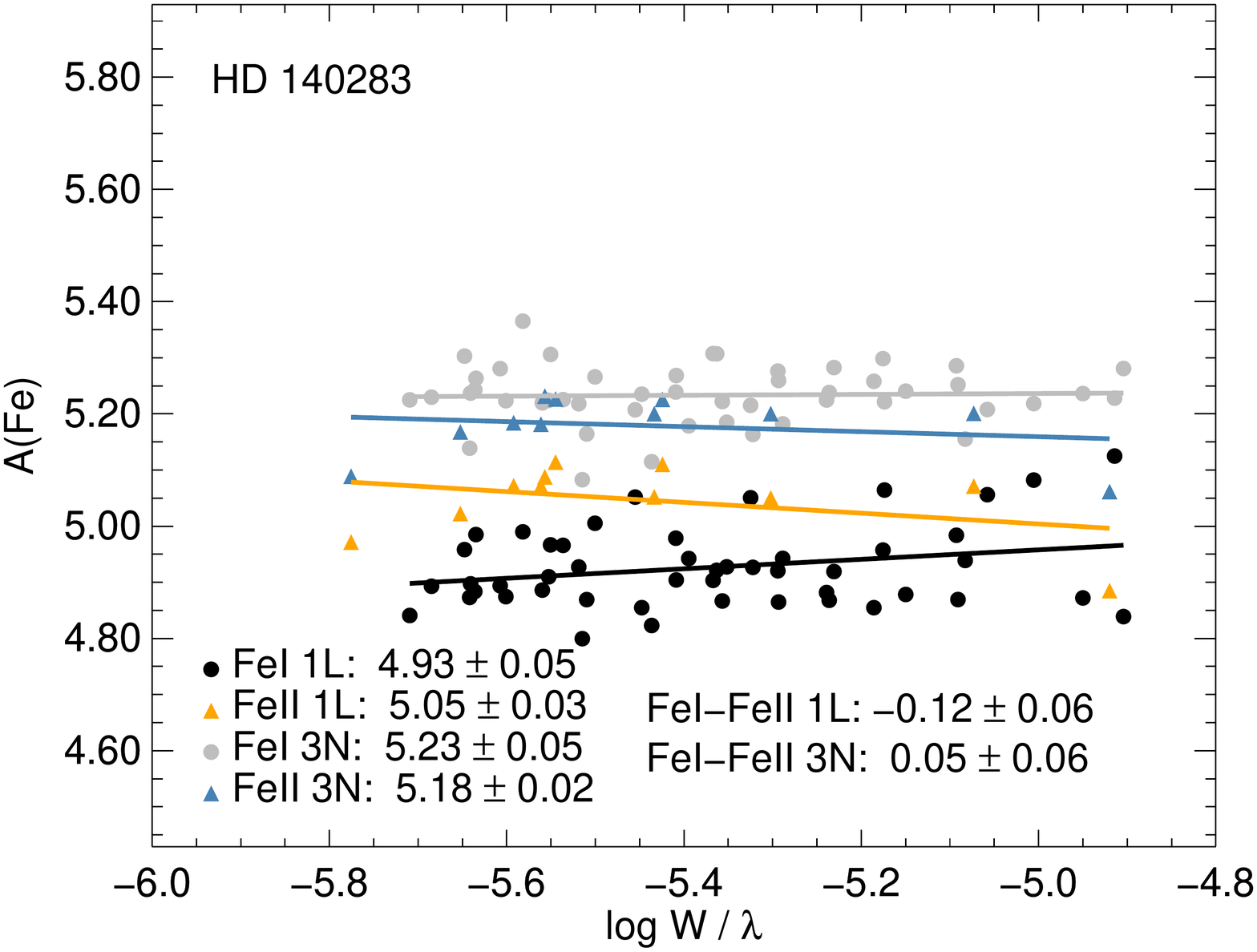}
        \caption{Line-by-line 
        $\lgeps{Fe}$ versus $\reqw$. 
        We note that 1D LTE and 3D non-LTE \ion{Fe}{I} and \ion{Fe}{II} lines
        are indicated separately,
        with least-squares fits being overdrawn.}
        \label{fig:abundances1}
    \end{center}
\end{figure*}

\begin{figure*}
    \begin{center}
        \includegraphics[scale=0.325]{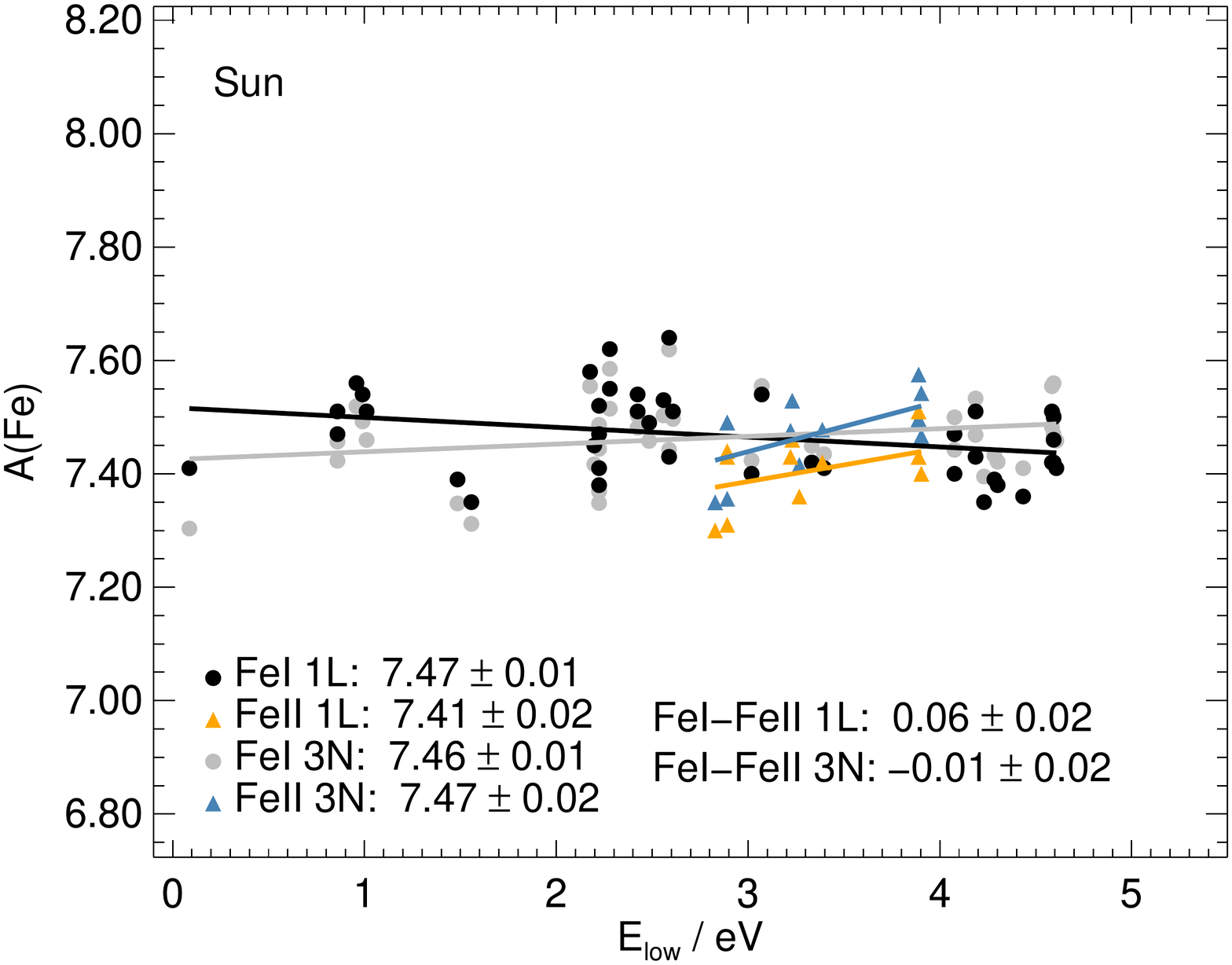}\includegraphics[scale=0.325]{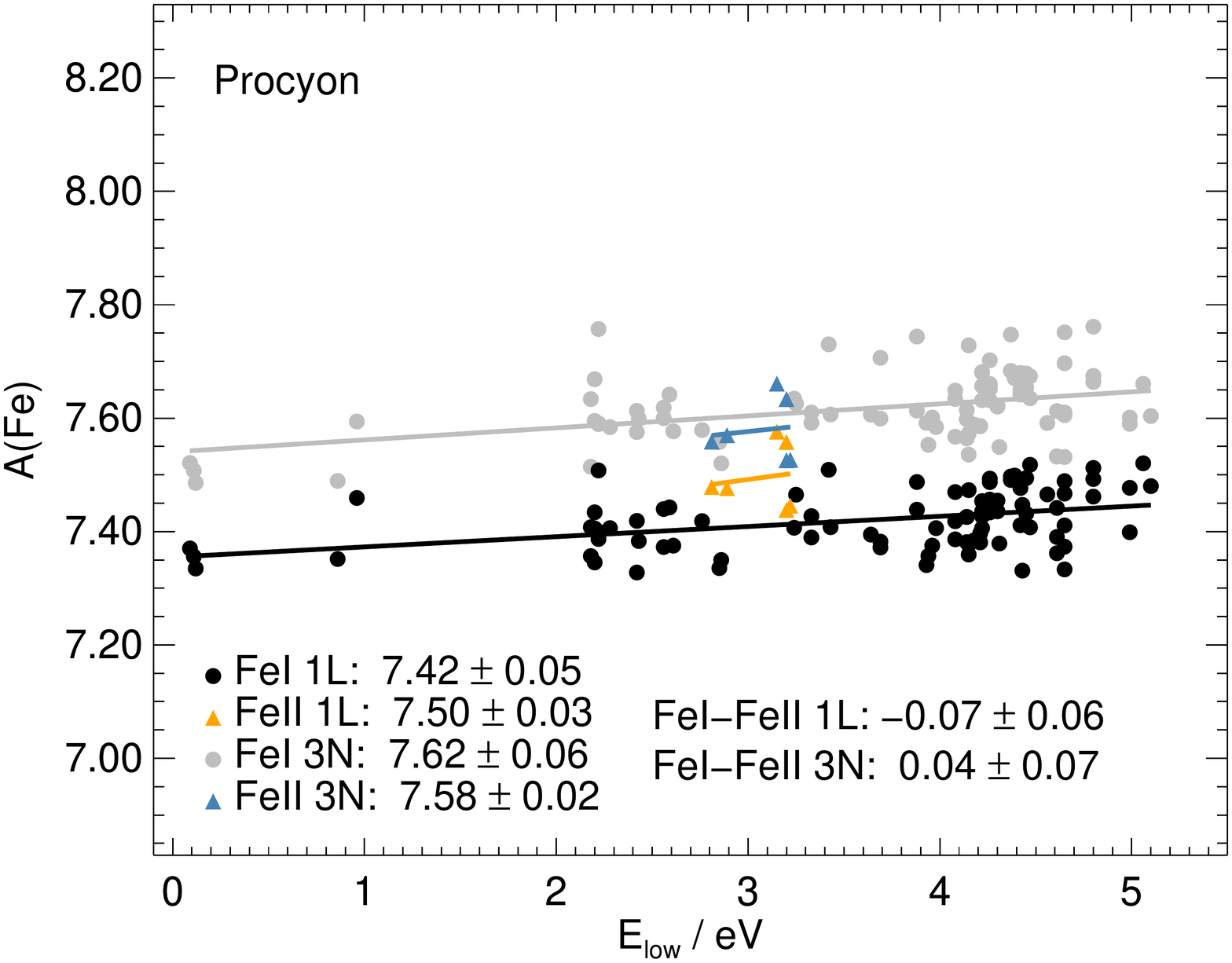}
        \includegraphics[scale=0.325]{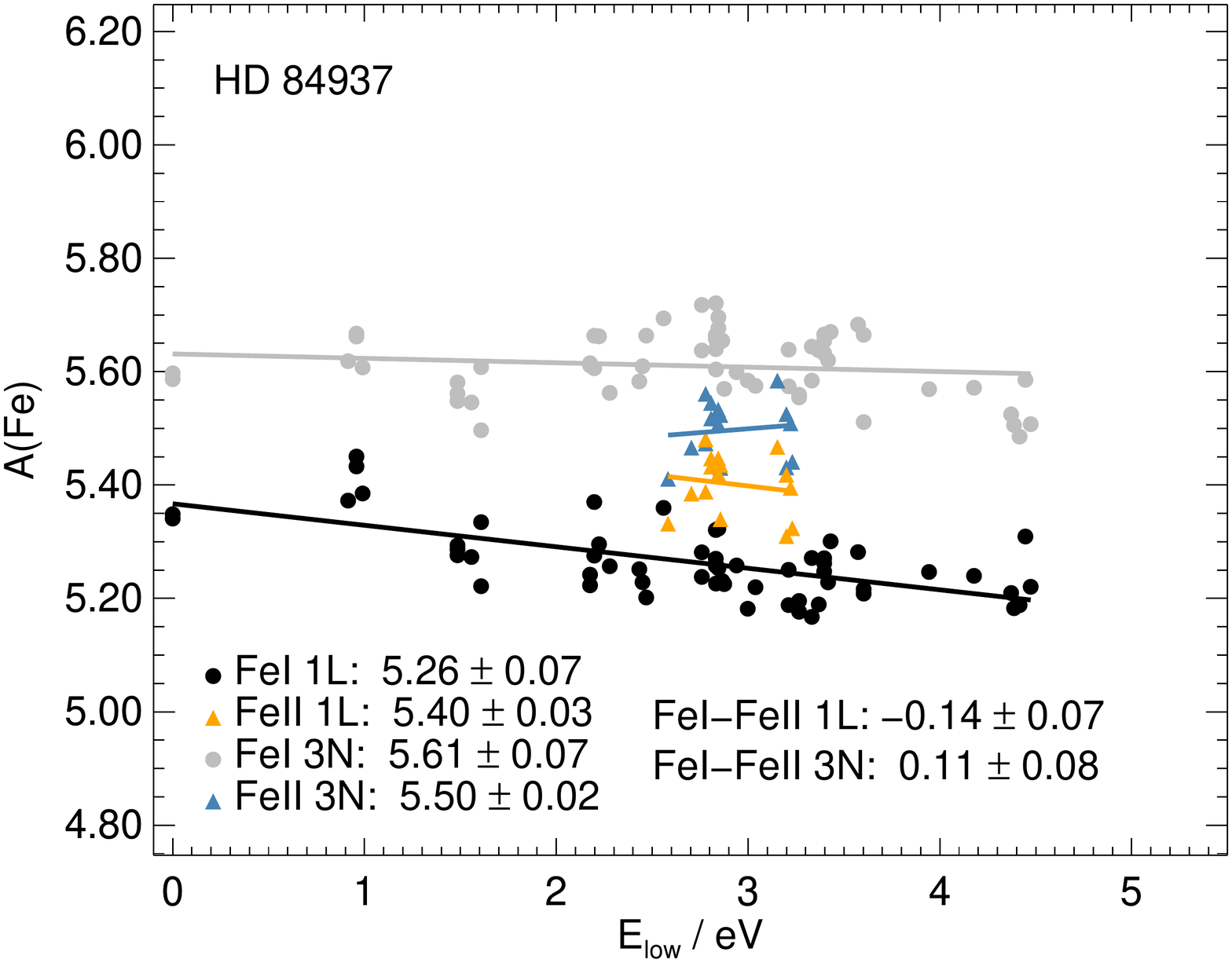}\includegraphics[scale=0.325]{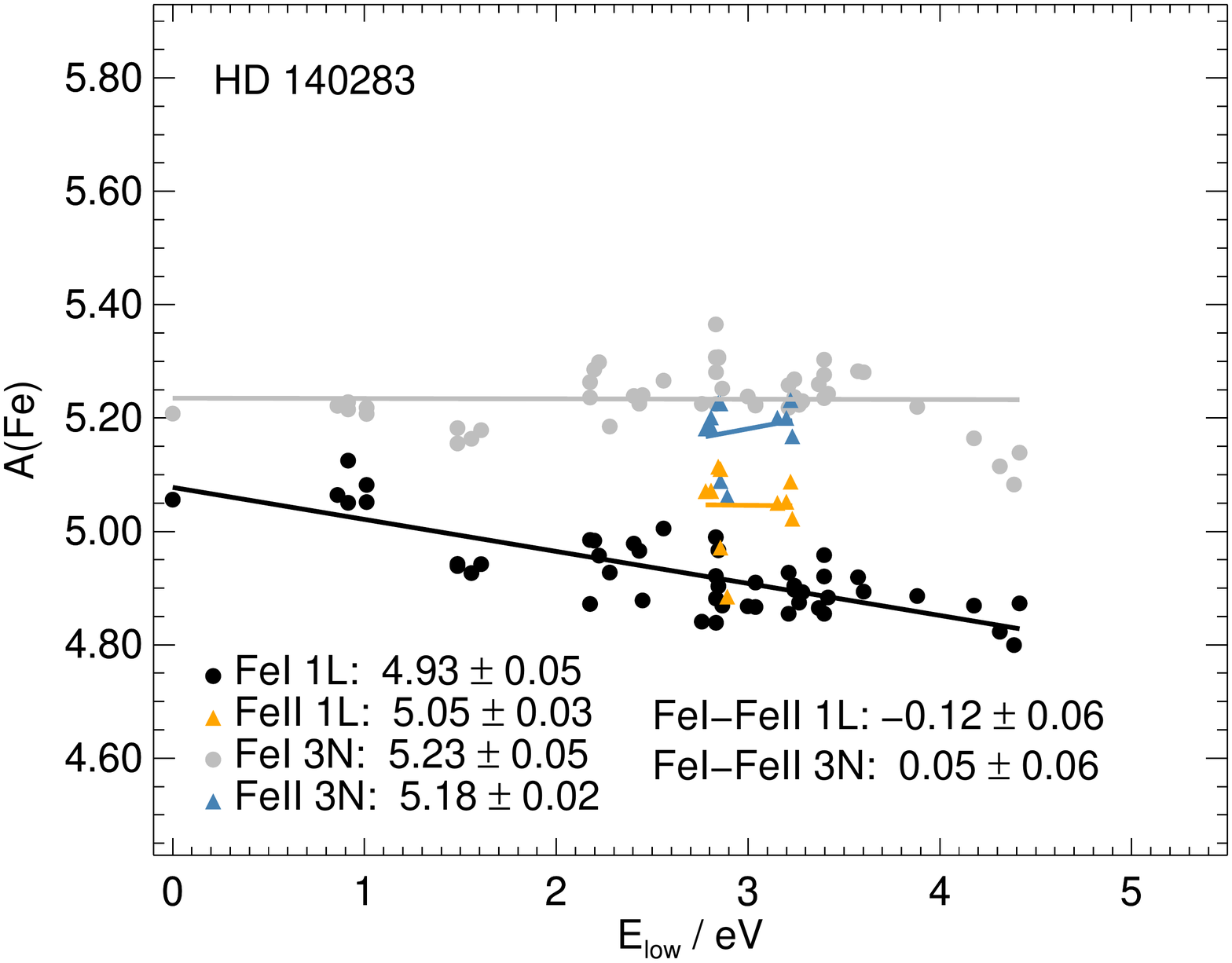}
        \caption{Line-by-line $\lgeps{Fe}$ versus $\elo$.
        We note that 1D LTE and 3D non-LTE \ion{Fe}{I} and \ion{Fe}{II} lines
        are indicated separately,
        with least-squares fits being overdrawn.}
        \label{fig:abundances2}
    \end{center}
\end{figure*}

\begin{figure*}
    \begin{center}
        \includegraphics[scale=0.325]{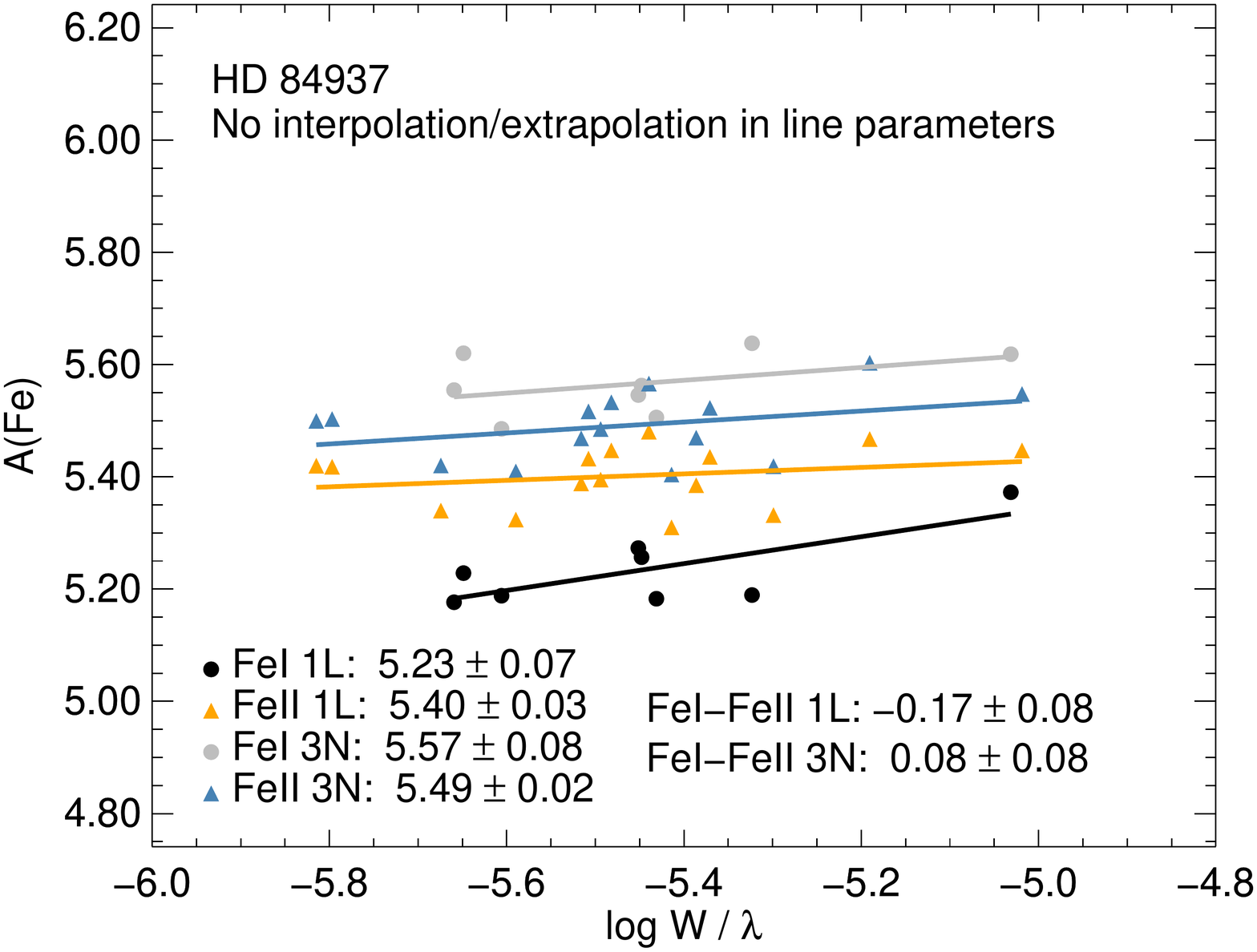}\includegraphics[scale=0.325]{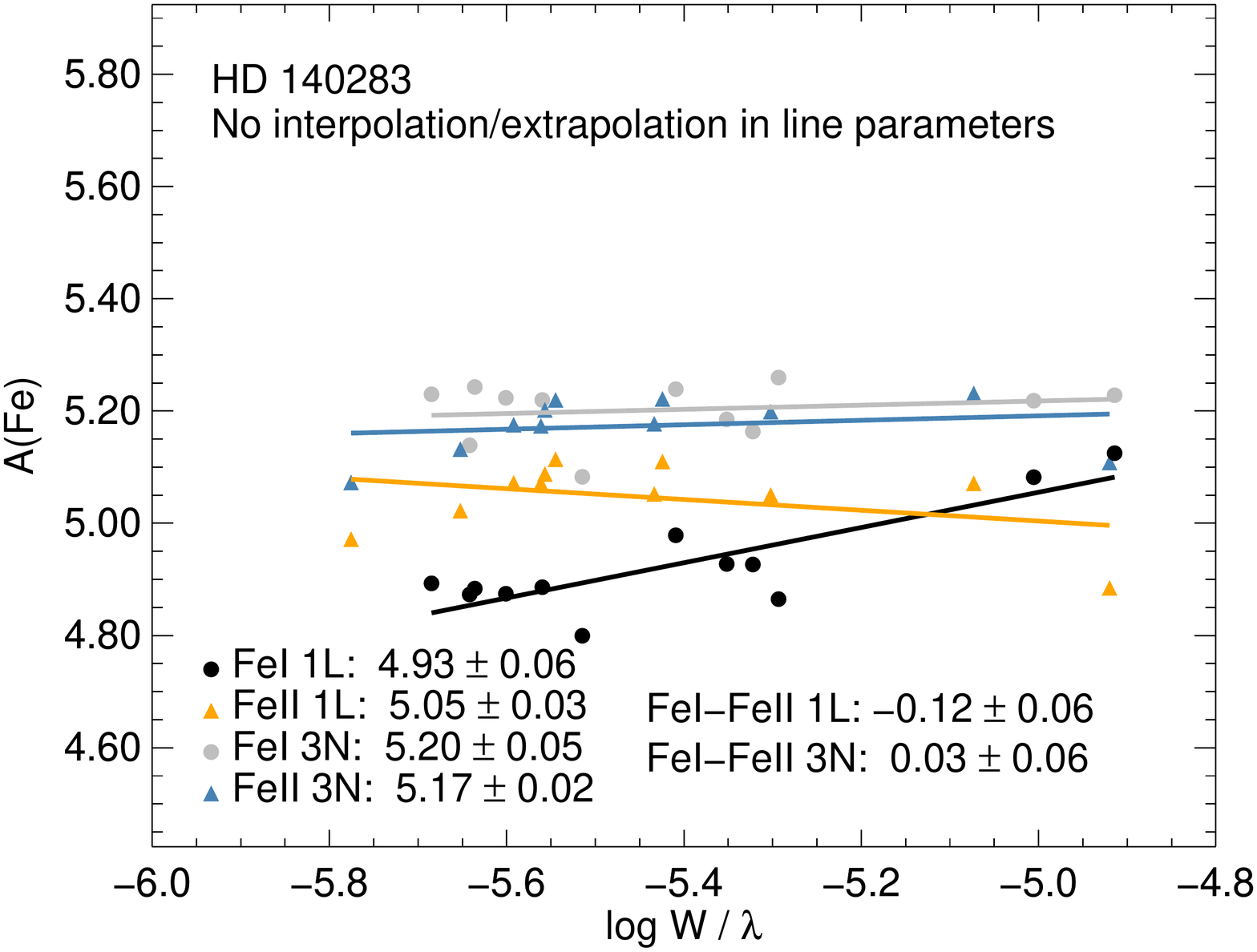}
        \includegraphics[scale=0.325]{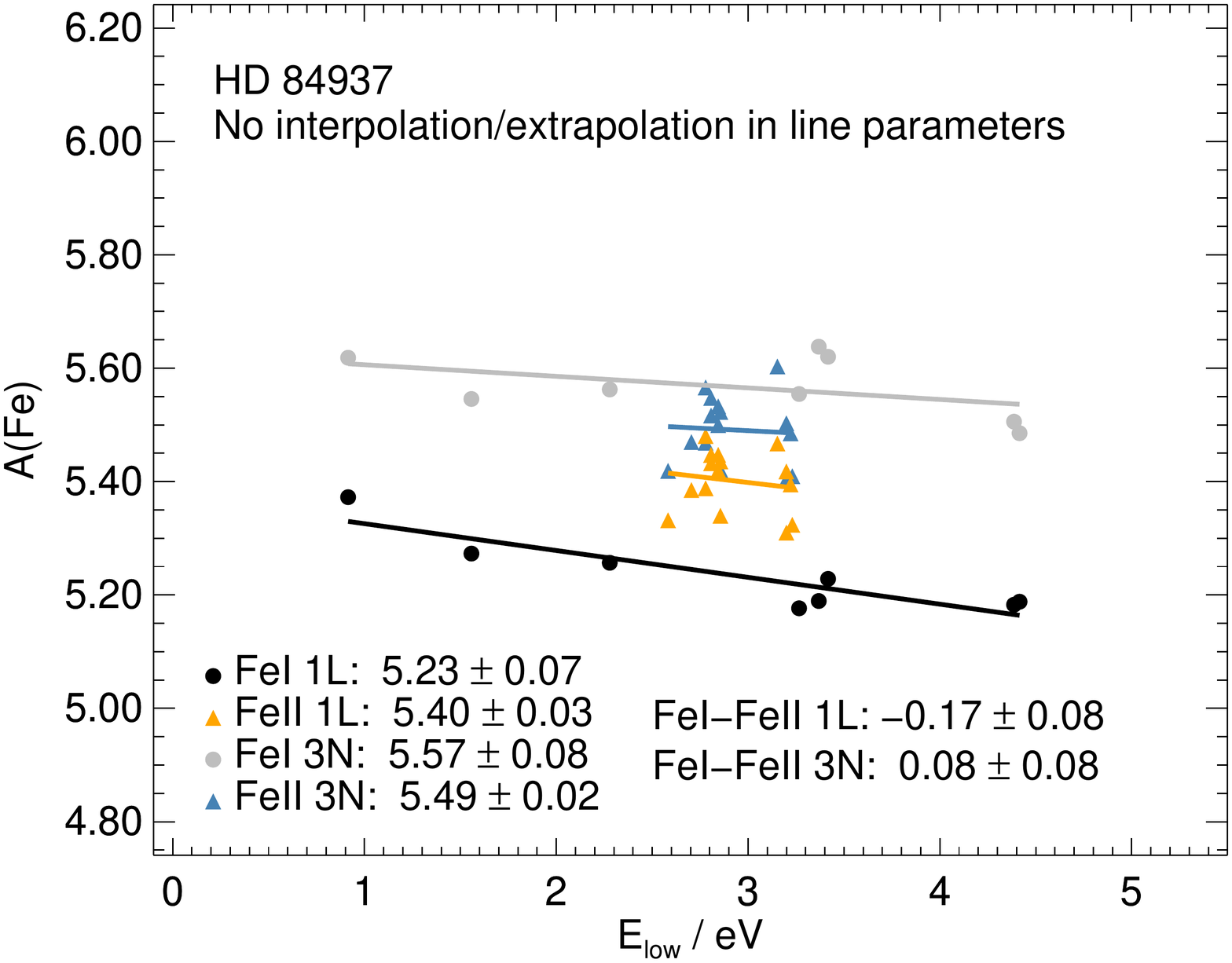}\includegraphics[scale=0.325]{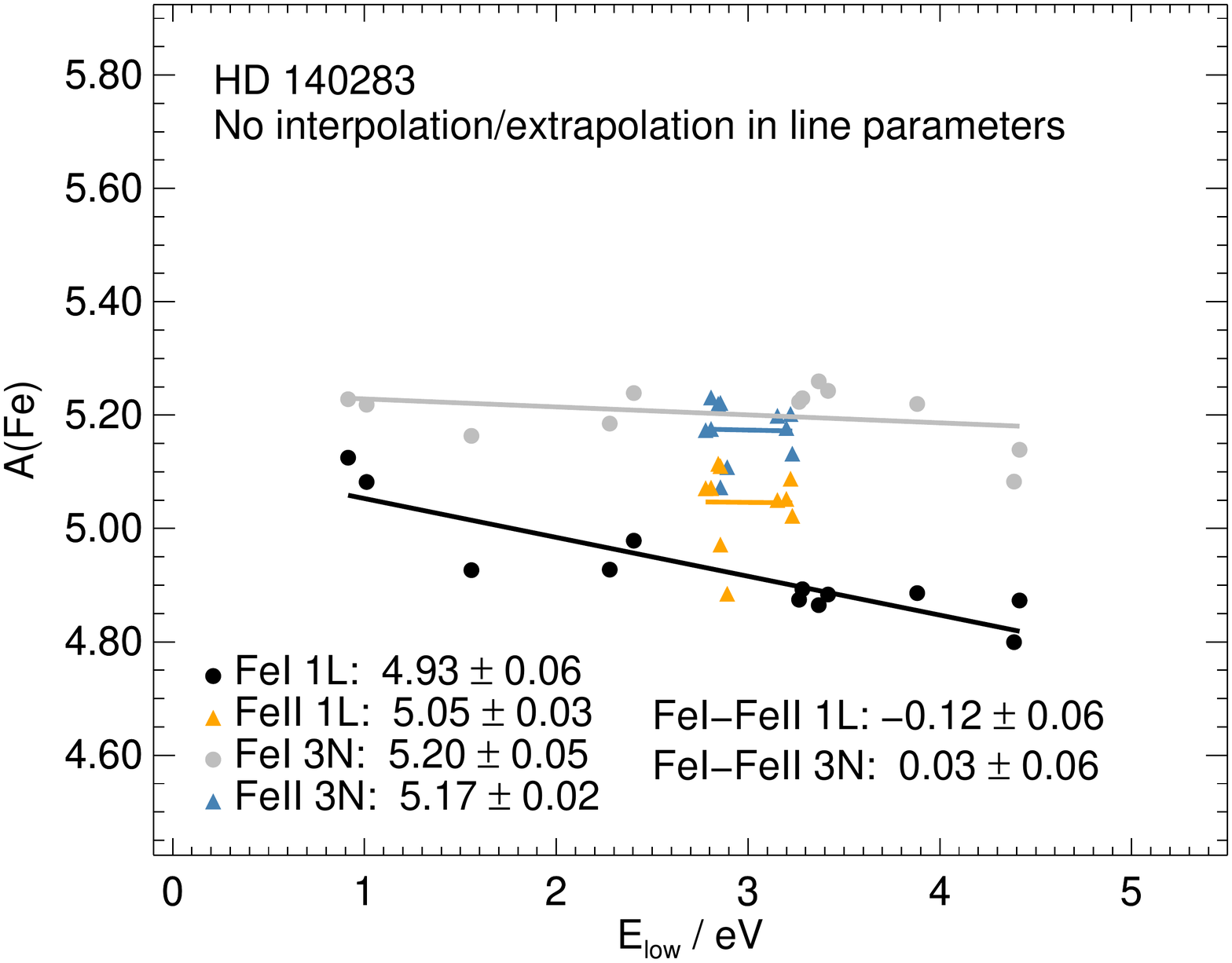}
        \caption{Same as \fig{fig:abundances1} and 
        \fig{fig:abundances2}, but for the metal-poor stars and restricted to the \ion{Fe}{I} lines that are included
        within the $\abdl$ grid (\fig{fig:golden}), and using 
        the extended 3D LTE grid from 
        \citet{2019A&A...630A.104A} for \ion{Fe}{II} lines.}
        \label{fig:abundances3}
    \end{center}
\end{figure*}

\begin{table*}
\begin{center}
\caption{$\lgeps{Fe}$ inferred from \ion{Fe}{I} and \ion{Fe}{II} lines in 1D LTE, 1D non-LTE, and 3D non-LTE. Errors reflect statistical and stellar parameter uncertainties (not model uncertainties).}
\label{tab:abundance}
\begin{tabular}{l | r r|  r r| r r}
\hline
\hline
\noalign{\smallskip}
\multirow{2}{*}{Name} & 
\multicolumn{2}{c|}{1D LTE} & 
\multicolumn{2}{c|}{1D non-LTE} & 
\multicolumn{2}{c}{3D non-LTE} \\ 
 & 
\multicolumn{1}{c}{\ion{Fe}{I}} & 
\multicolumn{1}{c|}{\ion{Fe}{II}} & 
\multicolumn{1}{c}{\ion{Fe}{I}} & 
\multicolumn{1}{c|}{\ion{Fe}{II}} & 
\multicolumn{1}{c}{\ion{Fe}{I}} & 
\multicolumn{1}{c}{\ion{Fe}{II}} \\ 
\noalign{\smallskip}
\hline
\hline
\noalign{\smallskip}
Sun & $ 7.47\pm 0.01$ &$ 7.41\pm 0.02$ &$ 7.45\pm 0.01$ &$ 7.41\pm 0.02$ &$ 7.46\pm 0.01$  & $ 7.47\pm 0.02$  \\
Procyon & $ 7.42\pm 0.05$ &$ 7.50\pm 0.03$ &$ 7.45\pm 0.06$ &$ 7.49\pm 0.03$ &$ 7.62\pm 0.06$  & $ 7.58\pm 0.02$  \\
HD 84937 & $ 5.26\pm 0.07$ &$ 5.40\pm 0.03$ &$ 5.43\pm 0.07$ &$ 5.38\pm 0.02$ &$ 5.61\pm 0.07$  & $ 5.50\pm 0.02$  \\
HD 140283 & $ 4.93\pm 0.05$ &$ 5.05\pm 0.03$ &$ 5.08\pm 0.05$ &$ 5.03\pm 0.03$ &$ 5.23\pm 0.05$  & $ 5.18\pm 0.02$  \\
\noalign{\smallskip}
\hline
\hline
\end{tabular}
\end{center}
\end{table*}

\begin{table}
\begin{center}
\caption{Ionisation imbalance $\mathrm{\lgeps{Fe}_{\ion{Fe}{I}}-\lgeps{Fe}_{\ion{Fe}{II}}}$ in 1D LTE, 1D non-LTE, and 3D non-LTE. Errors reflect statistical and stellar parameter uncertainties (not model uncertainties).}
\label{tab:imbalance}
\begin{tabular}{l | r r r}
\hline
\hline
\noalign{\smallskip}
Name & 
\multicolumn{1}{c}{1D LTE} & 
\multicolumn{1}{c}{1D non-LTE} & 
\multicolumn{1}{c}{3D non-LTE} \\ 
\noalign{\smallskip}
\hline
\hline
\noalign{\smallskip}
Sun & $ 0.06\pm 0.02$ &$ 0.04\pm 0.02$ &$-0.01\pm 0.02$ \\
Procyon & $-0.07\pm 0.06$ &$-0.04\pm 0.07$ &$ 0.04\pm 0.07$ \\
HD 84937 & $-0.14\pm 0.07$ &$ 0.05\pm 0.07$ &$ 0.11\pm 0.08$ \\
HD 140283 & $-0.12\pm 0.06$ &$ 0.05\pm 0.06$ &$ 0.05\pm 0.06$ \\
\noalign{\smallskip}
\hline
\hline
\end{tabular}
\end{center}
\end{table}

\subsection{Stellar parameters}
\label{gbs_stars}

This section demonstrates the
possible impact of $\abdl$ in practice.
For this purpose, as well as the Sun,
the standard stars Procyon (HD~61421), HD~84937, and HD~140283
were selected to be analysed. These stars
are interesting in this context because
they are located close to different edges of 
the grid of $\abdl$ in $\teff$ and $\xh{M}$.
Moreover, they are all in the 
original \gbs{} catalogue
\citep{2015A&A...582A..49H},
implying that they have relatively well-constrained
stellar parameters, and that they are of astrophysical importance
for calibrating and validating spectroscopic studies
\citep{2017A&A...598A...5P,2021MNRAS.506..150B}.
Lastly, the Sun \citep{2021A&A...653A.141A}
and the two metal-poor stars \citep{2016MNRAS.463.1518A}
have previously been studied using the same or
similar model atoms, and thus serve as a rough consistency
check of the grid-based approach adopted here.

The adopted stellar parameters of these stars are 
given in \tab{tab:stars}. 
Where possible, the parameters were updated 
to account for the most recent and reliable measurements.
These stars were analysed in slightly 
different ways, as discussed below (\sect{gbs_analysis}).

\subsection{Analysis}
\label{gbs_analysis}

The Sun was analysed in the simplest way.
Line-by-line 1D LTE $\lgeps{Fe}$ values measured in the 
solar flux atlas of \citet{1984sfat.book.....K}
were taken from \citet{2002ApJ...567..544A}.
The analysis was restricted to weak lines with
$\reqw<-4.9$ because partially saturated lines are more sensitive
to uncertainties in the equivalent widths
and also correspond to larger values of $\abdl$
(\sect{results_reqw_vturb}).
The $\lgeps{Fe}$ values for the \ion{Fe}{II} lines were
corrected for the more precise values
of $\lggf$ provided in \citet{2009A&A...497..611M}.
The 1D LTE values were then
corrected to obtain 3D non-LTE values,
using the interpolation routine described in \sect{interpolator}.
The interpolation routine uses
the 3D non-LTE $\lgeps{Fe}$ as an input parameter, and returns the
1D LTE versus 3D non-LTE abundance difference.
Since the 3D non-LTE $\lgeps{Fe}$ 
is initially unknown,
the interpolation routine was applied iteratively,
using the 1D LTE value as the first guess.

For Procyon (HD~61421), \citet{2002ApJ...567..544A} also provide 1D LTE $\lgeps{Fe}$.
However, their $\teff$ and $\lgg$ are smaller than those recommended by 
\citet{2012A&A...540A...5C} and \citet{2015A&A...582A..49H}.
Moreover, their \ion{Fe}{II} line list mostly consists of
partially saturated lines;
combined with their choice of $\vmic=2.2\,\kms$,
the corresponding $\abdl$ becomes large (\sect{results_reqw_vturb})
and more sensitive to the way in which the 1D LTE abundances were derived.
Instead, this star was analysed using the lines in Tables 4 and 5
of \citet{2014A&A...564A.133J}
(the same line list that is the basis of this study,
and is illustrated in \fig{fig:golden}).
The authors provide
up to five different measurements of the 
equivalent width for each line
(labelled \textit{epinarbo}, \textit{ucm}, \textit{porto}, \textit{bologna},
and \textit{ulb}).
For a given line, 1D LTE values of $\lgeps{Fe}$ were
determined based on each of these measurements separately via
spline interpolation of the theoretical grid of equivalent widths
based on \marcs{} model stellar atmospheres (\sect{method}).
The results from the (at most) five different equivalent widths
were then individually compared to the overall mean
1D LTE value from all of the \ion{Fe}{I} lines.
Those that were more than $2\,\sigma$ from the mean were
removed,
where $\sigma$ here is the standard deviation of $\lgeps{Fe}$
from the different \ion{Fe}{I} lines.
This sigma clipping was evaluated using the 1D LTE results;
the exact same clipping was then later adopted for the 3D non-LTE analysis.
The surviving data were then averaged together to get a final 1D LTE 
$\lgeps{Fe}$ for that particular star and from that particular 
\ion{Fe}{I} line.  This proceeded iteratively. 
An identical approach was adopted for \ion{Fe}{II} lines, except comparing 
them with the overall mean
1D LTE $\lgeps{Fe}$ from \ion{Fe}{II} lines.
Lastly, the 1D LTE values of $\lgeps{Fe}$
were corrected to obtain 3D non-LTE values,
using the interpolation routine described in \sect{interpolator}
in the same way as was done for the Sun,
albeit with $0$th-order extrapolation in $\teff$ and
$\lgeps{Fe}_{\text{3N}}$ (adopting the values at 
$\teff=6500\,\K$ and $\lgeps{Fe}_{\text{3N}}=7.5$).
As for the Sun, this analysis was restricted to weak lines with $\reqw<-4.9$.

The analyses of the metal-poor stars
HD~84937 and HD~140283 that were presented in \citet{2014A&A...564A.133J}
are based on a few lines, especially for \ion{Fe}{II}.
Here, these stars are instead analysed using
the lines, equivalent widths, and
theoretical 1D LTE grids presented in \citet{2016MNRAS.463.1518A}.
The stellar parameters were updated to those given in \tab{tab:stars}.
In all other ways, the analysis proceeded as described for Procyon.
In particular, the interpolation routine provided in this work
was iteratively used to correct the 1D LTE to 3D non-LTE values as before,
with $0$th-order extrapolation in $\lgg$ for
HD~140283 (adopting the value at $\lgg=4.0$).

Statistical uncertainties were calculated for each star
as the standard error in the mean for \ion{Fe}{I} and \ion{Fe}{II} separately.
These primarily 
reflect uncertainties in the equivalent width measurements
and adopted values of $\lggf$.
For Procyon, HD~84937, and HD~140283,
systematic uncertainties were estimated by repeating the above steps 
(though for the purposes of this study with a fixed selection of lines and equivalent widths 
in the case of Procyon), while
perturbing $\teff$, $\lgg$, and $\vmic$ individually
by the uncertainties listed in \tab{tab:stars}.
Systematic uncertainties were neglected for the Sun.
The final uncertainties were obtained by adding 
these difference sources of uncertainty together in quadrature.
These uncertainties are presented in \tab{tab:errors}.
As seen therein, the sensitivity to $\vmic$
reaches almost to zero for the 3D non-LTE runs for the Sun and
HD~84937 because these models
do not adopt any extra broadening parameters;
for Procyon and HD~140283, the uncertainty is slightly non-zero
because of the $0$th-order extrapolation in 
stellar parameters.

The 1D non-LTE analysis proceeded in almost the same way as the 3D
non-LTE one. The only difference was that the 1D non-LTE interpolation routine
was used instead of the 3D non-LTE one, as discussed in
\sect{interpolator}.

\subsection{Results}
\label{gbs_abundance}

The results for $\lgeps{Fe}$ are illustrated
in \fig{fig:abundances1} and \fig{fig:abundances2}.
Estimates of $\feh$ from the 
1D LTE, 1D non-LTE, and 
3D non-LTE models are
given in \tab{tab:stars}. These are based on the weighted means of
$\feh_{\text{\ion{Fe}{I}}}$ and $\feh_{\text{\ion{Fe}{II}}}$, 
with the uncertainties summarised 
\tab{tab:errors}, and adopting the
1D LTE, 1D non-LTE, and 3D non-LTE solar values 
(respectively) separately for the $\lgeps{Fe}_{\text{\ion{Fe}{I}}}$ and
$\lgeps{Fe}_{\text{\ion{Fe}{II}}}$ derived here and 
given in \tab{tab:abundance}.
In practice, the \ion{Fe}{II} lines
are given a much higher weight than the \ion{Fe}{I} lines
in these weighted means
because of the high
sensitivity of the \ion{Fe}{I} lines 
to $\teff$ (\tab{tab:errors}).

We make some general remarks here, before discussing
each star individually in
Sections \ref{gbs_abundance_sun}---\ref{gbs_abundance_hd140283}.
There are significant 3D non-LTE effects on $\ion{Fe}{I}$ or 
$\ion{Fe}{II}$, or both, for all four stars.
After averaging over lines, the 3D non-LTE effects act to increase 
$\lgeps{Fe}$ and correspond to negative values of $\abdl$ overall,
as is generally expected of \ion{Fe}{I} lines with $\elo>2\,\eV$ 
and of \ion{Fe}{II} lines (\sect{results_elo_fe2}).
For \ion{Fe}{I}, the largest difference is
for HD 84937 ($0.35\,\dex$),
while for \ion{Fe}{II} the largest difference
is for HD 140283 ($0.15\,\dex$).

Comparing the 1D LTE, 1D non-LTE, and 3D non-LTE
results helps disentangle the 3D effects and the non-LTE 
effects, to some extent.
For \ion{Fe}{II}, the line-averaged 
1D non-LTE results agree with the 1D LTE ones to $0.02\,\dex$
(\tab{tab:abundance});
these lines are primarily susceptible to 3D effects
(\sect{results_elo_fe2}).
There is more variation for \ion{Fe}{I}.
For the Sun and Procyon (HD~61421), 
the 1D LTE and 1D non-LTE results are within $0.03\,\dex$. 
This is due to the competition between over-ionisation and photon
losses in the 3D models at high $\xh{M}$
(\sect{results_elo_fe1}).
For the metal-poor stars HD 84937 and HD 140283,
the 1D non-LTE results 
are in between
the 1D LTE and 3D non-LTE ones, as the steeper
gradients in the 3D models enhance the non-LTE effects
\citep{2016MNRAS.463.1518A}.

Significant ionisation imbalances relative to the stipulated uncertainties may
reflect modelling errors since such errors were deliberately not taken into
account in \sect{gbs_analysis}.  
As noted in \sect{results_elo_fe2},
the 3D non-LTE effects on \ion{Fe}{I} lines of
high $\elo$ and on \ion{Fe}{II} lines 
usually have the same sign, which helps to counteract
ionisation imbalances in 1D LTE.
Despite this, in the current 1D LTE analysis,
ionisation balance is not met for any of the stars
as can be seen in \tab{tab:imbalance}
as well as \fig{fig:abundances1} and \fig{fig:abundances2}.
For Procyon, the imbalance in 1D LTE
is $-0.07\,\dex$, or $-1.2\,\sigma$;
whereas, for the other three stars, it amounts to $-2\,\sigma$.
For HD~84937 and HD~140283,
these 1D LTE imbalances correspond to underestimating
$\lgg$ by around $0.3\,\dex$.
In contrast, the 3D non-LTE models achieve ionisation balance for
the Sun, Procyon, and HD~140283 within the uncertainties.
Although HD~84937 shows an imbalance in 3D non-LTE 
that is $0.11\,\dex$, or $1.4\,\sigma$, this
is $0.03\dex$ less severe than that in 1D LTE.

Excitation imbalances in \ion{Fe}{I} lines 
(trends in $\lgeps{Fe}$ with $\elo$) may similarly reflect modelling errors.
For the Sun and Procyon, the 1D LTE
excitation imbalances amount to $0.09\,\dex$ 
as $\elo$ increases from $0\,\eV$ to $5\,\eV$.
In 3D non-LTE, this is reduced for the Sun
($0.07\,\dex$), but slightly increased for Procyon ($0.10\,\dex$).
For the two metal-poor stars,
the 1D LTE $\lgeps{Fe}$ changes by
$0.19$ for HD~84937, and $0.28\,\dex$ for HD~140283
as $\elo$ increases from $0\,\eV$ to $5\,\eV$;
flattening this slope would require one to reduce 
$\teff$ by $250\,\K$ and $300\,\K$, respectively.
In 3D non-LTE, these abundance slopes are markedly reduced.
They amount to just 
$0.04\,\dex$ across $5\,\eV$ for HD~84937, corresponding to around $50\,\K$
which is well within the formal uncertainty of $97\,\K$
(\tab{tab:stars}), while 
HD~140283 shows no significant trend at all.

Thus in terms of ionisation and excitation balance,
the 3D non-LTE models tend to outperform the 1D LTE ones.
The residual imbalances most likely reflect
remaining uncertainties in the 3D non-LTE models, 
possibly in the non-LTE model atom itself.
At low $\xh{M}$, different prescriptions of the inelastic hydrogen
collisions can change the inferred iron abundances
by $0.03\,\dex$ for dwarfs in 1D non-LTE
\citep{2019A&A...631A..43M}. 
These differences would be exacerbated
in the 3D non-LTE models of warmer metal-poor stars (such as HD~84937),
due to their steeper temperature gradients.
In any case, the residual imbalances are
typically smaller than the difference between
$\lgeps{Fe}$ inferred in 1D LTE or 1D non-LTE compared to
in 3D non-LTE.
The 3D non-LTE models may thus still lead to more accurate
determinations of $\lgeps{Fe}$ overall.

\subsubsection{Sun}
\label{gbs_abundance_sun}

For the Sun, \citet{2021A&A...653A.141A} determined
$\lgeps{Fe}=7.46\pm0.04$, with the stipulated uncertainties
being dominated by systematics in the models.
This is in excellent agreement with the 3D non-LTE value
determined here, $\lgeps{Fe}=7.47\pm0.01$, where systematic
modelling uncertainties have been neglected. 
This consistency is reassuring for the grid-based approach
adopted here, given 
that both studies use the same model atom. Nevertheless, 
it is preferable to use
the value of \citet{2021A&A...653A.141A} as
their analysis is superior to that presented here in a number of 
ways.  In particular,
their analysis was based on disc-centre intensity observations
rather than the disc-integrated solar flux used here;
a more up-to-date line list for \ion{Fe}{I}; 
and a tailored 3D \stagger{} model, thus avoiding
interpolation in stellar parameters
(with the interpolation routine adding some scatter to 
our results here; \sect{interpolator}).

\subsubsection{Procyon}
\label{gbs_abundance_procyon}

Procyon (HD~61421) is found to be $0.11\,\dex$ above solar.
This estimate is significantly 
larger than, to our knowledge, previous estimates based on
1D LTE and non-LTE \citep{2011A&A...528A..87M},
\mtd{} LTE and non-LTE \citep{2012MNRAS.427...27B},
and 3D LTE \citep{2002ApJ...567..544A}.
Different authors adopt different stellar parameters, $\vmic$,
line lists, and analysis approaches, so it is difficult to determine
exactly from where this discrepancy arises.
There is a slightly positive trend with $\reqw$ for 
\ion{Fe}{I} (\fig{fig:abundances1}),
which possibly reflects uncertainties in the upper layers
of the 3D model stellar atmospheres
\citep{2002ApJ...567..544A}.
Nevertheless, the weakest \ion{Fe}{I} lines agree well with the
\ion{Fe}{II} lines that form deeper in the atmosphere, for which
a positive trend with $\reqw$ is not seen,
and which are given the larger weight in the determination of $\feh$.

The 3D non-LTE estimate of $\feh$ would now imply that Procyon
is significantly $\upalpha$-poor, $\xfe{\upalpha}\approx-0.1$, 
at least when combined with 
1D non-LTE estimates of magnesium \citep{2017ApJ...847...15B}
and calcium \citep{2017A&A...605A..53M}.
This result may be in conflict with
previous 1D LTE studies of these elements in nearby FG-type dwarfs, which
indicate a plateau in $\upalpha$, or possibly even a slight upwards trend
as $\feh$ increases above solar metallicities 
(\citealt{2003A&A...410..527B}, Fig.~12 and Fig.~13).
However, this result would be consistent
with the negative 3D non-LTE trend found for
$\xfe{O}$ (Fig.~12 of \citealt{2019A&A...630A.104A}).
It would be interesting to revisit the abundances of these and 
other key elements in this star using tailored 3D non-LTE models.

As noted above, Procyon is around
$50\,\K$ outside of the $\abdl$ grid in $\teff$ and 
$0.15\,\dex$ outside of the $\abdl$ grid in $\lgeps{Fe}_{\text{3N}}$
(\sect{gbs_analysis}), and that
$0$th-order extrapolation was applied for these parameters.
The inspection of the trends for \ion{Fe}{I} indicates that
linear extrapolation in $\teff$ would lead to slightly larger 
abundance differences for \ion{Fe}{I}, while 
linear extrapolation in $\lgeps{Fe}$ would lead to slightly smaller
abundance differences for \ion{Fe}{I} and \ion{Fe}{II}.
These changes would be of the order $\pm0.02\,\dex$
at most for \ion{Fe}{I}, and somewhat less severe for \ion{Fe}{II}. These are 
not enough to explain the much larger value of $\feh$ found here.

\subsubsection{HD~84937}
\label{gbs_abundance_hd84937}

\citet{2016MNRAS.463.1518A} determined 
$\feh=-1.90\pm0.06$ from \ion{Fe}{I} and
$\feh=-1.97\pm0.02$ from \ion{Fe}{II} for HD~84937 in 3D non-LTE.
As for HD~140283 (\sect{gbs_abundance_hd140283}),
that study was based on an older version of the model atom
used here \citep{2017MNRAS.468.4311L},
as well as a tailored 3D model stellar atmosphere.
The authors 
also adopted a slightly smaller value of $\lgg$ ($4.06\,\dex$)
than that used here ($4.13\,\dex$; \citealt{2021A&A...650A.194G}).
The corresponding values found in this work are 
in good agreement within the uncertainties:
$-1.86\pm0.07$ and $-1.97\pm0.03$, respectively.

\citet{2016MNRAS.463.1518A} have also discussed how 1D LTE
severely underestimates $\lgeps{Fe}$ in HD~84937 relative
to 3D non-LTE. This is verified here, with 1D LTE 
giving $0.35\,\dex$ and $0.10\,\dex$ smaller values of 
$\lgeps{Fe}$ compared to 3D non-LTE from
\ion{Fe}{I} and \ion{Fe}{II}, respectively.
For \ion{Fe}{I},
the 1D non-LTE models give results that 
are improved compared to 1D LTE,
with $\lgeps{Fe}$ now only being $0.18\,\dex$ 
smaller than the 3D non-LTE result.
Thus for \ion{Fe}{I}
1D, non-LTE models should be used when 3D non-LTE models are unavailable.

As mentioned
above (\sect{gbs_abundance}), the ionisation balance
is not quite satisfactory at $+0.11\pm0.08$
($\lgeps{Fe}$ from \ion{Fe}{I} minus that from \ion{Fe}{II});
although, this is less severe than in 1D LTE at $-0.14\pm0.07$.
The sign of the imbalance flips, and 
thus it is possible that at low $\xh{M}$ the current models predict
slight departures that are too large
from LTE in \ion{Fe}{I}, perhaps due to
inelastic hydrogen collisions that are too inefficient. This would be exacerbated by
the steeper gradients in the metal-poor 3D model stellar atmospheres.
If this is indeed the case,
some of these model shortcomings would be masked by 
the shallower gradients in the metal-poor 1D models.

For this star,
the 1D non-LTE models give a good ionisation balance
of $0.05\pm0.07\,\dex$.
However, it should be noted that the residual imbalance in 3D non-LTE
of $0.11\,\dex$ is much smaller than
the difference in $\lgeps{Fe}$ between 1D non-LTE and 3D non-LTE,
which is $0.18\,\dex$ for \ion{Fe}{I}.
Thus the 3D non-LTE model may still give a more reliable result overall,
despite its remaining shortcomings.

\subsubsection{HD~140283}
\label{gbs_abundance_hd140283}

\citet{2016MNRAS.463.1518A} determined 
$\feh=-2.34\pm0.07$ from \ion{Fe}{I} and
$\feh=-2.28\pm0.04$ from \ion{Fe}{II} for HD~140283 in 3D non-LTE.
As for HD~84937 (\sect{gbs_abundance_hd84937}),
that study was based on an older model atom \citep{2017MNRAS.468.4311L}
and a tailored 3D model stellar atmosphere.
However, the authors adopted a much smaller $\teff$ ($5591\,\K$)
than that used here 
($5792\,\K$; see Fig.~2 of \citealt{2018MNRAS.475L..81K}
and the accompanying discussion).
The corresponding values found in this work are
$\feh=-2.24\pm0.05$ and $\feh=-2.29\pm0.03$, respectively.
For \ion{Fe}{II,} the result is in good agreement, being
relatively insensitive to differences in $\teff$
(\tab{tab:errors}), while much of the difference in \ion{Fe}{I}
can be attributed to the $200\,\K$ larger value of $\teff$.

Despite the different $\teff$, the large error in 1D LTE
found by \citet{2016MNRAS.463.1518A} is confirmed here, with 1D LTE 
giving 
$0.30\,\dex$ and $0.13\,\dex$ smaller values of $\lgeps{Fe}$ than 
3D non-LTE from \ion{Fe}{I} and \ion{Fe}{II}, respectively.
As for HD~84937 (\sect{gbs_abundance_hd84937}),
the 1D non-LTE models give results that are 
improved over 1D LTE,
with $\lgeps{Fe}$ now only being $0.15\,\dex$ 
smaller than the 3D non-LTE result. Thus for \ion{Fe}{I}
1D, non-LTE models should be used when 3D non-LTE models are unavailable.

While ionisation and excitation balance are achieved in 3D non-LTE
(\fig{fig:abundances1} and \fig{fig:abundances2}),
it should again be noted that HD~140283 is $0.35\,\dex$ outside of the
$\abdl$ grid in $\lgg$ (\sect{gbs_analysis}),
and so $0$th-order extrapolation was applied for this parameter.
Inspection of the trends for \ion{Fe}{I} indicates that
linear extrapolation in $\lgg$ would lead to slightly larger 
abundances from \ion{Fe}{I} and 
slightly smaller abundances from \ion{Fe}{II},
overall worsening the ionisation balance by around $0.03\,\dex$.
Therefore, if linear extrapolation is more valid than $0$th-order
extrapolation, these results could indicate that 
the departures from LTE in \ion{Fe}{I} are slightly over-estimated 
by the current models, at least at low $\xh{M}$.
This would be in line with the results for
HD~84937 (\sect{gbs_abundance_hd84937}).
As stated there, if this is indeed the case,
some of these model shortcomings would be masked by 
the shallower gradients in the 1D models.

\subsection{Interpolation and extrapolation in line parameters}
\label{gbs_extrapolation}

The $\abdl$ grid is unfortunately somewhat
limited in the number of lines, especially for \ion{Fe}{II} 
(\fig{fig:golden}). 
This is particularly problematic at low $\xh{M}$,
where many of the lines become too weak to 
be observed (lower right panel of \fig{fig:aberr_reqw2}).
Thus, the results shown in \fig{fig:abundances1} and 
\fig{fig:abundances2} for the metal-poor stars
clearly rely on extrapolation on an irregular grid of
line parameters.

To test the possible
impact of interpolation and extrapolation errors in line parameters,
the analysis of the two metal-poor stars was repeated 
with two modifications.
For \ion{Fe}{I} lines, the analysis was restricted to only those lines
in the grid (\sect{method_abundance}).
For \ion{Fe}{II} lines, the extended 3D LTE grid of
\citet{2019A&A...630A.104A} was used instead;
this contains all of the \ion{Fe}{II}
lines studied here.

The results of this
analysis are plotted in \fig{fig:abundances3}.
Compared to the original analysis,
the mean values of $\lgeps{Fe}$ from \ion{Fe}{I}
lines decreases by $0.03\,\dex$ and $0.04\,\dex$
for HD~84937 and HD~140283, respectively. It is not clear
if this is caused by extrapolation errors or if it is simply a selection
effect; nevertheless, the shift is
within the uncertainties due to $\teff$.
The 3D non-LTE ionisation balance
is further improved as a consequence of the shifts in \ion{Fe}{I};
although, the \ion{Fe}{I} excitation balance is slightly worsened.

Looking more closely at the \ion{Fe}{II} results,
the largest line-by-line abundance difference are $0.046\,\dex$ 
and the mean absolute abundance difference is $0.02\,\dex$ for HD~140283;
the differences are around half as severe for HD~84937. 
This is consistent with the expected 
errors discussed in \sect{interpolator}.
After averaging over \ion{Fe}{II} lines, the overall abundance difference 
between the two approaches was $-0.005\,\dex$, for both stars,
which may in part reflect non-LTE effects (\sect{results_elo_fe2}).

Overall, it appears that the main conclusions of this study
are not affected by the limited number of lines in the 
$\abdl$ grid.  Namely, the ionisation and excitation
balance are significantly improved in 3D (non-)LTE compared to 1D LTE.

\section{Conclusion}
\label{conclusion}

We have studied 1D LTE versus 3D non-LTE abundance differences
across an extended range of stellar parameters and 
\ion{Fe}{I} and \ion{Fe}{II} optical lines.
This work extends upon the 
grid of 1D non-LTE abundance corrections of 
\citet{2012MNRAS.427...50L},
the investigation of specific benchmark stars in
\citet{2016MNRAS.463.1518A} and \citet{2017MNRAS.468.4311L},
and the 3D LTE grid for \ion{Fe}{II} lines
across an even broader parameter
space in \citet{2019A&A...630A.104A}.

The results in the present study
as well as in those earlier works
clearly indicate that 1D LTE is a poor approximation
for \ion{Fe}{I} and \ion{Fe}{II}.
As expected, the errors tend to be most severe for warm, metal-poor dwarfs.
For HD~84937, the 1D LTE values of $\lgeps{Fe}$ are
$0.35\,\dex$ and $0.10\,\dex$ 
smaller than the 3D non-LTE ones
for \ion{Fe}{I} and \ion{Fe}{II}, respectively.
Similar values ($0.30\,\dex$ and $0.15\,\dex$) are
found for HD~140283.
However, the errors can also be significant for metal-rich dwarfs.
For Procyon (HD~61421), the 1D LTE values are $0.20\,\dex$ and
$0.08\,\dex$ smaller than the 3D non-LTE ones for \ion{Fe}{I} and \ion{Fe}{II},
respectively, when adopting the literature value
of $\vmic{}=1.66\,\kms$.

There are a number of limitations to this study.
First, the range of stellar parameters and line parameters
is unfortunately rather limited.  Future studies might focus
on extending the calculations to giants,
given their usefulness for probing the properties
of the Galaxy at larger distances, as well as for studying dwarf galaxies.
Such studies might also consider extending the line list
to include more useful lines at low $\xh{M}$,
particularly for \ion{Fe}{II}. The lines used by
\citet{2016MNRAS.463.1518A} in their analysis of metal-poor
dwarfs and giants could be useful for this purpose.
Secondly, as discussed above, there are some hints that the 3D non-LTE effects
are slightly overestimated towards smaller $\xh{M}$.
These residual errors in the models
are smaller than the differences between 1D LTE and 
3D non-LTE determinations of $\lgeps{Fe}$; as such,
these 3D non-LTE models should still give more realistic results overall.
Nevertheless, future work may look at refining the model iron atom
and the 3D model atmospheres to try to resolve these
remaining discrepancies.

We have made the interpolation routine for 1D LTE versus 3D non-LTE abundance
differences (\sect{interpolator}) 
publicly available\footnote{\url{https://github.com/sliljegren/1L-3NErrors}}.
We caution that this interpolation
routine is based on a model that covers a rather restricted range of 
3D models (\sect{method_atmosphere})
and optical lines (\sect{method_abundance}), and it might
not give reliable results for parameters far outside of this grid.
In particular, for \ion{Fe}{II} lines which form close to LTE 
when $\xh{M}\gtrsim-2$, the more
extended 3D LTE grid presented in \citet{2019A&A...630A.104A} could
also be useful.

Finally, following \citet{2020A&A...642A..62A}, we have 
constructed new grids of 1D non-LTE departure
coefficients and made them publicly 
available\footnote{\url{https://zenodo.org/record/7088951}}.
This is an update to the grid constructed by
\citet{2016MNRAS.463.1518A} that was used in
the second and third data releases of GALAH
\citep{2018MNRAS.478.4513B,2021MNRAS.506..150B}.
The new data cover dwarfs and giants of spectral types
FGKM (Fig.~1 of \citealt{2020A&A...642A..62A}).  
They can be used readily with the 1D spectral synthesis code \sme{}
\citep{2017A&A...597A..16P,2022arXiv220611070L}, and could also be adapted to
be used with \synspec{}
\citep{2021arXiv210402829H,2022ApJ...928..173O} and more recently
\turbo{} \citep{2022arXiv220600967G}.

\begin{acknowledgements}
    The authors thank Bengt Gustafsson and the anonymous referee for their
    comments and suggestions on the manuscript.  AMA acknowledges support from
    the Swedish Research Council (VR 2020-03940).  This research was supported
    by computational resources provided by the Australian Government through the
    National Computational Infrastructure (NCI) under the National Computational
    Merit    Allocation Scheme and the ANU Merit Allocation Scheme (project
    y89).  Some of the computations were also enabled by resources provided by
    the Swedish National Infrastructure for Computing (SNIC) at the
    Multidisciplinary Center for Advanced Computational Science (UPPMAX) and at
    the High Performance Computing Center North (HPC2N) partially funded by the
    Swedish Research Council through grant agreement no.  2018-05973. 
\end{acknowledgements}

\bibliographystyle{aa} 
\bibliography{bibl.bib}

\end{document}